\documentclass[twocolumn]{aastex63}
\submitjournal{ApJ}
\usepackage{threeparttablex}
\usepackage{setspace}
\usepackage{catchfile}
\usepackage{graphicx}
\usepackage{subfig}
\usepackage{float}
\usepackage{amsmath}
\hypersetup{colorlinks,linkcolor={blue},citecolor={blue},urlcolor={blue}} 

\DeclareUnicodeCharacter{2212}{-}

\def\lapp{\ifmmode\stackrel{<}{_{\sim}}\else$\stackrel{<}{_{\sim}}$\fi}
\def\gapp{\ifmmode\stackrel{>}{_{\sim}}\else$\stackrel{>}{_{\sim}}$\fi}

\newcommand{\tempo}{{\texttt{TEMPO}}}
\newcommand{\presto}{{\texttt{PRESTO}}}

\newcommand{\degrees}{^{\circ}}

\begin{document}

\title{Study of 72 pulsars discovered in the PALFA survey: Timing analysis, glitch activity, emission variability, and a pulsar in an eccentric binary }
\correspondingauthor{E.~Parent}
\author[0000-0002-0430-6504]{E.~Parent}\email{parente@physics.mcgill.ca}
\affiliation{Dept.~of Physics and McGill Space Institute, McGill Univ., Montreal, QC H3A 2T8, Canada}

\author{H.~Sewalls}
\affiliation{Dept.~of Physics and McGill Space Institute, McGill Univ., Montreal, QC H3A 2T8, Canada}

\author[0000-0003-1307-9435]{P. C. C.~Freire}
\affiliation{Max-Planck-Institut f\"{u}r Radioastronomie, Auf dem H\"{u}gel 69, Bonn 53121, Germany}

\author{T.~Matheny}
\affiliation{Department of Physics and Astronomy, West Virginia University, Morgantown, WV 26506-6315, USA }
\affiliation{Center for Gravitational Waves and Cosmology, Chestnut Ridge Research Building, Morgantown, WV 26505, USA}

\author{A. G.~Lyne}
\affiliation{Jodrell Bank Centre for Astrophysics, School of Physics and Astronomy, University of Manchester, Manchester, M13 9PL, UK}

\author[0000-0002-8509-5947]{B.~B.~P.~Perera}
\affiliation{Arecibo Observatory, HC3 Box 53995, Arecibo, PR 00612, USA}

\author{F.~Cardoso}
\affiliation{Department of Physics and Astronomy, West Virginia University, Morgantown, WV 26506-6315, USA }
\affiliation{Center for Gravitational Waves and Cosmology, Chestnut Ridge Research Building, Morgantown, WV 26505, USA}

\author[0000-0001-7697-7422]{M.~A.~McLaughlin}
\affiliation{Department of Physics and Astronomy, West Virginia University, Morgantown, WV 26506-6315, USA }
\affiliation{Center for Gravitational Waves and Cosmology, Chestnut Ridge Research Building, Morgantown, WV 26505, USA}

\author[0000-0003-4285-6256]{B.~Allen}
\affiliation{Max-Planck-Institut für Gravitationsphysik (Albert-Einstein-Institut), D-30167 Hannover, Germany}
\affiliation{Department of Physics, University of Wisconsin - Milwaukee, Milwaukee WI 53211, USA}

\author[0000-0001-6341-7178]{A.~Brazier}
\affiliation{Cornell Center for Astrophysics and Planetary Science, Ithaca, NY 14853, USA}

\author[0000-0002-1873-3718]{F.~Camilo}
\affiliation{South African Radio Astronomy Observatory, Observatory, 7925, South Africa}

\author[0000-0002-2878-1502]{S.~Chatterjee}
\affiliation{Cornell Center for Astrophysics and Planetary Science, Ithaca, NY 14853, USA}

\author[0000-0002-4049-1882]{J. M.~Cordes}
\affiliation{Cornell Center for Astrophysics and Planetary Science, Ithaca, NY 14853, USA}

\author[0000-0002-2578-0360]{F.~Crawford}
\affiliation{Department of Physics and Astronomy, Franklin and Marshall College, Lancaster, PA 17604-3003, USA}

\author[0000-0003-1226-0793]{J. S.~Deneva}
\affiliation{George Mason University, resident at the Naval Research Laboratory, Washington, DC 20375}

\author[0000-0003-4098-5222]{F.~A.~Dong}
\affiliation{Dept. of Physics and Astronomy, University of British Columbia, 6224 Agricultural Road, Vancouver, BC V6T 1Z1, Canada}

\author[0000-0002-2223-1235]{R. D.~Ferdman}
\affiliation{School of Chemistry, University of East Anglia, Norwich Research Park, Norwich NR4 7TJ, UK}

\author[0000-0001-8384-5049]{E.~Fonseca}
\affiliation{Department of Physics and Astronomy, West Virginia University, Morgantown, WV 26506-6315, USA }
\affiliation{Center for Gravitational Waves and Cosmology, Chestnut Ridge Research Building, Morgantown, WV 26505, USA}
\affiliation{Dept.~of Physics and McGill Space Institute, McGill Univ., Montreal, QC H3A 2T8, Canada}

\author[0000-0003-2317-1446]{J. W. T.~Hessels}
\affiliation{ASTRON, the Netherlands Institute for Radio Astronomy, Oude Hoogeveensedijk 4, 7991 PD Dwingeloo, The Netherlands}
\affiliation{Anton Pannekoek Institute for Astronomy, University of Amsterdam, Postbus 94249, 1090 GE Amsterdam, The Netherlands}

\author[0000-0001-9345-0307]{V.~M.~Kaspi}
\affiliation{Dept.~of Physics and McGill Space Institute, McGill Univ., Montreal, QC H3A 2T8, Canada}

\author[0000-0003-3168-0929]{B.~Knispel}
\affiliation{Max-Planck-Institut für Gravitationsphysik (Albert-Einstein-Institut), D-30167 Hannover, Germany}

\author[0000-0001-8503-6958]{J.~van Leeuwen} 
\affiliation{ASTRON, the Netherlands Institute for Radio Astronomy, Oude Hoogeveensedijk 4, 7991 PD Dwingeloo, The Netherlands}
\affiliation{Anton Pannekoek Institute for Astronomy, University of Amsterdam, Postbus 94249, 1090 GE Amsterdam, The Netherlands}

\author[0000-0001-5229-7430]{R.~S.~Lynch}
\affiliation{Green Bank Observatory, P.O. Box 2, Green Bank, WV 24494, USA}

\author[0000-0001-8845-1225]{B.~M.~Meyers}
\affiliation{Dept. of Physics and Astronomy, University of British Columbia, 6224 Agricultural Road, Vancouver, BC V6T 1Z1, Canada}

\author[0000-0002-2885-8485]{J.~W.~McKee}
\affiliation{Canadian Institute for Theoretical Astrophysics, University of Toronto, 60 Saint George Street, Toronto, ON M5S 3H8, Canada}

\author{M.~B.~Mickaliger}
\affiliation{Jodrell Bank Centre for Astrophysics, School of Physics and Astronomy, University of Manchester, Manchester, M13 9PL, UK}

\author[0000-0003-3367-1073]{C.~Patel}
\affiliation{Dept.~of Physics and McGill Space Institute, McGill Univ., Montreal, QC H3A 2T8, Canada}
\affiliation{Dunlap Institute for Astronomy \& Astrophysics, University of Toronto, 50 St. George Street, Toronto, ON M5S 3H4, Canada}

\author[0000-0001-5799-9714]{S.~M.~Ransom}
\affiliation{National Radio Astronomy Observatory, 520 Edgemont Rd., Charlottesville, VA 22903, USA}

\author{A.~Rochon}
\affiliation{Dept.~of Physics and McGill Space Institute, McGill Univ., Montreal, QC H3A 2T8, Canada}

\author[0000-0002-7374-7119]{P.~Scholz}
\affiliation{Dunlap Institute for Astronomy \& Astrophysics, University of Toronto, 50 St. George Street, Toronto, ON M5S 3H4, Canada}

\author[0000-0001-9784-8670]{I.~H.~Stairs}
\affiliation{Dept. of Physics and Astronomy, University of British Columbia, 6224 Agricultural Road, Vancouver, BC V6T 1Z1, Canada}

\author[0000-0001-9242-7041]{B.~W.~Stappers}
\affiliation{Jodrell Bank Centre for Astrophysics, School of Physics and Astronomy, University of Manchester, Manchester, M13 9PL, UK}

\author[0000-0001-7509-0117]{C.~M.~Tan}
\affiliation{Dept.~of Physics and McGill Space Institute, McGill Univ., Montreal, QC H3A 2T8, Canada}

\author[0000-0001-5105-4058]{W.~W.~Zhu}
\affiliation{CAS Key Laboratory of FAST, NAOC, Chinese Academy of Sciences, Beijing 100101, China}

\begin{abstract}
We present new discoveries and results from long-term timing of 72 pulsars discovered in the Arecibo PALFA survey, including precise determination of astrometric and spin parameters, and flux density and scatter broadening measurements at 1.4\,GHz.  Notable discoveries include two young pulsars (characteristic ages $\sim$30\,kyr) with no apparent supernova remnant associations, three mode changing, 12 nulling and two intermittent pulsars. We detected eight glitches in five pulsars. Among them is PSR~J1939+2609, an apparently old pulsar (characteristic age $\sim$1\,Gy), and PSR~J1954+2529, which likely belongs to a newly-emerging class of binary pulsars.  The latter is the only pulsar among the 72 that is clearly not isolated: a non-recycled neutron star with a 931\,ms spin period in an eccentric ($e\,=\,0.114$) wide ($P_b\,=\,82.7\,$d) orbit with a companion of undetermined nature having a minimum mass of $\sim 0.6\,M_{\odot}$. Since operations at Arecibo ceased in 2020 August, we give a final tally of PALFA sky coverage, and compare its 207 pulsar discoveries to the known population. On average, they are 50\% more distant than other Galactic plane radio pulsars; PALFA millisecond pulsars (MSP) have twice the dispersion measure per unit spin period than the known population of MSP in the Plane. The four intermittent pulsars discovered by PALFA more than double the population of such objects, which should help to improve our understanding of pulsar magnetosphere physics. The statistics for these, RRATS, and nulling pulsars suggest that there are many more of these objects in the Galaxy than was previously thought.\\
\end{abstract}

\section{Introduction} \label{sec:intro}
The observed population of radio pulsars currently numbers over 3000\footnote{According to version 1.65 of the ATNF Catalog \citep{mht+05}  available here:  \url{www.atnf.csiro.au/research/pulsar/psrcat/}.}. Approximately 500 of them have periods less than 100\,ms, 80\% of which are millisecond pulsars (MSPs) with periods $\lesssim$\,30\,ms, whereas the other 20\% are young or partially recycled neutron stars. The remaining objects are so-called ``normal'' pulsars. While discovering MSPs or partially recycled objects is important for fundamental physics experiments and neutron-star mass measurements (e.g., \citealt{nano11yrSB,afw+13,msf+15,agh+18, fcp+21}) and studies of binary evolution (see recent review by \citealt{dt20}), expanding the known population of normal pulsars is essential for understanding the neutron star population in terms of birth rates, magnetic fields, spatial distribution, and similar statistics (e.g. \citealt{fk06}), for probing the electron density (see review by \citealt{w96}) and magnetic fields (see review by \citealt{h17}) of the interstellar medium, and to gain insight into pulsar emission processes and associated plasma physics (see review by \citealt{cb17}). 

The wealth of astrophysical studies emerging from pulsar astronomy is often made possible by long-term monitoring and timing of newly discovered sources. Pulsar timing builds upon their remarkable rotational stability and consists of developing a mathematical model that accurately and precisely predicts the time of a pulse emitted by a pulsar when detected on Earth. This leads directly to high-precision measurements of spin, astrometric, dispersion and (if applicable) binary parameters. 

The Pulsar Arecibo L-band Feed Array (PALFA\footnote{\url{www.naic.edu/alfa/pulsar/}}) was a Galactic plane survey for radio pulsars and fast transients conducted from 2004 to 2020 at 1.4\,GHz with the Arecibo William E. Gordon 305-m telescope at the Arecibo observatory (AO) in Puerto Rico, USA. Extensive descriptions of the survey methodology can be found in the literature (e.g., \citealt{cfl+06,slm+14,lbh+15,pkr+18,pab+18}). Thanks to Arecibo's large collecting area and the high time and frequency resolution of PALFA data, the survey was particularly prolific in discovering highly dispersed MSPs \citep{crl+08,dfc+12,csl+12,akc+13,skl+15,kls+15,sab+16,pkr+19}. It was also among the best surveys for finding compact binaries \citep{pml+21} and/or highly accelerated systems, as evidenced by the discovery of three compact double neutron star systems (DNSs): PSR~J1906+0746, the youngest DNS known \citep{lfl+06,lks+15}; PSR~J1913+1102, a member of a new population of merging DNSs with large mass asymmetries \citep{lfa+16,ffp+20} and PSR~J1946+2052, the most compact DNS known in the Galaxy \citep{sfc+18}. PALFA was also the first pulsar survey to make use of volunteer distributed computing to search for compact binary systems \citep{akc+13}, which is the most computationally-intensive regions of parameter space.
Besides MSPs, the survey found several rotating radio transients  (RRATs; \citealt{dcm+09,pab+18}) and has  significantly contributed to our understanding of fast radio bursts (FRBs) with the discovery of the first repeating FRB 121102 \citep{sch+14,ssh+16,ssh+16b}.

In total, the survey discovered 207 pulsars\footnote{All PALFA discoveries can be found here: \url{www.naic.edu/~palfa/newpulsars/}.}, 46 of which are MSPs. 
PALFA has also discovered many slow pulsars: timing solutions for 66 of them were presented in \cite{nab+13}, \cite{lsf+17} and \cite{lsb+17}, with the latter describing two ``intermittent'' pulsars \citep{klo+06} discovered in the survey.

In this work, we present the results of several years of follow-up timing observations of an additional 72 long-period ($>100\,$ms) pulsars found by the PALFA survey. One pulsar, PSR~J1930$+$1722, was co-discovered in 2013 by the Parkes Northern Galactic Plane survey \citep{lcm13}.
Of the 72 sources being studied here, 32 were presented in some of the aforementioned PALFA publications, but only basic parameters were provided, with no timing solutions reported for any of them. These are, together with their previous names, listed in Table~\ref{tab:discoveries}. 
The remaining pulsars are new discoveries presented here for the first time. In addition to these discoveries, we present a set of 23 new pulsars that are not studied in detail in this work; they are listed, with some basic parameters, in Table~\ref{tab:new} and will be described in detail elsewhere. 

In Section \ref{sec:obs}, we give an overview of the discovery and timing observations, and we describe our data analysis, timing procedure and results in Section \ref{sec:timing}. Individual sources of interest are discussed in Section \ref{sec:psrs}.  Section \ref{sec:fermi} presents a search for gamma-ray pulsations in \textit{Fermi} data associated with our pulsars. We compare the properties of pulsars found by PALFA to those of the observed population of radio pulsars in the Galactic plane in Section \ref{sec:discussion}. Given the (early) termination of the PALFA survey in 2020 August, we provide an update on the survey status and sky coverage in Section \ref{sec:srv_status}. Finally, we summarize our results in Section \ref{sec:conclusion}.
\begin{spacing}{0.25}
\begin{table*}
  \centering
  {\scriptsize
  \caption{List of pulsar discoveries. For each source, 
  we indicate the previous name and respective reference,  the definitive name from the timing position obtained in this work, and the pipeline used for the discovery. The discovery dates provided here refer to the date when pulsars were identified in the data. PALFA references are, by order of publication:
   1: \cite{zbm+14},
   2: \cite{lbh+15},
   3: \cite{pkr+18},
   4: \cite{pab+18}, the non-PALFA reference is 5: \cite{lcm13}. If no reference is indicated, this work. 
   The pipelines used are: Quicklook \citep{s13}, Einstein@Home (E@H; \citealt{akc+13}) and the main PRESTO-based pipeline \citep{lbh+15},
   which includes a Fourier-domain periodicity search (FFT), a Fast Folding Algorithm (FFA; \citealt{pkr+18}) and a single-pulse search (SP; \citealt{dcm+09,pab+18}).\label{tab:discoveries} } 
  \setlength{\tabcolsep}{4.5mm}
  \begin{tabular}{l l l l l }
  \hline
Previous name(s) & Ref. & Name & Disc. date & Pipeline(s)  \\ [-0.3em]
 & \\  
\hline\hline
 & & J0608+1635 & 2013 04 18  & FFT, SP \\ [-0.3em] 
J1843+01 & (4) & J1843+0119 & 2017 10 01 & SP \\ [-0.3em] 
 & & J1849+0106 & 2013 11 25 & FFT, SP \\ [-0.3em] 
 & & J1849+0430 & 2016 10 14 & Quicklook \\ [-0.3em] 
 & & J1851+0241 & 2012 03 27 & E@H\\ [-0.3em] 
J1852+0000 & (3) & J1852$-$0000 & 2015 06 04 & FFT, Quicklook \\ [-0.3em] 
 & & J1853+0029 & 2015 09 23 & E@H \\ [-0.3em] 
J1853+03 & (2) & J1853+0259 & 2011 12 15 & FFT, SP \\ [-0.3em] 
J1853+04 & (4) & J1853+0427 & 2015 12 10 & SP \\ [-0.3em] 
J1854+00 & (1,2) & J1854+0050 & 2013 09 15 & FFT, SP \\ [-0.3em] 
 & & J1855+0306 & 2012 02 27 & E@H \\ [-0.3em] 
 & & J1855+0626 & 2018 06 08 & FFT\\ [-0.3em] 
J1856+0911 / J1856+09 & (3,4) & J1856+0912 & 2016 06 20 & FFA, SP\\ [-0.3em] 
J1858+02 & (2) & J1858+0239 & 2010 01 06 & SP \\ [-0.3em] 
 & & J1859+0345 & 2013 07 29 & E@H\\ [-0.3em] 
J1901+11 & (4) & J1902+1141 & 2017 05 25 & Quicklook, SP \\ [-0.3em] 
J1902+02 & (2) & J1902+0235 & 2014 02 24 & FFT \\ [-0.3em] 
J1903+04 / J1903+0415 & (1,2) & J1903+0415 & 2013 09 05 & FFT, SP\\ [-0.3em] 
 & & J1903+0912 & 2015 08 26 & Quicklook \\ [-0.3em] 
 & & J1904+0056 & 2018 09 19 & FFT \\ [-0.3em] 
 & & J1905+1034 & 2011 10 30 & FFT, SP \\ [-0.3em] 
J1906+0725 & (2) & J1906+0724 & 2013 09 10 & FFT \\ [-0.3em] 
 & & J1907+0833 & 2011 04 30 & Quicklook \\ [-0.3em] 
J1907+05 & (2) & J1908+0558 & 2014 07 27 & FFT \\ [-0.3em] 
Cand. J1908+13 & (4) & J1908+1351 & 2017 07 27 & SP \\ [-0.3em] 
 & & J1909+1205 & 2011 12 14 & E@H \\ [-0.3em] 
 & & J1910+0435 & 2015 09 28 & Quicklook \\ [-0.3em] 
 & & J1910+0710 & 2014 10 15 & E@H \\ [-0.3em] 
 & & J1910+1017 & 2011 11 04 & E@H \\ [-0.3em] 
J1910+1027 & (2) & J1910+1026 & 2011 11 05 & FFT \\ [-0.3em] 
J1911+09 & (2) & J1911+0921 & 2011 11 14 & FFT \\ [-0.3em] 
 & & J1911+0925 & 2012 07 23 & E@H \\ [-0.3em] 
J1911+10 & (2) & J1911+1051 & 2014 09 09 & FFT \\ [-0.3em] 
 & & J1911+1301 & 2015 12 08 & FFT \\ [-0.3em] 
 & & J1911+1336 & 2017 06 15 & FFT \\ [-0.3em] 
 & & J1913+0523 & 2017 10 10 & FFT \\ [-0.3em] 
J1913+1103 & (2) & J1913+11025$^*$ & 2011 09 15 & FFT \\ [-0.3em] 
 & & J1914+0625 & 2014 04 07 & FFT, SP \\ [-0.3em] 
 & & J1914+0805 & 2017 05 18 & FFT \\ [-0.3em] 
 & & J1914+0838 & 2012 10 09 & FFT \\ [-0.3em] 
 & & J1914+1428 & 2012 01 09 & E@H \\ [-0.3em] 
 & & J1915+0639 & 2014 01 24 & FFT, SP \\ [-0.3em] 
J1915+1144 & (2) & J1915+1145 & 2011 09 15 & FFT \\ [-0.3em] 
J1915+1149 & (2) & J1915+1150 & 2012 09 24 & FFT \\ [-0.3em] 
J1918+1310 & (2) & J1918+1311 & 2012 09 12 & FFT  \\ [-0.3em] 
 & & J1921+0921 & 2016 07 18 & Quicklook \\ [-0.3em] 
J1921+16 & (2) & J1921+1630 & 2014 02 07 & FFT \\ [-0.3em] 
J1924+1628 & (2) & J1924+1628 & 2014 01 14 & FFT \\ [-0.3em] 
J1924+17 & (2) & J1924+1713 & 2011 11 07 & FFT \\ [-0.3em] 
 & & J1924+1917 & 2017 19 19 & FFT \\ [-0.3em] 
J1926+1613 & (2) & J1926+1614 & 2014 03 10 & FFT \\ [-0.3em] 
 & & J1928+1725 & 2013 11 12 & FFT, SP\\ [-0.3em] 
J1930+14 & (2) & J1930+1408 & 2011 09 15 & FFT \\ [-0.3em] 
J1930+17 & (1,5) & J1930+1722 & - & - \\ [-0.3em] 
 & & J1931+1817 & 2015 02 21 & Quicklook \\ [-0.3em] 
J1934+19 & (2) & J1934+1926 & 2011 09 15 & FFT\\ [-0.3em] 
 & & J1935+1829 & 2015 09 25 & FFT \\ [-0.3em] 
J1936+20 & (2) & J1936+2042 & 2013 09 08 & FFT \\ [-0.3em] 
 & & J1938+2659 & 2018 11 17 & Quicklook \\ [-0.3em] 
 & & J1939+2609 & 2018 11 17 & Quicklook \\ [-0.3em] 
 & & J1948+1808 & 2018 12 13 & Quicklook \\ [-0.3em] 
 & & J1948+2819 & 2015 01 05 & E@H \\ [-0.3em] 
J1950+3000 & (3) & J1950+3001 & 2015 05 28 & Quicklook, FFA\\ [-0.3em] 
 & & J1952+2513 & 2011 10 27 & E@H \\ [-0.3em] 
J1952+3022 / J1952+30 & (3,4) & J1952+3021 & 2015 07 20 & FFA, SP\\ [-0.3em] 
 & & J1953+2819 & 2015 01 02 & E@H \\ [-0.3em] 
 & & J1954+2529 & 2014 10 24 & Quicklook \\ [-0.3em] 
 & & J1955+2930 & 2015 06 22 & E@H \\ [-0.3em] 
J1958+30 & (4) & J1958+3033 & 2015 09 03 & SP \\ [-0.3em] 
J2000+2921 / J2000+29 & (3,4) & J2000+2920 & 2015 11 24 & FFA, SP \\ [-0.3em] 
 & & J2003+2916 & 2015 01 25 & Quicklook \\ [-0.3em] 
 & & J2008+3139 & 2015 06 19 & Quicklook \\ 
\hline \\[-1.8em]
  \end{tabular}
  \tablenotetext{\scriptsize{*}}{\scriptsize Extra digit was added to the name to differentiate this pulsar from PSR~J1913+1102, a PALFA-discovered double neutron star system \citep{lfa+16,ffp+20}.}
  }
 
\end{table*}
\end{spacing}

%
\section{Observations} \label{sec:obs}
\subsection{Survey observations and discoveries} \label{sec:obs-disc}
Survey data were collected with the ALFA receivers, which consisted of seven beams with half power widths of 3.6$'$, and data were recorded with the Mock spectrometers\footnote{\url{www.naic.edu/~astro/mock.shtml}}. This backend processed two overlapping 172-MHz bands which, once combined and the two polarizations summed, provided 323\,MHz of bandwidth centered at 1375.5\,MHz and 960 frequency channels sampled every 64\,$\mu$s. PALFA targeted two regions of the Galactic plane ($| b |\,<\,5\degrees$): the ``inner'' Galaxy ($32\degrees\,\lesssim\,l\,\lesssim\,77\degrees$) and the ``outer'' Galaxy ($168\degrees\,\lesssim\,l\,\lesssim\,214\degrees$).  Integration times were 268\,s and 180\,s for inner and outer Galaxy observations, respectively. Additional details on survey observations for PALFA, including the strategy for surveying the inner and outer Galaxy regions, are provided in Section \ref{sec:srv_status}. 

Survey data have been processed and searched by three independent pipelines. The first is a reduced-resolution ``Quicklook'' pipeline performed in near real-time on site that enables rapid discovery of bright, nearby pulsars \citep{s13}. The second is a full-resolution \texttt{PRESTO}\footnote{\url{https://github.com/scottransom/presto}}-based pipeline \citep{presto}. The latter processes data on the B\'eluga supercluster, a Compute Canada/Calcul Qu\'ebec facility hosted at the \'Ecole de technologie sup\'erieure in Montr\'eal. It searches for dispersed periodic signals in the Fourier domain as well as in the time domain with a Fast-Folding Algorithm \citep{pkr+18}. Sporadic pulses such as those emitted by RRATs and FRBs are searched for with a single-pulse pipeline \citep{pab+18}. Lastly, data are searched for pulsars, particularly those in compact orbits, using the Einstein@Home\footnote{\url{https://einsteinathome.org/}} pipeline (E@H) described in \cite{akc+13}. 

Approximately 40\% of the 72 objects we study in this work were first found by either the Quicklook or E@H pipelines. The remaining sources were only found by the full-resolution pipeline. The pipeline search algorithm that identified each source is listed in Table~\ref{tab:discoveries} along with the date of discovery. The pulsars' time-integrated pulse profiles are shown in Figure~\ref{fig:profiles}. 

\begin{figure*}
    \begin{center}
        \centerline{\includegraphics[scale=0.77]{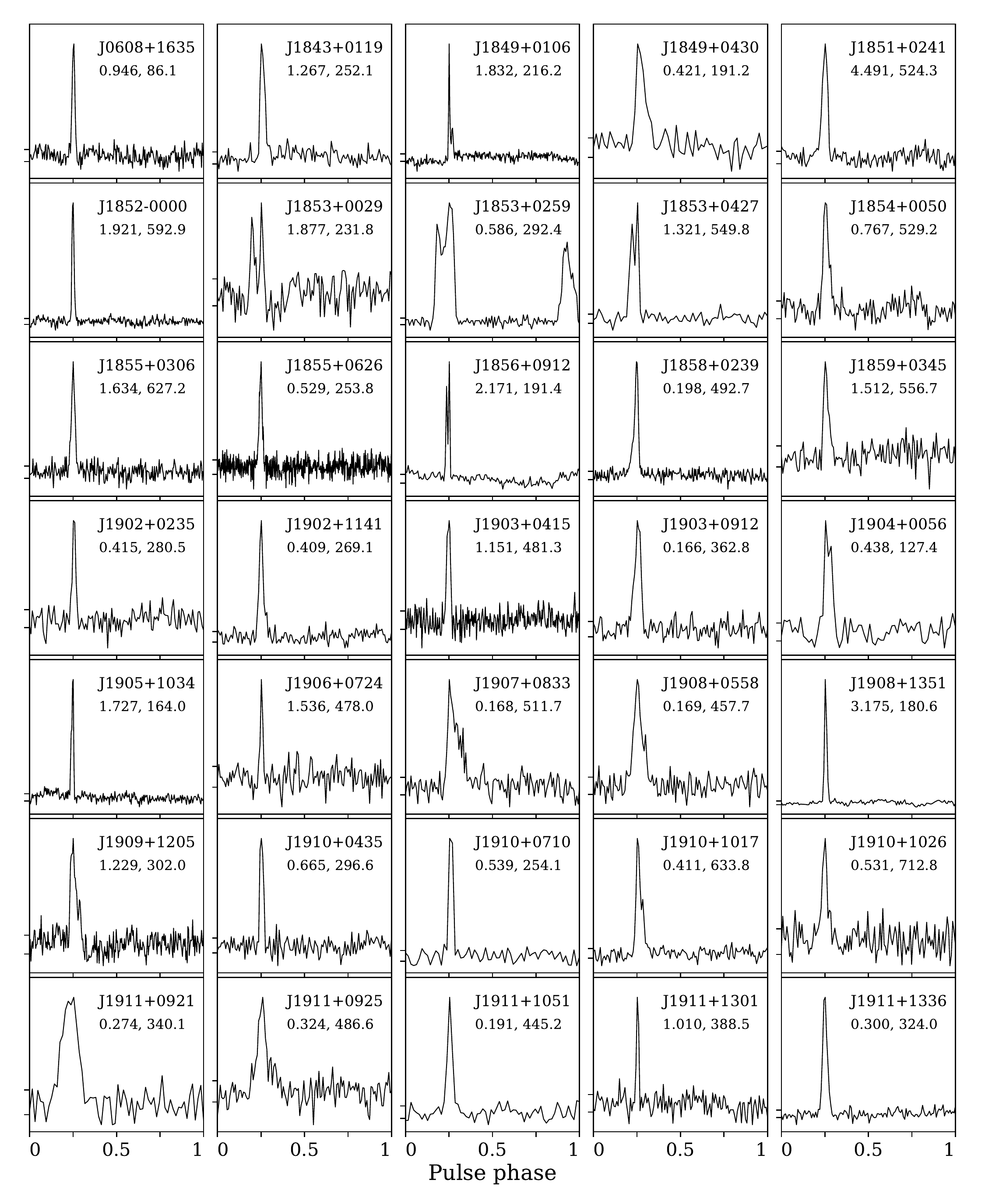}}
    \end{center}
    \vspace*{-10mm}
        \caption{Integrated pulse profiles, generated from AO data at 1.4\,GHz, from which template profiles were generated. The name, spin period (s) and DM (pc\,cm$^{-3}$) of each pulsar are listed above their corresponding profile. Depending on the pulsar brightness, the pulse phase resolution varies between 64 and 512 bins. 
        Tick marks on the vertical axis of each panel represent the root-mean-square intensity level of the off-source region. 
        We note that only the single-peak emission mode of PSR~J1858$+$0239 and the normal mode of PSR~J1914$+$0625 are plotted here. Other variants of their average profiles are shown in Sections \ref{sec:1858} and \ref{sec:1914}, respectively.  }
\end{figure*}
\begin{figure*}\ContinuedFloat
    \begin{center}
        \centerline{\includegraphics[scale=0.77]{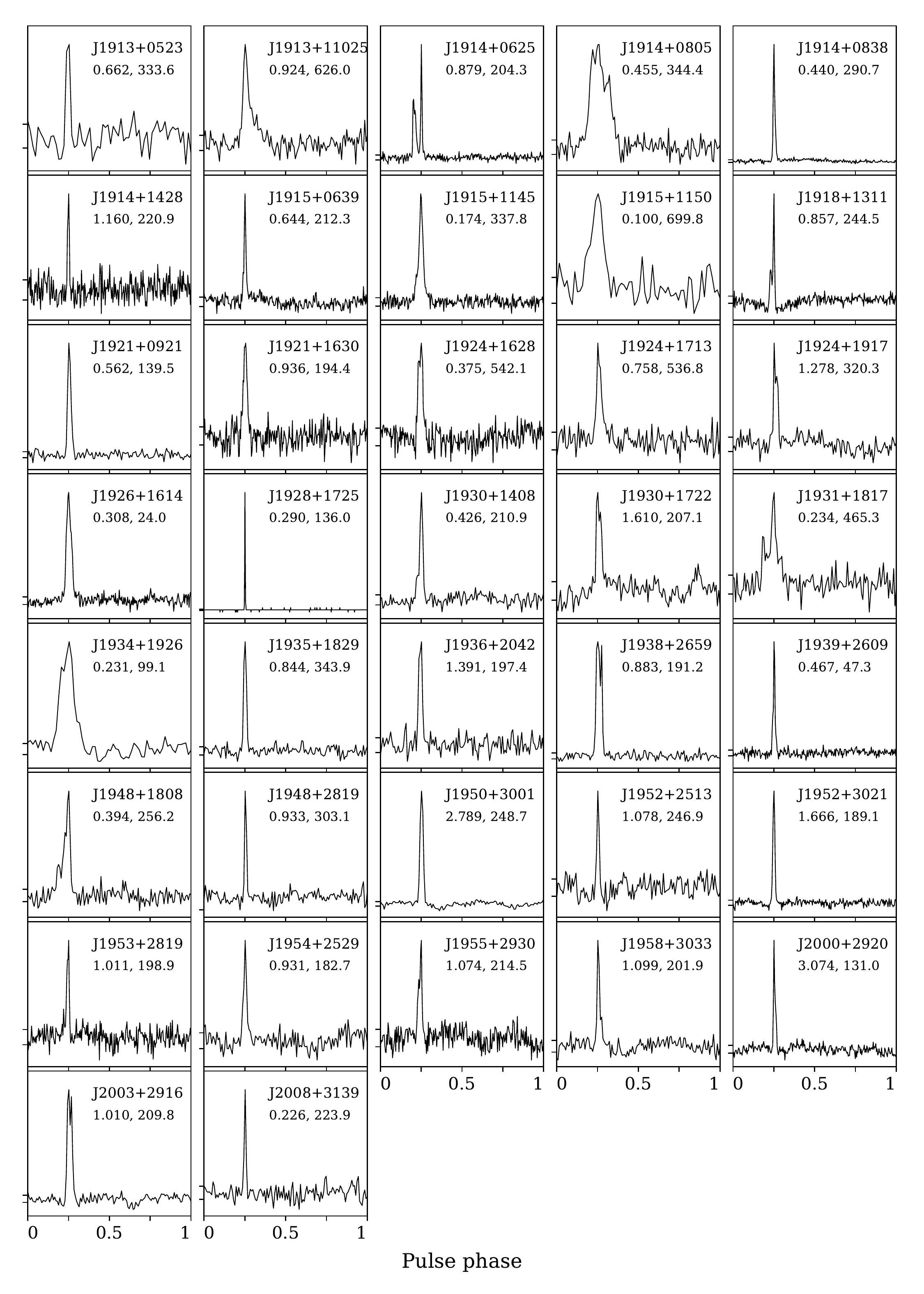}}
    \end{center}
    \vspace*{-10mm}
    \caption{\textit{(continued)} }
    \label{fig:profiles}
\end{figure*}

\subsubsection{Additional pulsar discoveries} \label{sec:morepsr}
\begin{figure}
    \begin{center}
        \centerline{\includegraphics[scale=0.65]{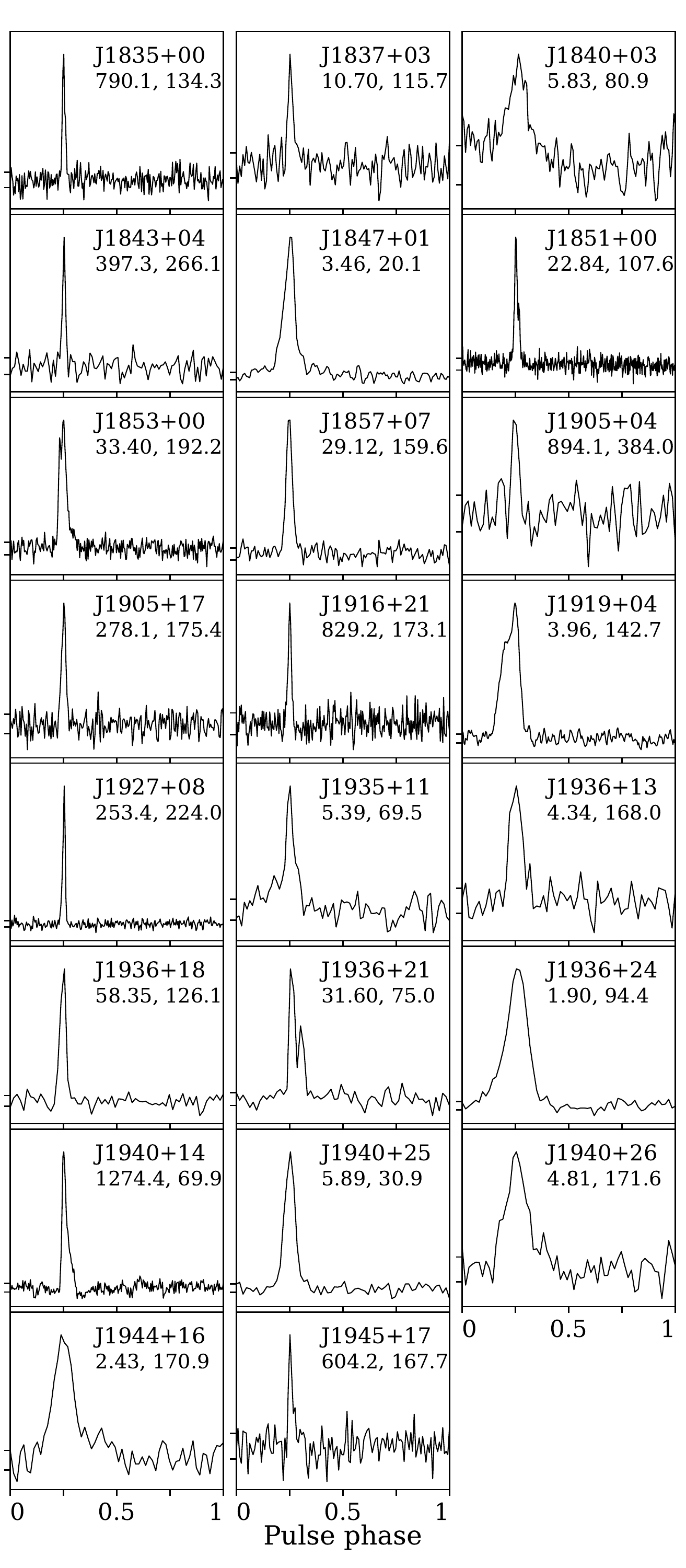}}
    \end{center}
    \vspace*{-10mm}
        \caption{Integrated pulse profiles of 23 additional pulsars recently discovered in the PALFA survey, generated from AO data at 1.4\,GHz.
        Similarly to Figure~\ref{fig:profiles}, the name, spin period (ms) and DM (pc\,cm$^{-3}$) of each pulsar are listed above their corresponding profile and tick marks on the vertical axis represent the root-mean-square intensity level of the off-source region.} 
   \label{fig:newpsrs}
\end{figure}

We also report on the recent discovery of 23 additional pulsars in the PALFA survey. Discovery parameters and estimates of their pulsed flux densities are listed in Table~\ref{tab:new}, and their pulse profiles  are shown in Figure~\ref{fig:newpsrs}.  Supplemental information such as diagnostic plots and the center position of the ALFA beam in which the pulsars were discovered can be found on the survey discovery page. Timing results and/or other notable properties will be reported in future publications (Parent et al. in prep, Haniewicz et al. in prep, Doskoch et al. in prep) -- the analyses and results presented in the remaining sections of this paper pertain to the aforementioned set of 72 pulsars (Table~\ref{tab:discoveries} and Figure~\ref{fig:profiles}).

\begin{table}[ht!]
  \centering
  \caption{Pulsars recently discovered by PALFA. Average pulsed flux densities at 1400\,MHz, $S_{1400}$, were estimated using the approach described in Section \ref{sec:flux}. 
  } 
  \setlength{\tabcolsep}{4.5mm}
  \begin{tabular}{l r r r }
\hline
PSR      & $P$  & DM              & $S_{1400}$  \\ 
(J2000)  & (ms) & (pc\,cm$^{-1}$) & ($\mu$Jy)   \\ 
 \hline \hline
J1835+00 &  790.1	&  134.3 	 &  80(20)  \\[-0.2em] 
J1837+03 &  10.70	&  115.7 	 &  57(15)  \\[-0.2em]  
J1840+03 &  5.83	&  80.9 	 &  150(40)  \\[-0.2em] 
J1843+04 &  397.3	&  266.1 	 &  55(15)  \\[-0.2em] 
J1847+01$^\dagger$ &  3.46	&  20.1 	 &  90(20)  \\[-0.2em] 
J1851+00$^\dagger$ &  22.84	&  107.6 	 &  67(16)  \\[-0.2em]  
J1853+00$^\dagger$ &  33.40	&  192.2 	 &  79(19)  \\[-0.2em]  
J1857+07 &  29.12	&  159.6 	 &  52(14)  \\[-0.2em]  
J1905+04 &  894.1	&  384.0 	 &  45(12)  \\[-0.2em]  
J1905+17 &  278.1	&  175.4 	 &  74(19)  \\[-0.2em]  
J1916+21 &  829.2	&  173.1 	 &  51(14)  \\[-0.2em]  
J1919+04 &  3.96	&  142.7 	 &  110(30)  \\[-0.2em]  
J1927+08 &  253.4	&  224.0 	 &  160(40)  \\[-0.2em]  
J1935+11 &  5.39	&  69.5 	 &  90(20)  \\[-0.2em]  
J1936+13 &  4.34	&  168.0 	 &  65(17)  \\[-0.2em] 
J1936+18$^\dagger$ &  58.35	&  126.1 	 &  38(11)  \\[-0.2em]  
J1936+21$^\dagger$ &  31.60	&  75.0 	 &  45(12)  \\[-0.2em]  
J1936+24 &  1.90	&  94.4 	 &  120(30)  \\[-0.2em]  
J1940+14 &  1274.4	&  69.9 	 &  130(50)  \\[-0.2em]  
J1940+25 &  5.89	&  30.9 	 &  70(20)  \\[-0.2em] 
J1940+26$^\dagger$ &  4.81	&  171.6 	 &  29(9)  \\[-0.2em]  
J1944+16 &  2.43	&  170.9 	 &  170(50)  \\[-0.2em]  
J1945+17 &  604.2	&  167.7 	 &  48(13)  \\  
\hline
  \end{tabular}
  \tablenotetext{\dagger}{Pulsars re-detected by the FAST GPPS survey; \cite{hww+21} provide higher precision on the position and/or DM parameters.\\}
  \label{tab:new}
\end{table}

\subsection{Timing observations}
Timing observations were largely conducted at AO and, for the 33 strongest sources, with the 76-m Lovell Telescope at Jodrell Bank Observatory (JBO) in Macclesfield, UK. Until 2014, some limited amount of timing data was collected for six of our pulsars 
with the 100-m Green Bank Telescope (GBT) at the Green Bank Observatory (GBO) in West Virginia, US. In 2020 May, we began follow-up observations on five other sources with the Canadian Hydrogen Intensity Mapping Experiment (CHIME) telescope\footnote{\url{https://chime-experiment.ca/en}} \citep{chimepsr}, located at the Dominion Radio Astrophysical Observatory (DRAO) in British Columbia, Canada. More information on observations for individual sources such as observations sites, cadence, and time span of observations can be inferred from plots introduced in Section \ref{sec:psrtiming}.

At AO, timing data were collected with the L-Wide receiver and recorded with the PUPPI\footnote{\url{https://www.naic.edu/puppi-observing/}} (Puerto Rican Ultimate Pulsar Processing Instrument) backend at a central frequency of 1380\,MHz with a nominal 800-MHz bandwidth split into 4096 channels sampled every 40.96\,$\mu$s. Following RFI excision, the usable bandwidth was typically $\sim$600\,MHz and integration lengths ranged from 300 to 1500\,s, depending upon the pulsar brightness. Initially, data were recorded in incoherent search mode but as a pulsar ephemeris improved, we switched to fold-mode observations where data were coherently dedispersed at the pulsar DM and folded into 10-s subintegrations in real-time modulo the instantaneous pulsar period, producing time-integrated pulse profiles for all frequency channels.

Timing data obtained with the Lovell Telescope were processed using a Digital Filterbank (DFB) which Nyquist samples a 512\,MHz band at 8-bit resolution and channelised it into 1024 channels using a polyphase filter. After RFI cleaning, the resultant band spans the range 1350 to 1700\,MHz. The data are incoherently dedispersed and folded into 10\,s long sub-integrations with 1024 pulse phase bins. Observation durations ranged from 30 to 60\,minutes depending on the source flux density. More information can be found in \cite{lsb+17}.

Observations with GBT were recorded with the Green Bank Ultimate Pulsar Processing Instrument (GUPPI, \citealt{drd+08}) backend, which processes 800\,MHz of bandwidth centered at 1.5\,GHz. Coherent-dedispersion mode data were recorded into 512 frequency channels and folded into subintegrations every 10\,s.  

As for CHIME observations, the stationary instrument operates in the 400-800\,MHz frequency range and beamformed data were collected with the CHIME/Pulsar backend. The latter discretizes the observing band into 1024 baseband channels which are then coherently dedispersed. The ``Digital  Signal  Processing  for  Pulsars'' (DSPSR) suite\footnote{\url{http://dspsr.sourceforge.net/}} is used to fold the data into 10 or 30\,s subintegrations, depending on the source's DM. The duration of any given observation at CHIME is limited by the transit time as set by the declination, $\delta$, of each pulsar. On average, the five objects  we followed up with CHIME (all with $16\degrees\,\lesssim\,\delta\,\lesssim\,27\degrees$) were observed for $\sim$\,15\,minutes each per day. More information on the CHIME/Pulsar system can be found in \cite{chimepsr}. 

\section{Data Analysis} \label{sec:timing}
\subsection{Timing Analysis} \label{sec:psrtiming}
In order to calculate the times of arrival (TOAs) of radio pulses in our data sets, we first excised RFI. Where only AO search-mode data were available, the cleaned raw data were folded at the topocentric period and dispersion measure (DM) that yielded the strongest detection with \presto's \texttt{prepfold} tool. We then produced a standard profile template by fitting one or more Gaussian components to the integrated profile that was detected with the highest signal-to-noise ratio (S/N). TOAs were then extracted by cross-correlating the folded data with the standard template using the \texttt{get\_TOAs.py} program from \presto, which fits for a linear phase gradient in the Fourier domain to determine the shifts between the profiles and the standard template \citep{t92}. 

When analyzing fold-mode data, we constructed improved, high-S/N standard templates by summing in phase the pulse profiles from multiple observations with the \texttt{psradd} tool from the  \texttt{PSRCHIVE}\footnote{\url{http://psrchive.sourceforge.net/}} software package. As profiles were being combined, weights were applied based on the signal-to-noise ratio of the pulsar signal in each data set. We then set profile baselines to zero to create “noise-free” templates before smoothing the final profiles with \texttt{PSRCHIVE}'s \texttt{psrsmooth} tool. TOAs were then extracted using the Fourier phase gradient approach with \texttt{pat}, also from the \texttt{PSRCHIVE} package. Given the difference in observing frequencies and the time delays introduced by backend systems, a separate standard profile template was generated to analyse data collected at each site.

The timing analysis was carried out with the \tempo\, software package\footnote{\url{http://tempo.sourceforge.net/}}, which implements a $\chi^2$ minimization technique to compute the best-fit parameters of a timing model. The JPL DE436 planetary ephemeris\footnote{\url{https://naif.jpl.nasa.gov/pub/naif/JUNO/kernels/spk/de436s.bsp.lbl}} and the UTC(NIST) time standard\footnote{\url{https://www.nist.gov/}} were used. 
The basic timing model for each source was parameterized by the pulsar period and first period derivative, $P$ and $\dot{P}$, right ascension, declination, and DM. To fit for DM, we extracted TOAs at different frequencies for a number of epochs using between two and eight frequency subbands per epoch, depending on the signal strength. Arbitrary time offsets between TOAs collected at different observatories were also allowed (where applicable) as timing parameters. Orbital parameters were also included in the timing model of the binary pulsar PSR~J1954+2529, which we describe later in Section~\ref{sec:1954}. In some cases, we additionally fit for higher-order frequency derivatives to reduce scatter due to the timing noise, an effect that arises from spin irregularities that are intrinsic to the pulsar \citep{mt77}. For the five pulsars that displayed glitch activity (Section \ref{sec:glitch}), glitch parameters were also computed with \tempo. 

Best-fit timing parameters are provided in Table~\ref{tab:timing} and corresponding post-fit residuals, with RMS values that range from 86\,$\mu$s to 5.4\,ms, are shown in Figure~\ref{fig:residuals}. Other measured and inferred pulsar properties are listed in Table~\ref{tab:derived}. 

Coherent solutions were obtained for all but three pulsars: PSRs~J1855$+$0626, J1858$+$0239 and J1928$+$1725. The first pulsar, PSR~J1855$+$0626, is an intermittent pulsar with too few detections to enable phase connection. We discuss its intermittency and emission properties in Section \ref{sec:1855}. PSR~J1858$+$0239 is a pulsar whose unstable average profiles appeared to show evidence of mode-changing behavior, which we discuss in Section \ref{sec:modechg}. For that source, timing observations were conducted solely at AO and we monitored the pulsar for $\sim$1.5\,years. We were able to phase connect a subset of TOAs from $\sim$\,10 consecutive epochs within a dense-observing timing campaign (spanning approximately one month) during which time the average pulse profile remained fairly stable and could thus be analysed with one common standard profile. Within that TOA set, we detected a significant spin-down rate (see $\dot{P}$ reported in Table~\ref{tab:timing}). However, when we attempted to connect time gaps between other consecutive pairs of closely spaced observations in which the pulsar displayed similar average profiles, we could not produce a consistent solution. Although we clearly observed a change in the observed spin frequency, an accurate $\dot{P}$  measurement requires that the timing coherence extends over at least one year in order to break the covariance between spin and astrometric parameters. Furthermore, we suspect that magnetospheric activity may coexist with torque variability \citep{klo+06}, thus the putative $\dot{P}$ reported in Table~\ref{tab:timing} may not be representative of the pulsar's normal spin-down (see Section \ref{sec:1858}). Dedicated, long-integration and regular observations would be required to obtain an accurate description of this pulsar's rotation. Finally, the last unsolved source, PSR~J1928$+$1725, is a RRAT that could only be timed through its single pulses. All timing observations were conducted at AO. A number of bright, narrow single pulses were detected but at a very irregular rate, with clusters of pulses being emitted within a few 15-minutes integrations and no detectable emission in several data sets, leaving large time gaps between detections. We extracted topocentric TOAs from each single pulse and attempted to achieve phase connection but were unsuccessful. Similarly to PSR~J1858$+$0239, we suspect that PSR~J1928$+$1725 experiences large torque variability.  We discuss in more detail the properties of this RRAT in Section \ref{sec:1928} and the evidence for the possible changes in spin-down rate.
%
\begin{spacing}{0.5}
\begin{table*}
  \centering
  {\footnotesize
  \caption{Best-fit parameters of the timing models. Numbers in parentheses are the 1$\sigma$ uncertainties on the last digit reported by \tempo, after weighting the TOAs such that $\chi^2$ = 1. Best-fit glitch and orbital (PSR~J1954+2529) parameters are provided separately in Sections~\ref{sec:glitch} and \ref{sec:1954}, respectively. \label{tab:timing} } 
  \setlength{\tabcolsep}{1.5mm}
  \begin{tabular}{l l l l l l c c c}
  \hline
PSR & R.A. (J2000) & Dec. (J2000) & $P$ & $\dot{P}$ & DM & Epoch & Data span & $\sigma_{res}$ \\[-0.3em]
(J2000) & (h:m:s) & (d:m:s) & (s) & (10$^{-15}$) & (pc cm$^{-3}$)  & (MJD) & (yr) & (ms) \\ 
\hline\hline
J0608$+$1635 & 06:08:51.662(5) & $+$16:35:09.4(4) & 0.945844752002(3) & 13.51042(18) & 86.08(3) & 57584.00 & 9.3 & 0.63\\[-0.35em] 
J1843$+$0119 & 18:43:23.90(3) & $+$01:19:27.1(8) & 1.26699835538(3) & 3.758(7) & 252.1(7)$^c$ & 58575.00 & 1.9 & 1.85\\[-0.35em] 
J1849$+$0106 & 18:49:55.404(7) & $+$01:06:22.6(2) & 1.83225931855(9) & 17.0080(9) & 216.24(15) & 57016.00 & 3.0 & 0.42\\[-0.35em] 
J1849$+$0430 & 18:49:40.44(2) & $+$04:30:36.8(6) & 0.42112580396(9) & 0.1137(11) & 191(1)$^c$ & 58073.00 & 2.3 & 1.99\\[-0.35em] 
J1851$+$0241 & 18:51:20.34(3) & $+$02:41:20.0(9) & 4.4913183586(5) & 22.568(14) & 524.3(7) & 56240.00 & 2.2 & 2.52\\[-0.35em] 
J1852$-$0000 & 18:52:40.167(9) & --00:00:25.5(3) & 1.92066632921(2) & 251.9666(3) & 593(1)$^c$ & 58205.00 & 5.7 & 1.29\\[-0.35em] 
J1853$+$0029 & 18:53:17.745(17) & $+$00:29:23.8(7) & 1.8767576226(3) & 2.431(2) & 232(4)$^c$ & 57826.00 & 3.5 & 2.38\\[-0.35em] 
J1853$+$0259 & 18:53:14.979(8) & $+$02:59:47.9(3) & 0.585552887667(3) & 0.11225(9) & 292.4(4) & 57093.00 & 6.8 & 0.97\\[-0.35em] 
J1853$+$0427 & 18:53:47.007(13) & $+$04:27:41.2(5) & 1.32065850582(15) & 2.645(1) & 550(2)$^c$ & 57890.00 & 3.3 & 1.06\\[-0.35em] 
J1854$+$0050 & 18:54:43.47(2) & $+$00:50:17.8(6) & 0.76727953408(2) & 0.5775(18) & 529.2(7) & 56854.00 & 1.9 & 1.08\\[-0.35em] 
J1855$+$0306 & 18:55:38.30(2) & $+$03:06:22.7(7) & 1.6335656928(3) & 7.0029(4) & 627(2) & 57548.00 & 8.6 & 2.74\\[-0.35em] 
J1855$+$0626$^a$ & 18:55:25 & $+$06:26:53 & 0.5288321(9) &     & 253.8(2) & 58375.00 & 0.7 &    \\[-0.35em] 
J1856$+$0912 & 18:56:33.40(1) & $+$09:12:29.7(3) & 2.1707012972(15) & 2.6283(15) & 191.4(4) & 58135.00 & 3.3 & 1.44\\[-0.35em] 
J1858$+$0239$^b$ & 18:58:18 & $+$02:39:52 & 0.197644188243(13) & 14(1) & 492.71(2) & 56441.00 & 1.1 &    \\[-0.35em] 
J1859$+$0345 & 18:59:12.71(2) & $+$03:45:57(1) & 1.51150850359(2) & 0.6478(9) & 557(6)$^c$ & 57354.00 & 6.1 & 5.42\\[-0.35em] 
J1902$+$0235 & 19:02:31.062(4) & $+$02:35:14.75(11) & 0.415394227732(2) & 0.0948(2) & 280.49(14) & 56835.00 & 2.0 & 0.34\\[-0.35em] 
J1902$+$1141 & 19:02:02.201(5) & $+$11:41:05.50(9) & 0.40914018296(6) & 2.59192(13) & 269.12(17) & 58377.00 & 2.5 & 0.42\\[-0.35em] 
J1903$+$0415 & 19:03:28.321(17) & $+$04:15:07.6(6) & 1.15139859175(13) & 0.2268(5) & 481(3)$^c$ & 57450.00 & 5.4 & 2.77\\[-0.35em] 
J1903$+$0912 & 19:03:42.101(5) & $+$09:12:41.69(16) & 0.166314477824(3) & 14.8383(2) & 362.8(1) & 57605.00 & 2.0 & 0.33\\[-0.35em] 
J1904$+$0056 & 19:04:07.06(3) & $+$00:56:59(1) & 0.43808945697(2) & 0.004(3) & 127(2) & 58677.00 & 1.7 & 3.81\\[-0.35em] 
J1905$+$1034 & 19:05:20.625(14) & $+$10:34:27.7(4) & 1.72681020359(3) & 20.6980(8) & 164.0(6)$^c$ & 57737.00 & 8.2 & 1.17\\[-0.35em] 
J1906$+$0724 & 19:06:22.577(18) & $+$07:24:22.8(6) & 1.5364901376(2) & 2.9990(6) & 478(4)$^c$ & 57080.00 & 7.7 & 3.09\\[-0.35em] 
J1907$+$0833 & 19:07:57.044(3) & $+$08:33:59.99(7) & 0.167627579462(16) & 3.69542(6) & 511.68(16) & 56161.00 & 2.6 & 0.29\\[-0.35em] 
J1908$+$0558 & 19:08:01.997(6) & $+$05:58:33.94(18) & 0.168677558616(3) & 2.2796(3) & 457.7(2) & 57632.00 & 4.5 & 0.65\\[-0.35em] 
J1908$+$1351 & 19:08:35.31(4) & $+$13:51:40(1) & 3.174831829(8) & 3.7(2) & 180.59(19) & 58489.00 & 3.0 & 0.55\\[-0.35em] 
J1909$+$1205 & 19:09:51.47(5) & $+$12:05:47(2) & 1.229312421(2) & 3.40(5) & 302(1)$^c$ & 55897.00 & 3.2 & 1.39\\[-0.35em] 
J1910$+$0435 & 19:10:11.072(6) & $+$04:35:29.5(2) & 0.664679416494(9) & 17.2366(3) & 297(1) & 57862.00 & 3.2 & 0.87\\[-0.35em] 
J1910$+$0710 & 19:10:13.873(15) & $+$07:10:46.4(5) & 0.53864678794(3) & 0.205(2) & 254.1(4) & 57671.00 & 5.5 & 0.64\\[-0.35em] 
J1910$+$1017 & 19:10:26.124(2) & $+$10:17:54.09(6) & 0.411158865683(4) & 5.41520(7) & 633.83(16) & 55938.00 & 3.6 & 0.22\\[-0.35em] 
J1910$+$1026 & 19:10:48.753(9) & $+$10:26:52.5(5) & 0.53149303397(2) & 257.067(4) & 712.8(5) & 56334.00 & 1.4 & 1.03\\[-0.35em] 
J1911$+$0921 & 19:11:46.487(9) & $+$09:21:56.8(3) & 0.273706758194(2) & 0.01751(18) & 340.1(6) & 56041.00 & 3.3 & 1.19\\[-0.35em] 
J1911$+$0925 & 19:11:59.472(12) & $+$09:25:32.2(6) & 0.323857547341(5) & 3.5480(7) & 486.6(5) & 55982.00 & 3.0 & 1.03\\[-0.35em] 
J1911$+$1051 & 19:11:42.262(8) & $+$10:51:26.74(16) & 0.190872844929(12) & 12.16555(16) & 445.21(17) & 56578.00 & 6.1 & 0.40\\[-0.35em] 
J1911$+$1301 & 19:11:31.739(9) & $+$13:01:26.6(3) & 1.01046173336(9) & 1.8899(6) & 389(2) & 57915.00 & 3.5 & 1.52\\[-0.35em] 
J1911$+$1336 & 19:11:59.541(3) & $+$13:36:55.00(7) & 0.299992040976(1) & 0.15860(7) & 323.95(14) & 58410.00 & 3.0 & 0.43\\[-0.35em] 
J1913$+$0523 & 19:13:22.721(7) & $+$05:23:58.8(3) & 0.661997424287(7) & 1.7969(3) & 333.6(3) & 58290.00 & 3.8 & 1.31\\[-0.35em] 
J1913$+$11025 & 19:13:42.715(8) & $+$11:02:58.8(2) & 0.923871917718(11) & 0.3404(17) & 626.0(4) & 56356.00 & 1.6 & 1.13\\[-0.35em] 
J1914$+$0625 & 19:14:08.359(3) & $+$06:25:00.97(5) & 0.878889431192(9) & 0.4531(11) & 204.33(5) & 56964.00 & 3.1 & 0.15\\[-0.35em] 
J1914$+$0805 & 19:14:05.508(9) & $+$08:05:12.7(2) & 0.455499390131(6) & 0.0302(7) & 344.4(2) & 58528.00 & 1.6 & 1.03\\[-0.35em] 
J1914$+$0838 & 19:14:26.4506(14) & $+$08:38:45.14(3) & 0.440039882669(4) & 0.586395(4) & 290.70(11) & 56825.00 & 9.2 & 0.14\\[-0.35em] 
J1914$+$1428 & 19:14:53.946(8) & $+$14:28:46.2(2) & 1.15951978505(6) & 2.1814(6) & 220.9(4) & 56000.00 & 3.4 & 0.90\\[-0.35em] 
J1915$+$0639 & 19:15:54.327(2) & $+$06:39:46.21(4) & 0.64414015325(3) & 1.8435(4) & 212.32(5) & 57374.00 & 5.4 & 0.13\\[-0.35em] 
J1915$+$1145 & 19:15:33.1231(8) & $+$11:45:40.98(2) & 0.173647195715(2) & 0.01531(3) & 337.78(3) & 56367.00 & 1.6 & 0.09\\[-0.35em] 
J1915$+$1150 & 19:15:16.61(6) & $+$11:50:35.4(9) & 0.10004095461(3) & 13.671(1) & 699.83(12) & 55927.00 & 3.2 & 0.28\\[-0.35em] 
J1918$+$1311 & 19:18:46.220(5) & $+$13:11:24.51(13) & 0.856748867762(6) & 2.2579(6) & 244.5(3) & 56866.00 & 1.9 & 0.61\\[-0.35em] 
J1921$+$0921 & 19:21:53.487(9) & $+$09:21:30.2(2) & 0.562302288458(14) & 9.576(1) & 139.5(2) & 58119.00 & 3.1 & 0.39\\[-0.35em] 
J1921$+$1630 & 19:21:00.142(18) & $+$16:30:55.8(9) & 0.93644800775(19) & 22.345(2) & 194.4(5) & 56891.00 & 1.7 & 1.36\\[-0.35em] 
J1924$+$1628 & 19:24:44.090(6) & $+$16:28:37.69(16) & 0.375082251011(14) & 0.32096(6) & 542(1)$^c$ & 57514.00 & 5.1 & 1.19\\[-0.35em] 
J1924$+$1713 & 19:24:32.517(19) & $+$17:13:33.0(3) & 0.758433236391(9) & 0.1130(17) & 536.8(6) & 56055.00 & 3.2 & 1.19\\[-0.35em] 
J1924$+$1917 & 19:24:26.21(4) & $+$19:17:24.4(3) & 1.27794162459(4) & 0.199(8) & 320.3(6) & 58528.00 & 1.6 & 1.72\\[-0.35em] 
J1926$+$1614 & 19:26:50.202(3) & $+$16:14:18.77(9) & 0.308305907254(15) & 0.03353(4) & 24.02(18) & 57589.00 & 4.7 & 0.40\\[-0.35em] 
J1928$+$1725$^a$ & 19:28:52 & $+$17:25:29 & 0.28983833(8) &     & 135.96(12) & 57696.00 & 5.3 &    \\[-0.35em] 
J1930$+$1408 & 19:30:18.9526(18) & $+$14:08:55.39(5) & 0.425720327378(5) & 0.00190(1) & 210.87(13) & 56885.00 & 8.6 & 0.21\\[-0.35em] 
J1930$+$1722 & 19:30:30.11(2) & $+$17:22:53.2(6) & 1.60970633781(2) & 0.8808(7) & 207(7)$^c$ & 57656.00 & 6.2 & 4.10\\[-0.35em] 
J1931$+$1817 & 19:31:52.739(4) & $+$18:17:00.77(8) & 0.234131440128(17) & 107.3637(5) & 465(1) & 58156.00 & 6.0 & 0.72\\[-0.35em] 
J1934$+$1926 & 19:34:21.651(6) & $+$19:26:35.31(14) & 0.230984425819(16) & 0.00268(8) & 99.1(3) & 55977.00 & 3.6 & 0.80\\[-0.35em] 
J1935$+$1829 & 19:35:42.91(1) & $+$18:29:28.1(3) & 0.843547910278(9) & 2.3207(4) & 343.9(6) & 57852.00 & 3.8 & 1.28\\[-0.35em] 
J1936$+$2042 & 19:36:27.42(2) & $+$20:42:04.5(4) & 1.39072342303(15) & 49.3744(14) & 197.4(5)$^c$ & 56065.00 & 5.8 & 1.26\\[-0.35em] 
J1938$+$2659 & 19:38:39.175(6) & $+$26:59:14.96(14) & 0.883331781241(9) & 3.2275(5) & 191.20(5) & 58829.00 & 2.3 & 0.67\\[-0.35em] 
J1939$+$2609 & 19:39:42.3413(14) & $+$26:09:36.39(4) & 0.466962555351(7) & 0.0052(4) & 47.30(6) & 58823.00 & 2.3 & 0.11\\[-0.35em] 
J1948$+$1808 & 19:48:22.129(4) & $+$18:08:30.15(8) & 0.394354427486(4) & 0.2271(2) & 256.16(7) & 58861.00 & 2.3 & 0.51\\[-0.35em] 
J1948$+$2819 & 19:48:38.39(1) & $+$28:19:20.06(19) & 0.932692952758(16) & 61.26032(13) & 303(2) & 58102.00 & 6.3 & 1.70\\[-0.35em] 
J1950$+$3001 & 19:50:53.68(2) & $+$30:01:42.7(3) & 2.78891789352(6) & 149.013(1) & 249(2) & 58162.00 & 5.5 & 1.85\\[-0.35em] 
J1952$+$2513 & 19:52:20.738(15) & $+$25:13:44.1(7) & 1.07764729476(8) & 1.0648(17) & 246.9(5) & 55731.00 & 4.3 & 1.19\\[-0.35em] 
J1952$+$3021 & 19:52:19.696(11) & $+$30:21:20.0(4) & 1.66566523108(16) & 10.8257(9) & 189.1(4) & 57821.00 & 3.6 & 1.08\\[-0.35em] 
J1953$+$2819 & 19:53:35.238(11) & $+$28:19:39.52(19) & 1.01100245325(5) & 2.14126(17) & 199(2) & 58094.00 & 6.2 & 1.76\\[-0.35em] 
J1954$+$2529 & 19:54:19.716(3) & $+$25:29:27.34(7) & 0.931210094606(4) & 1.25870(7) & 182.70(16) & 58104.00 & 6.5 & 0.73\\[-0.35em] 
J1955$+$2930 & 19:55:07.527(9) & $+$29:30:49.80(16) & 1.07387774187(7) & 3.3574(4) & 214(1) & 57822.00 & 3.7 & 1.29\\[-0.35em] 
J1958$+$3033 & 19:58:06.82(1) & $+$30:33:52.7(2) & 1.09858060946(1) & 6.4576(6) & 201.9(4) & 57683.00 & 2.7 & 0.95\\[-0.35em] 
J2000$+$2920 & 20:00:16.508(6) & $+$29:20:07.47(12) & 3.07378325868(3) & 37.4364(4) & 131(1)$^c$ & 58196.00 & 5.7 & 1.00\\[-0.35em] 
J2003$+$2916 & 20:03:03.194(6) & $+$29:16:00.96(11) & 1.0098766696(5) & 0.3410(2) & 210(1) & 57767.00 & 4.0 & 0.92\\[-0.35em] 
J2008$+$3139 & 20:08:39.9874(13) & $+$31:39:27.36(2) & 0.226118635651(2) & 0.082764(11) & 223.9(2) & 57831.00 & 3.6 & 0.18\\
\hline \\[-2.0em]
  \end{tabular}
\tablenotetext{a}{Unsolved timing model.\\[-0.8em]}\tablenotetext{b}{Partially solved timing model.\\[-0.8em]}\tablenotetext{c}{Improved precision from re-detection in the FAST GPPS survey can be found in \cite{hww+21}.}
  }
\end{table*}
\end{spacing}
\begin{spacing}{0.5}
\begin{table*}
  \centering
  {\footnotesize
  \caption{Measured and derived pulsar properties. The average pulsed flux density at 1400\,MHz, $S_{1400}$, and the pulse full width at half maximum, $W50$, were calculated from the ALFA discovery data. The $D_{\textrm{NE2001}}$ and $D_{\textrm{YMW16}}$ parameters are the DM-estimated distances of the pulsars predicted by the NE2001 and YMW16 models, respectively. Numbers in parentheses are 1$\sigma$ uncertainties on the last digit. } 
  \setlength{\tabcolsep}{2.5mm}
  \begin{tabular}{p{1.4cm} r r r r r r r c c c }
\hline
PSR & $\ell$ &  $b$ & $D_{\textrm{NE2001}}$ & $D_{\textrm{YMW16}}$  && $W50$ & $S_{1400}$ & log($\tau_c$) & log(B) & log($\dot{E}$) \\[-0.35em] 
(J2000) & ($^\circ$) & ($^\circ$) & (kpc) & (kpc) && (ms) & ($\mu$Jy) & (yr) & (G) & (erg s$^{-1}$) \\ 
\hline\hline
J0608$+$1635 & 193.366 & --1.569 & 2.4 & 1.7 && 17(4) & 140(40)  & 6.0 & 12.6 & 32.8 \\[-0.35em] 
J1843$+$0119 & 33.195 & 2.392 & 6.1 & 6.1 && 4(1) & 63(16)  & 6.7 & 12.3 & 31.9 \\[-0.35em] 
J1849$+$0106 & 33.744 & 0.841 & 5.5 & 4.6 && 13(7) & 22(6)  & 6.2 & 12.8 & 32.0 \\[-0.35em] 
J1849$+$0430 & 36.748 & 2.446 & 5.1 & 6.0 && 26(7) & 110(30)  & 7.8 & 11.3 & 31.8 \\[-0.35em] 
J1851$+$0241 & 35.314 & 1.248 & 9.7 & 11.3 && 6(4) & 56(14)  & 6.5 & 13.0 & 31.0 \\[-0.35em] 
J1852$-$0000 & 33.066 & --0.277 & 8.2 & 5.8 && 27(8) & 115(19)  & 5.1 & 13.3 & 33.1 \\[-0.35em] 
J1853$+$0029 & 33.580 & --0.19 & 5.5 & 4.1 && 75(15) & 74(19)  & 7.1 & 12.3 & 31.2 \\[-0.35em] 
J1853$+$0259 & 35.806 & 0.963 & 6.3 & 5.8 && 51(5) & 220(40)  & 7.9 & 11.4 & 31.3 \\[-0.35em] 
J1853$+$0427 & 37.170 & 1.512 & 10.7 & 15.7 && 7(2) & 100(30)  & 6.9 & 12.3 & 31.7 \\[-0.35em] 
J1854$+$0050 & 34.053 & --0.349 & 7.9 & 5.7 && 21(6) & 48(8)  & 7.3 & 11.8 & 31.7 \\[-0.35em] 
J1855$+$0306 & 36.175 & 0.483 & 9.5 & 7.4 && 39(6) & 35(6)  & 6.6 & 12.5 & 31.8 \\[-0.35em] 
J1855$+$0626$^a$ & 39.125 & 2.055 & 6.5 & 9.3 && 10(1) & 130(30)  &   &   &   \\[-0.35em] 
J1856$+$0912 & 41.715 & 3.058 & 5.9 & 9.6 && 53(17) & 40(11)  & 7.1 & 12.4 & 31.0 \\[-0.35em] 
J1858$+$0239$^b$ & 36.085 & --0.31 & 7.9 & 5.9 && 11(2) & 140(40)  & 5.3 & 12.2 & 34.9 \\[-0.35em] 
J1859$+$0345 & 37.169 & --0.01 & 8.6 & 6.2 && 31(12) & 90(20)  & 7.6 & 12.0 & 30.9 \\[-0.35em] 
J1902$+$0235 & 36.498 & --1.282 & 8.2 & 13.9 && 9(3) & 40(11)  & 7.8 & 11.3 & 31.7 \\[-0.35em] 
J1902$+$1141 & 44.537 & 2.984 & 5.9 & 6.1 && 12(3) & 120(30)  & 6.4 & 12.0 & 33.2 \\[-0.35em] 
J1903$+$0415 & 38.088 & --0.732 & 8.1 & 7.6 && 30(5) & 72(12)  & 7.9 & 11.7 & 30.8 \\[-0.35em] 
J1903$+$0912 & 42.522 & 1.49 & 8.5 & 11.8 && 6(3) & 90(20)  & 5.2 & 12.2 & 35.1 \\[-0.35em] 
J1904$+$0056 & 35.224 & --2.387 & 2.8 & 4.3 && 21(7) & 45(12)  & 9.2 & 10.6 & 30.3 \\[-0.35em] 
J1905$+$1034 & 43.919 & 1.755 & 5.0 & 6.9 && 26(7) & 40(11)  & 6.1 & 12.8 & 32.2 \\[-0.35em] 
J1906$+$0724 & 41.222 & 0.074 & 8.4 & 6.9 && 23(12) & 45(11)  & 6.9 & 12.3 & 31.5 \\[-0.35em] 
J1907$+$0833 & 42.431 & 0.262 & 9.2 & 8.3 && 15(1) & 100(17)  & 5.9 & 11.9 & 34.5 \\[-0.35em] 
J1908$+$0558 & 40.141 & --0.949 & 8.7 & 10.6 && 6(1) & 50(13)  & 6.1 & 11.8 & 34.3 \\[-0.35em] 
J1908$+$1351 & 47.206 & 2.558 & 6.1 & 6.9 && 5(2) & 22(6)  & 7.1 & 12.5 & 30.7 \\[-0.35em] 
J1909$+$1205 & 45.780 & 1.472 & 7.9 & 10.1 && 47(5) & 100(30)  & 6.8 & 12.3 & 31.9 \\[-0.35em] 
J1910$+$0435 & 39.158 & --2.062 & 7.3 & 11.3 && 15(5) & 90(20)  & 5.8 & 12.5 & 33.4 \\[-0.35em] 
J1910$+$0710 & 41.461 & --0.879 & 6.1 & 6.0 && 16(8) & 32(5)  & 7.6 & 11.5 & 31.7 \\[-0.35em] 
J1910$+$1017 & 44.250 & 0.517 & 12.5 & 13.7 && 11(6) & 37(9)  & 6.1 & 12.2 & 33.5 \\[-0.35em] 
J1910$+$1026 & 44.426 & 0.503 & 14.9 & 15.8 && 13(4) & 58(15)  & 4.5 & 13.1 & 34.8 \\[-0.35em] 
J1911$+$0921 & 43.576 & --0.207 & 5.0 & 6.4 && 15(4) & 113(18)  & 8.4 & 10.8 & 31.5 \\[-0.35em] 
J1911$+$0925 & 43.654 & --0.227 & 6.9 & 8.1 && 19(3) & 90(20)  & 6.2 & 12.0 & 33.6 \\[-0.35em] 
J1911$+$1051 & 44.890 & 0.499 & 9.0 & 10.1 && 7(3) & 51(13)  & 5.4 & 12.2 & 34.8 \\[-0.35em] 
J1911$+$1301 & 46.792 & 1.538 & 10.5 & 11.6 && 16(8) & 51(13)  & 6.9 & 12.1 & 31.9 \\[-0.35em] 
J1911$+$1336 & 47.369 & 1.712 & 9.2 & 10.4 && 8(2) & 72(19)  & 7.5 & 11.3 & 32.4 \\[-0.35em] 
J1913$+$0523 & 40.243 & --2.395 & 8.7 & 16.1 && 2(1) & 25(7)  & 6.8 & 12.0 & 32.4 \\[-0.35em] 
J1913$+$11025 & 45.289 & 0.151 & 13.2 & 11.4 && 31(7) & 140(40)  & 7.6 & 11.8 & 31.2 \\[-0.35em] 
J1914$+$0625 & 41.233 & --2.092 & 5.8 & 8.3 && 49(3) & 60(16)  & 7.5 & 11.8 & 31.4 \\[-0.35em] 
J1914$+$0805 & 42.708 & --1.307 & 8.1 & 10.7 && 29(4) & 160(40)  & 8.4 & 11.1 & 31.1 \\[-0.35em] 
J1914$+$0838 & 43.243 & --1.124 & 7.1 & 8.1 && 4(2) & 280(50)  & 7.1 & 11.7 & 32.4 \\[-0.35em] 
J1914$+$1428 & 48.461 & 1.487 & 6.9 & 6.5 && 22(9) & 57(15)  & 6.9 & 12.2 & 31.7 \\[-0.35em] 
J1915$+$0639 & 41.655 & --2.366 & 6.1 & 8.9 && 8(3) & 41(11)  & 6.7 & 12.0 & 32.4 \\[-0.35em] 
J1915$+$1145 & 46.129 & 0.082 & 8.0 & 7.2 && 5.6(7) & 90(20)  & 8.3 & 10.7 & 32.1 \\[-0.35em] 
J1915$+$1150 & 46.170 & 0.18 & 16.9 & 14.0 && 12(2) & 47(12)  & 5.1 & 12.1 & 35.7 \\[-0.35em] 
J1918$+$1311 & 47.759 & 0.055 & 6.8 & 6.2 && 7(3) & 54(14)  & 6.8 & 12.1 & 32.2 \\[-0.35em] 
J1921$+$0921 & 44.733 & --2.418 & 4.8 & 6.1 && 10(4) & 58(15)  & 6.0 & 12.4 & 33.3 \\[-0.35em] 
J1921$+$1630 & 50.949 & 1.138 & 6.4 & 5.1 && 16(4) & 48(13)  & 5.8 & 12.7 & 33.0 \\[-0.35em] 
J1924$+$1628 & 51.337 & 0.331 & 14.8 & 10.5 && 13(2) & 110(30)  & 7.3 & 11.5 & 32.4 \\[-0.35em] 
J1924$+$1713 & 51.975 & 0.726 & 15.6 & 10.9 && 20(6) & 36(6)  & 8.0 & 11.5 & 31.0 \\[-0.35em] 
J1924$+$1917 & 53.782 & 1.725 & 9.8 & 8.6 && 4(1) & 39(11)  & 8.0 & 11.7 & 30.6 \\[-0.35em] 
J1926$+$1614 & 51.367 & --0.226 & 1.9 & 1.3 && 8(1) & 90(20)  & 8.2 & 11.0 & 31.7 \\[-0.35em] 
J1928$+$1725$^a$ & 52.641 & --0.087 & 4.9 & 3.7 && 1.1(6) & 1400(400)  &   &   &   \\[-0.35em] 
J1930$+$1408 & 49.932 & --1.962 & 6.9 & 5.9 && 8(3) & 51(13)  & 9.6 & 10.5 & 30.0 \\[-0.35em] 
J1930$+$1722 & 52.790 & --0.45 & 6.6 & 4.7 && 59(13) & 90(30)  & 7.5 & 12.1 & 30.9 \\[-0.35em] 
J1931$+$1817 & 53.739 & --0.303 & 12.9 & 9.6 && 18(2) & 150(40)  & 4.5 & 12.7 & 35.5 \\[-0.35em] 
J1934$+$1926 & 55.038 & --0.256 & 4.1 & 3.2 && 22(4) & 130(30)  & 9.1 & 10.4 & 30.9 \\[-0.35em] 
J1935$+$1829 & 54.361 & --0.999 & 9.9 & 8.6 && 16(7) & 37(11)  & 6.8 & 12.2 & 32.2 \\[-0.35em] 
J1936$+$2042 & 56.376 & --0.074 & 6.6 & 5.0 && 40(11) & 45(12)  & 5.6 & 12.9 & 32.9 \\[-0.35em] 
J1938$+$2659 & 62.107 & 2.562 & 7.2 & 8.6 && 26(7) & 80(20)  & 6.6 & 12.2 & 32.3 \\[-0.35em] 
J1939$+$2609 & 61.501 & 1.951 & 3.1 & 2.7 && 4(2) & 150(40)  & 9.2 & 10.7 & 30.3 \\[-0.35em] 
J1948$+$1808 & 55.544 & --3.782 & 9.4 & 11.3 && 15(3) & 80(20)  & 7.4 & 11.5 & 32.2 \\[-0.35em] 
J1948$+$2819 & 64.366 & 1.307 & 9.7 & 11.0 && 13(7) & 46(8)  & 5.4 & 12.9 & 33.5 \\[-0.35em] 
J1950$+$3001 & 66.087 & 1.749 & 8.3 & 8.7 && 6(2) & 240(60)  & 5.5 & 13.3 & 32.4 \\[-0.35em] 
J1952$+$2513 & 62.125 & --0.98 & 8.1 & 8.4 && 19(8) & 31(8)  & 7.2 & 12.0 & 31.5 \\[-0.35em] 
J1952$+$3021 & 66.526 & 1.65 & 7.0 & 7.5 && 26(13) & 23(6)  & 6.4 & 12.6 & 32.0 \\[-0.35em] 
J1953$+$2819 & 64.926 & 0.374 & 7.1 & 7.6 && 19(4) & 47(8)  & 6.9 & 12.2 & 31.9 \\[-0.35em] 
J1954$+$2529 & 62.580 & --1.229 & 6.8 & 7.8 && 16(7) & 59(16)  & 7.1 & 12.0 & 31.8 \\[-0.35em] 
J1955$+$2930 & 66.116 & 0.697 & 7.4 & 7.6 && 24(4) & 72(19)  & 6.7 & 12.3 & 32.0 \\[-0.35em] 
J1958$+$3033 & 67.350 & 0.691 & 7.1 & 7.3 && 13(9) & 32(9)  & 6.4 & 12.4 & 32.3 \\[-0.35em] 
J2000$+$2920 & 66.549 & --0.353 & 5.4 & 6.7 && 26(12) & 100(30)  & 6.1 & 13.0 & 31.7 \\[-0.35em] 
J2003$+$2916 & 66.812 & --0.903 & 7.3 & 7.5 && 35(8) & 200(50)  & 7.7 & 11.8 & 31.1 \\[-0.35em] 
J2008$+$3139 & 69.478 & --0.639 & 7.3 & 7.2 && 3(2) & 43(11)  & 7.6 & 11.1 & 32.5 \\  
\hline \\[-3.0em]
  \end{tabular}
  \tablenotetext{a}{Unsolved timing model.\\[-0.8em]}\tablenotetext{b}{Partially solved timing model.}
  \label{tab:derived}
  }
\end{table*}
\end{spacing}

\begin{figure*}
    \begin{center}
        \centerline{\includegraphics[scale=0.535]{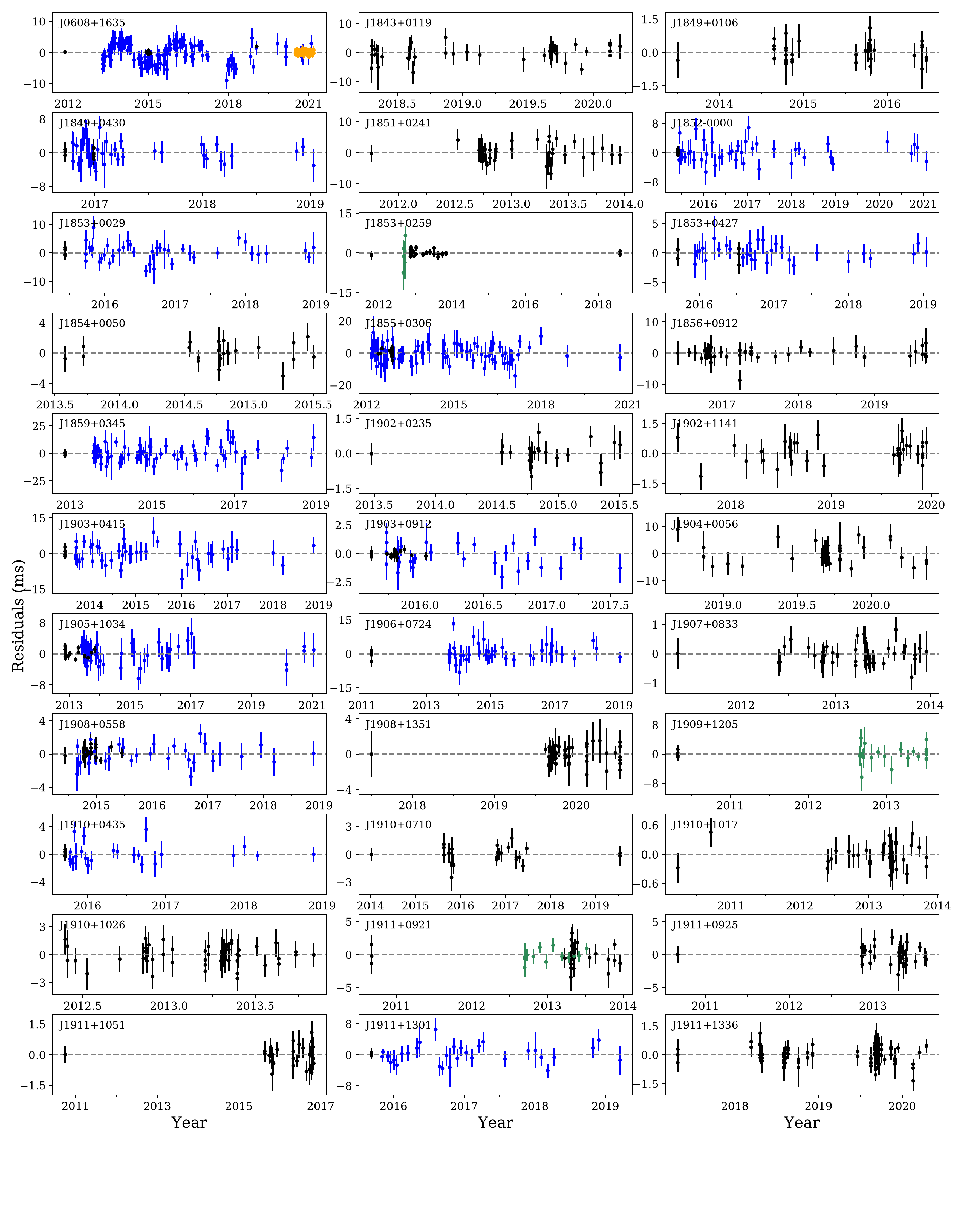}}
    \end{center}
    \vspace*{-28mm}
        \caption{Post-fit timing residuals of pulsars with phase-connected timing solutions. TOAs collected at AO are shown in black, at JBO in blue, at GBO in green and with CHIME/Pulsar in orange. We note that the features in the JBO timing residuals of PSR~J0608+1635 are likely due to the combined effect of timing noise and the use of a slightly inaccurate initial ephemeris when observing the pulsar in fold mode at JBO. The features are inconsistent with an orbital motion.  }
\end{figure*}
\begin{figure*}\ContinuedFloat
    \begin{center}
        \centerline{\includegraphics[scale=0.53]{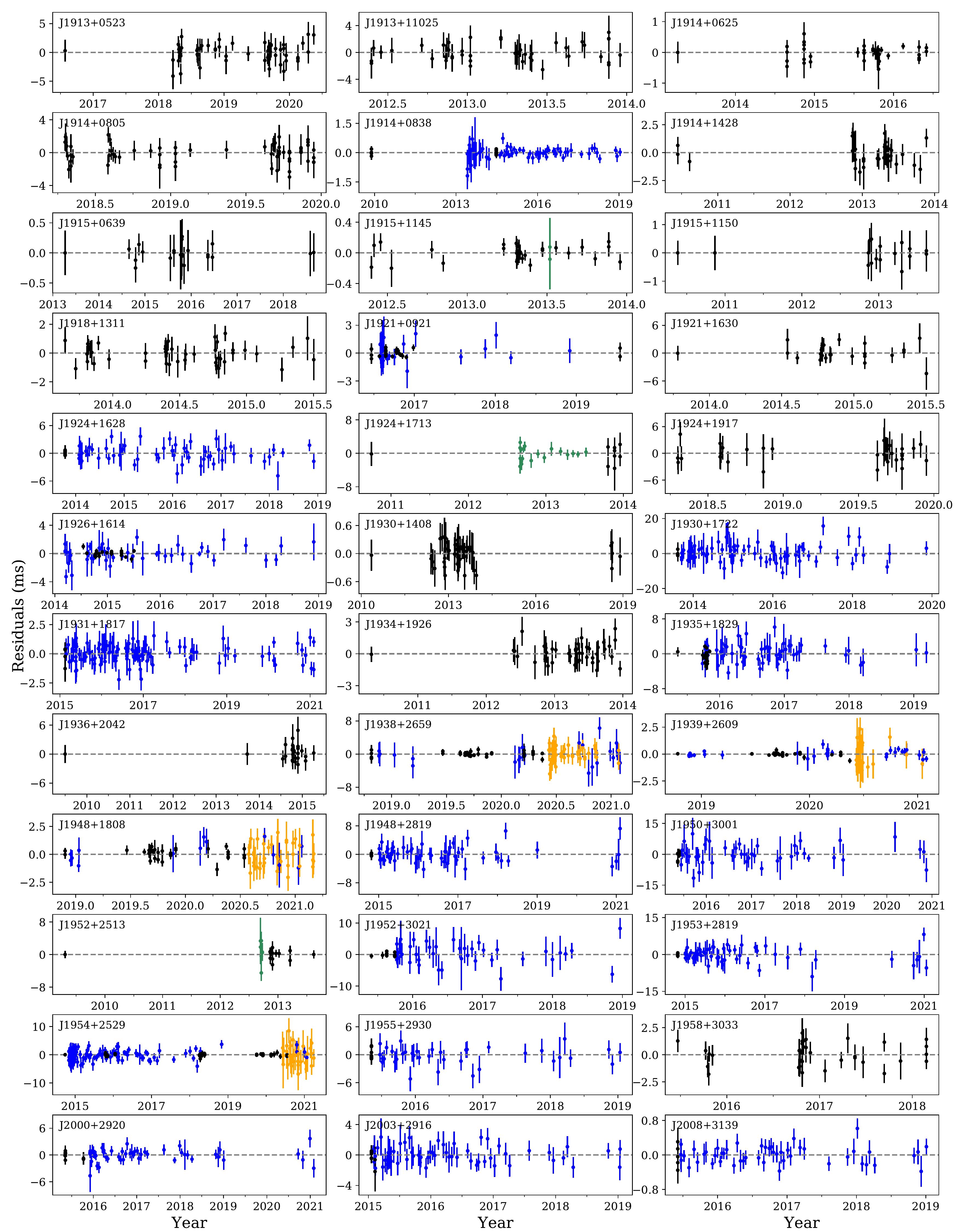}}
    \end{center}
    \vspace*{-8mm}
    \caption{\textit{(continued)} }
    \label{fig:residuals}
\end{figure*}
\subsection{Flux Density Calculations} \label{sec:flux}
We estimated the average pulsed flux densities at 1400\,MHz, $S_{1400}$, by calibrating the ALFA discovery data using the radiometer equation \citep{dtw+85}. The ALFA receiver provides a nominal bandwidth of 323\,MHz, however effective bandwidths typically range between 260\,MHz and 300\,MHz following RFI excision. The latter, narrower bandwidths were used in the calibration procedure.
Sky temperatures at the pulsar positions were estimated by extrapolating the 408-MHz all-sky map from \cite{rdb+15} to 1400\,MHz, assuming a spectral index\footnote{Defined as  $S_{\nu}\propto\nu^{\alpha}$, where $S_{\nu}$ is the flux density at frequency $\nu$ and $\alpha$ is the spectral index.} of --2.7 \citep{rdb+15}. System temperatures of the ALFA receiver typically varied between 28\,K and 32\,K. In estimating the flux density, we assumed a system temperature of 30\,K. We used a value of 9\,K\,Jy$^{-1}$ for the gain of the central ALFA beam, and scaled the gain of the outer six beams to be 79\% that of the central beam \citep{cfl+06}. The data analysis procedure discards any rotation-independent radio flux from the pulsar.

Our average pulsed flux density measurements are reported in Table~\ref{tab:derived}. Considerable systematic uncertainties affect our measurements, arising notably from fluctuations in the system temperature,  reductions in the effective gain due to variations in the receiver response and positional offsets between the true position of the pulsars (i.e., timing positions) and the beam center positions. These uncertainties are included in the error estimates reported in Table~\ref{tab:derived}. The same approach was used to estimate the pulsed flux density of the 23 additional pulsar discoveries reported in Table~\ref{tab:new} of Section~\ref{sec:morepsr}.

We note here that the pulse profiles shown in Figure~\ref{fig:profiles}, most of which were created from data collected with the 800-MHz L-Wide receiver at Arecibo, are not the profiles that were used to estimate the average pulsed flux densities. Due to the difference in the spectral response of the ALFA and L-Wide receivers and because pulsars generally have power-law spectra and their pulse profiles evolve with observing frequency, the profiles in Figure~\ref{fig:profiles} cannot be directly normalized to the flux densities reported in Table~\ref{tab:derived}.

\subsection{Interstellar Scattering} \label{sec:scat}
\begin{figure*}[t]
	\centering
	\vspace{-3.0mm}
    \includegraphics[width=1.01\textwidth,height=7.5cm]{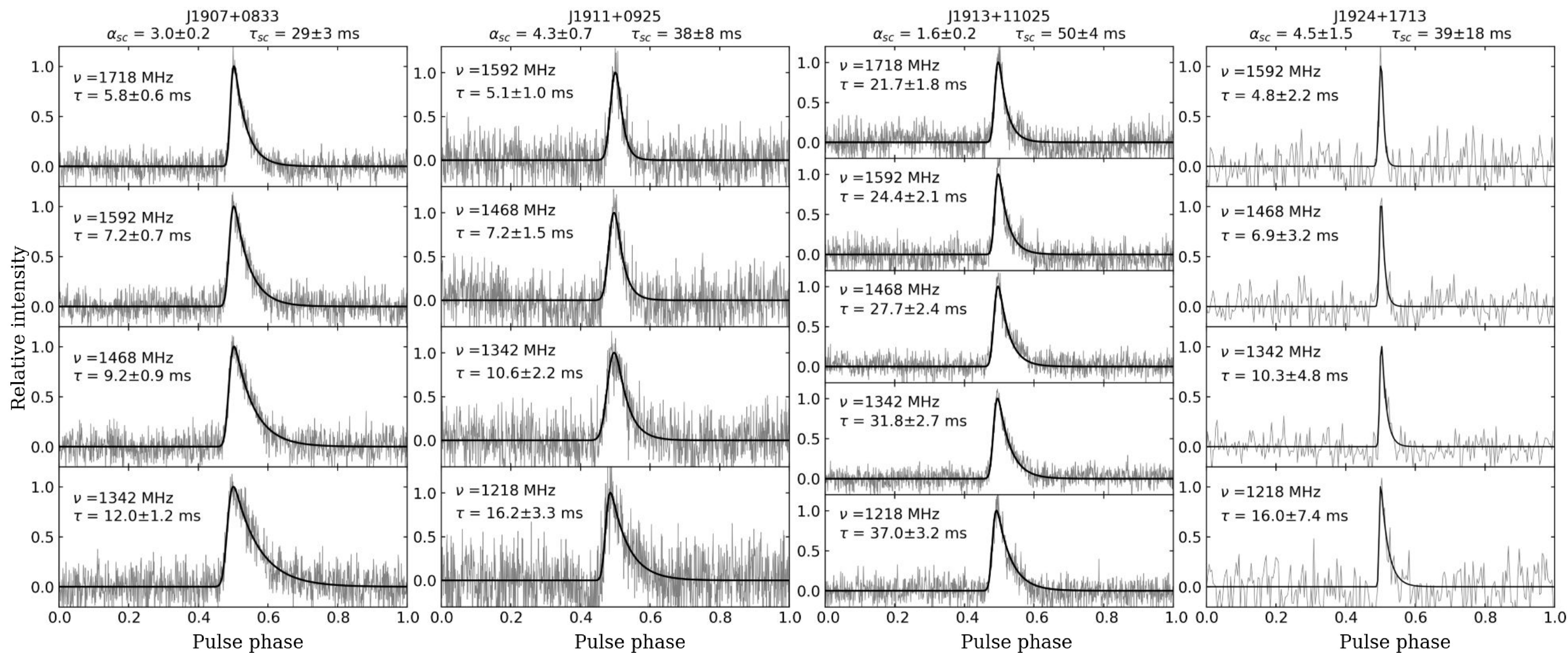}
    \caption{ Pulse profiles (grey) in different frequency subbands for the four pulsars that display scatter-broadened profiles, and best-fit pulse broadening functions (solid black) of the subbanded profiles (see text for model description). Pulsar names, best-fit scattering spectral indices and characteristic timescales at 1GHz are listed above each panel (also in Table~\ref{tab:scattering}). Within each frequency sub-panel, we provide the central observing frequency ($\nu$) of the subband and the scattering time ($\tau$) at that frequency.   }
    \label{fig:scatprofs}
\end{figure*} 
Integrated pulse profiles were examined to identify asymmetric broadening that would be indicative of frequency-dependent scattering by the turbulent interstellar medium. We attempted to quantify ISM scattering in each source by fitting a temporal pulse-broadening function to the observed pulse profile in different frequency bands. In our model, we assume that the radio waves are scattered isotropically by a single thin scattering screen. The pulse broadening function associated with this assumption takes the form $\tau_{sc}^{-1} e^{-1/\tau_{sc}}$, where $\tau_{sc}$ is the characteristic scattering time which scales with observing frequency as $\tau_{sc}(\nu)\,\propto\,\nu^{-\alpha_{sc}}$. 

In the latter expression, $\alpha_{sc}$ is the scattering spectral index. We simultaneously fit the subbanded pulse profiles as a single-component Gaussian convolved with the scattering broadening function, and compared the goodness of the fit to that of a Gaussian mixture that includes up to three components. For several of our low-S/N pulsars, we did not fit subbanded profiles but instead profiles integrated over the entire observing band.

Only four pulsars display pulse shapes that are best described by the scatter-broadening model. We show in Figure~\ref{fig:scatprofs} the fits to the subbanded pulse profiles, and best-fit scattering spectral indices $\alpha_{sc}$ and broadening timescales $\tau_{sc}$ at 1\,GHz are provided in Table~\ref{tab:scattering} along with the scattering timescales predicted by the NE2001 \citep{cl02} Galactic electron density model. Another model that has been widely used is the YMW16 \citep{ymw16} model, but unlike the NE2001 model, it does not use scattering as a modeling parameter. Instead, it estimates $\tau_{sc}$ at 1\,GHz for a given DM value based on the empirical scaling between scattering timescale and DM obtained by \cite{kmn+15}. We include those estimates in Table~\ref{tab:scattering} as well. We note significant discrepancies (by up to two orders in magnitude) between the NE2001 predictions and our measurements. These inconsistencies could be attributed to unmodeled foreground structures such as \ion{H}{2} regions. New distance and scattering measurements along various lines of sight are thus valuable to construct more complete models of the Galactic electron density in the future. 
\begin{table}[b]
  \vspace*{3.0mm}
  \centering
  {\small
  \caption{Best-fit scattering parameters for the four pulsars with pulse profiles showing measurable scatter broadening. The scattering timescales $\tau_{sc}$ listed in the fourth column were scaled to 1\,GHz using the pulsar's scattering spectral index $\alpha_{sc}$. Scattering timescales at 1\,GHz predicted by the NE2001 model ($\tau_{\rm NE2001}$) and the scaling relation obtained by \cite{kmn+15} ($\tau_{\rm kmn+15}$) are provided in the last two columns.} 
  \setlength{\tabcolsep}{1.0mm}
  \begin{tabular}{l c c c c c}
\hline
PSR      & DM  & $\alpha_{sc}$ & $\tau_{sc}$ & $\tau_{\rm NE2001}$ & $\tau_{\rm kmn+15}$ \\
(J2000)  & (pc cm$^{-3}$)&          & (ms)        & (ms)                & (ms) \\
\hline \hline
J1907+0833	& 511.68(16)& 3.0(2)    	&	29(3)	&	5.3	&	19.0 \\[-0.06em]  
J1911+0925	& 486.6(5)	& 4.3(7)    	&	38(8)	&	569	&	15.4 \\[-0.06em]  
J1913+11025	& 626.0(4)	& 1.6(2)    	&	50(4)	&	4.0	&	44.3 \\[-0.06em]  
J1924+1713	& 536.8(6)	& 4.5(1.5)    	&	39(18)	&	0.3	&	23.3 \\[-0.06em] 
\hline\\
\end{tabular}
\label{tab:scattering}
}
\end{table}

PSRs~J1911$+$0925 and J1924$+$1713 have scattering indices consistent with both an isotropic scattering mechanism ($\alpha_{sc}$\,=\,4, \citealt{c70}) and Kolmogorov turbulence in a cold plasma ($\alpha_{sc}$\,=\,4.4, \citealt{lj76,r77}), but PSRs~J1907$+$0833 ($\alpha_{sc}$\,=\,3.0) and J1913$+$11025 ($\alpha_{sc}$\,=\,1.6) have flatter scattering spectra than the aforementioned theoretical models. Anomalous scattering arising from structures in the ISM \citep{cl01,rjt+09} or anisotropic scattering mechanisms \citep{smc+01,tbw13} could explain the lower $\alpha_{sc}$ values. PSRs J1907$+$0833 and J1913$+$11025 also have large DM-estimated distances, both exceeding 8\,kpc. As such, their low $\alpha_{sc}$ could also be explained by the presence of multiple scattering screens along their long lines of sight. Dedicated observations would be required to probe the nature of the scattering observed in these pulsars -- our flux- and band-limited data sets do not allow us to distinguish between models.

\begin{figure}[b]
	\centering
	\vspace*{5.0mm}
    \includegraphics[scale=0.33]{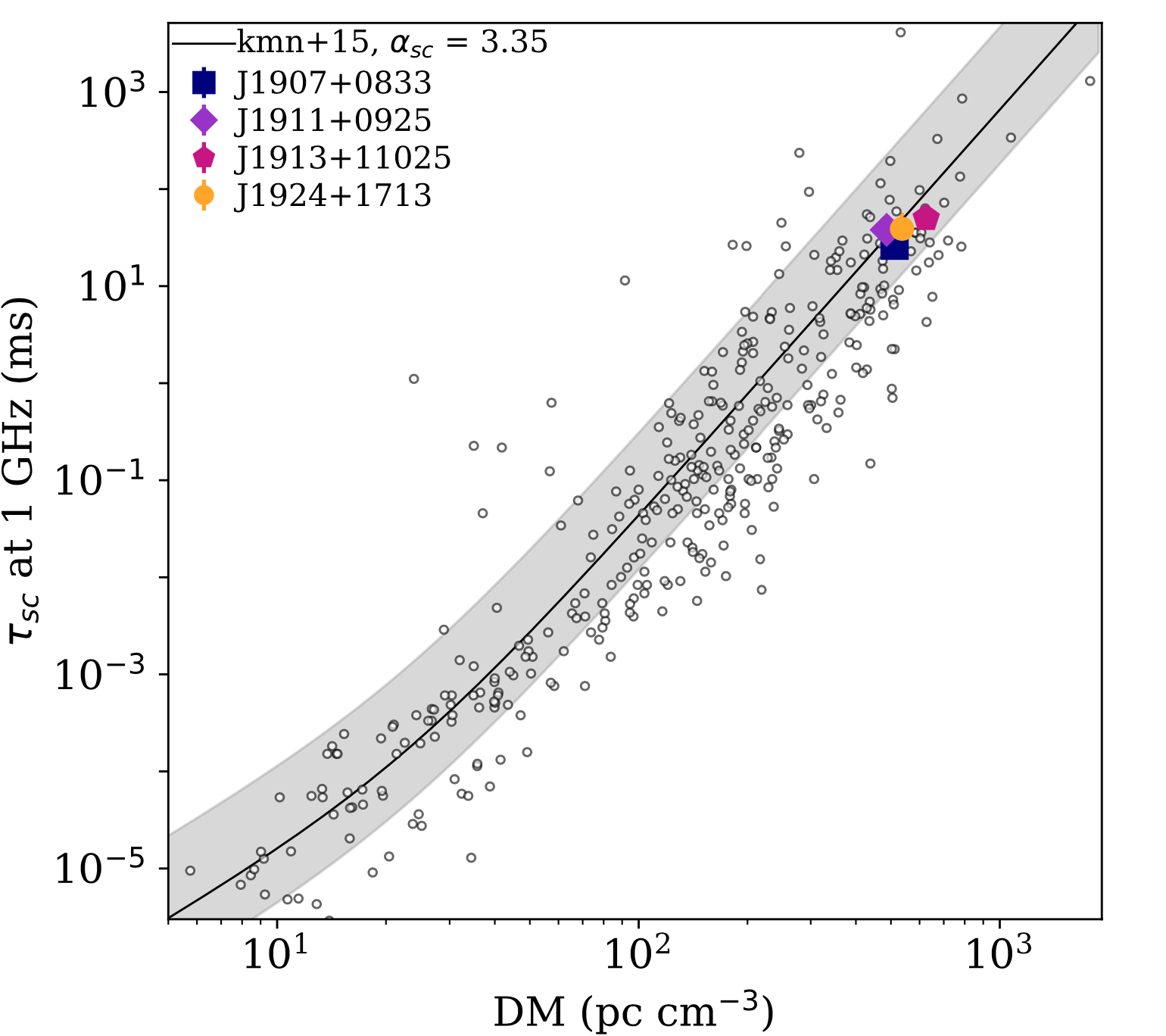}
    \vspace*{-3.0mm}
    \caption{ Scattering timescales $\tau_{sc}$ at 1\,GHz versus DM for objects with reported $\tau_{sc}$ measurements in the ATNF catalog (black circles). The four pulsars presented in this work for which we measured scatter broadening are highlighted as color-filled points.
    The solid line corresponds to the best-fit scaling relation between $\tau_{sc}$ and DM obtained by \cite{kmn+15} at 327\,MHz, scaled to 1\,GHz using a scattering spectral index of 3.35, which is the average $\alpha_{sc}$ value of our scattered pulsars, while the shaded region is the same relation, also scaled to 1\,GHz, but using the range of measured $\alpha_{sc}$  (Table~\ref{tab:scattering}). }
    \label{fig:scattering}
\end{figure}

Figure~\ref{fig:scattering} shows the distribution of $\tau_{sc}$ at 1\,GHz as a function of DM values. For comparison, we additionally plot measurements reported in the ATNF catalog (version 1.65) for other known pulsars. The best-fit solution to the empirical scaling between $\tau_{sc}$ and DM from \cite{kmn+15} at 327\,MHz, scaled to a frequency of 1\,GHz using the average value of our best-fit $\alpha_{sc}$ measurements, is also shown in Figure~\ref{fig:scattering} (solid line). We see that our sources, which have high DMs and large $\tau_{sc}$, are consistent with the scaling relation from \cite{kmn+15} for an average $\alpha_{sc}$ of 3.35. 

We also note that two low-DM sources, PSRs J1926$+$1614 (DM\,=\,24.0\,pc\,cm$^{-3}$) and J1939$+$2609 (DM\,=\,47.3\,pc\,cm$^{-3}$) displayed obvious signs of scintillation features in their spectra. From epoch to epoch, we observed fluctuations in the  flux densities of both pulsars ranging from roughly 20 to 200\,$\mu$Jy. Measurement of the scintillation bandwidth, and hence the scattering delay, for these and other low-DM PALFA sources could place complementary constraints to our pulsar broadening measurements for high-DM pulsars. This analysis is however beyond the scope of this paper.

\section{Pulsar properties} \label{sec:psrs}
\begin{figure*}[ht]
	\begin{center}
	\vspace{-4.0mm}
	   \includegraphics[scale=0.53]{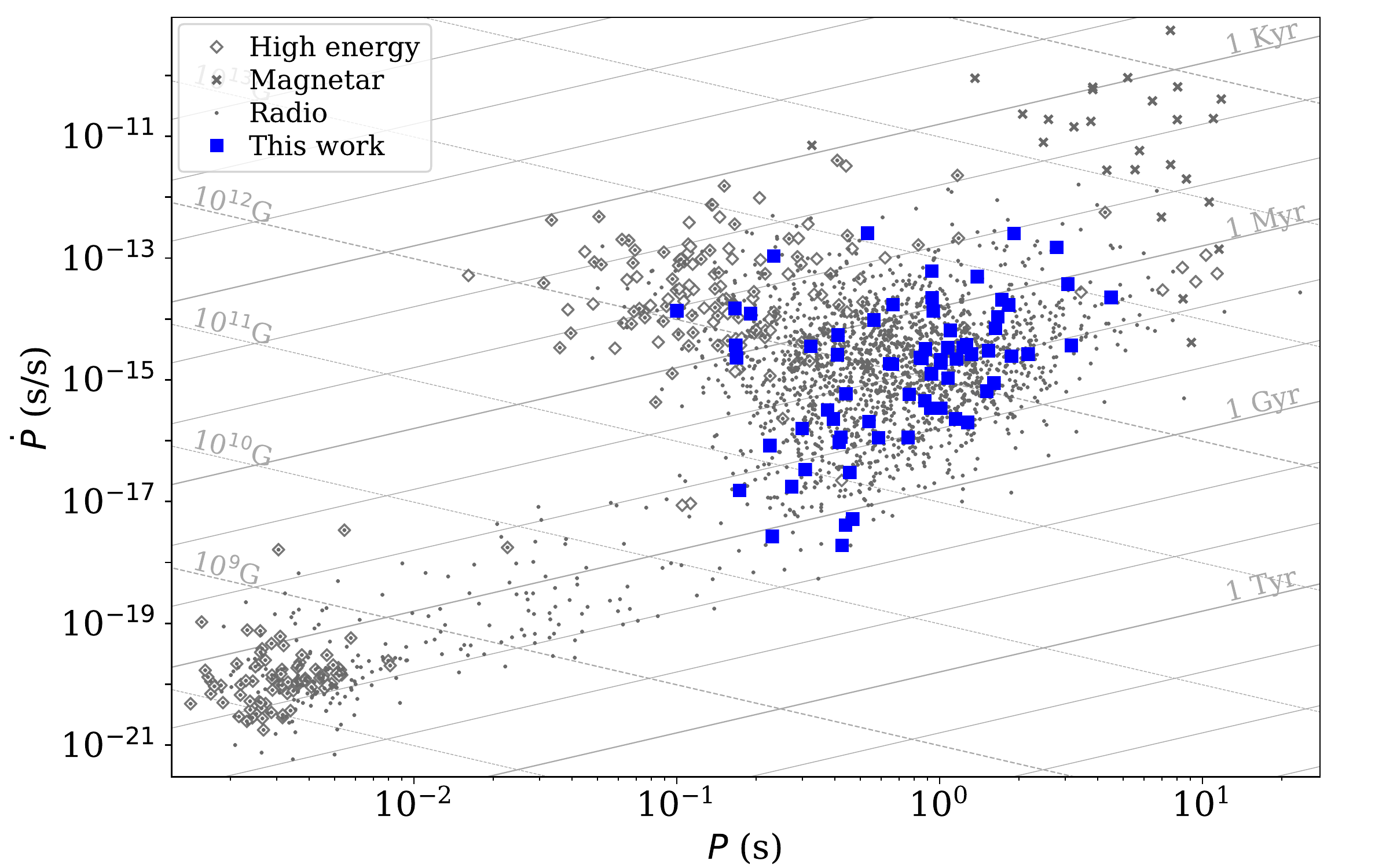} 
	\end{center}
    \vspace*{-5.5mm}
    \caption{ $P-\dot{P}$ diagram showing the periods and period derivatives of known pulsars, separated based on the nature of their emission: grey points are rotation-powered pulsars emitting in radio,  diamonds are pulsars emitting at high energies (gamma and/or X-rays), and crosses are magnetars. Pulsars presented in this paper are shown as blue squares. Lines of constant magnetic field (dashed) and characteristic age (solid) are overlaid. $P$ and $\dot{P}$ values for pulsars other than from this work were taken from the ATNF Catalog and the McGill Magnetar Catalog\textsuperscript{a} \citep{ok14}. }
    \label{fig:ppdot}
    \vspace{-0.50mm}
    \small\textsuperscript{a} \url{www.physics.mcgill.ca/~pulsar/magnetar/main.html}\\
\end{figure*}

The pulsars we present in this work are typical in their spin-down properties and representative of the broader population. Figure~\ref{fig:ppdot} shows the positions of the 68 pulsars with well measured $\dot{P}$ on a $P-\dot{P}$ diagram compared to the rest of the known pulsar population. Here, we discuss individual sources having interesting properties, including pulsars that exhibit considerable variability in their radio emission such as mode changing (Section \ref{sec:modechg}) and nulling (Section \ref{sec:nullers}, and extreme manifestations thereof in Sections \ref{sec:intermit} and \ref{sec:1928}). Our timing analysis has also revealed glitch activity in five pulsars (Section \ref{sec:glitch}), two of which are young pulsars with characteristic ages $\sim$\,30\,kyr (Section \ref{sec:young}). Another interesting source is PSR~J1954$+$2529, an unusual, non-recycled pulsar in an eccentric binary system (Section \ref{sec:1954}) which also exhibited a glitch. 

\subsection{Mode-changing pulsars} \label{sec:modechg}
Mode changing (or switching) is a type of discontinuous transition in the radio emission where the average pulse profile abruptly switches between two or more quasi-stable states \citep{b70b}. It is a broadband phenomenon \citep{bms+82} that occurs on variable timescales. Changes in the radio beam emission pattern, and hence in the observed pulse profile, are believed to be the result of a global redistribution in the magnetosphere currents and/or magnetic fields. Mode changing has been recognized in roughly two dozens pulsars thus far (see e.g., \citealt{wmj07,lhk+10,nwm+20}). Below we
describe the emission of three pulsars that displayed mode-changing behavior.

\subsubsection{PSR~J1853+0259} \label{sec:1853}
\begin{figure}[tb]
	\centering
    \includegraphics[scale=0.483]{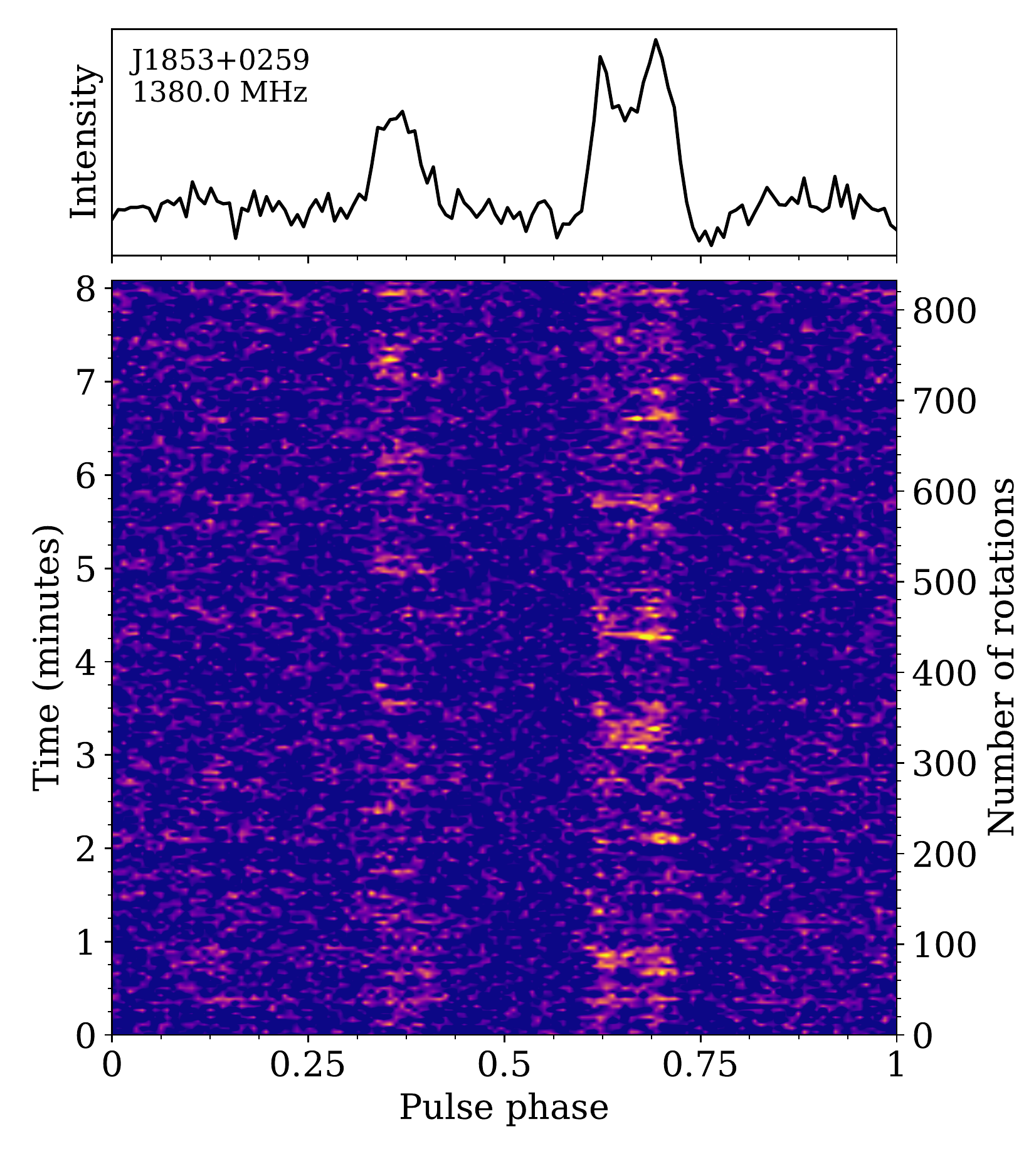}
    \vspace*{-4.3mm}
    \caption{Integrated pulse profile (\textit{top panel}) and intensity as a function of time and phase (\textit{bottom panel}), showing temporal intensity variations in the various profile components of PSR~J1853$+$0259. Individual rows in the time versus phase panel have been averaged over four pulsar rotations. The data shown here were taken with the L-Wide backend at AO on 2018 August 6.}
    \label{fig:J1853}
\end{figure}

PSR~J1853$+$0259 ($P$\,=\,585.6\,ms) has a complex pulse profile structure (shown in Figure~\ref{fig:J1853}). It exhibits two main components separated by 107$\degrees$ in pulsar rotation phase. The leading component is a weaker, single-peaked pulse of width $W50$\,=\,26\,ms (near pulse phase 0.35 in Figure~\ref{fig:J1853}), while the main component has a double-peak pulse shape of width $W50$=62\,ms (near phase 0.65 in Figure~\ref{fig:J1853}) and is about two times brighter than the leading component. 

The emission of PSR~J1853$+$0259 is notable in that it shows an uncommonly high degree of variability in its profile shape and intensity from one cycle to the next, to the point where it is comparable to a mode-changing behavior. For approximately 70\% of the time, the pulsar is in a transition state where emission is detectable in both components, albeit over a range of phase, shape and intensity. The second most common state, observed 25\% of the time, is the one where the pulsar shows emission only in the main component (phase 0.65). The remaining 5\% of the time, emission is present only in the leading pulse.  However, the timescales associated with that mode are too short -- lasting on average for a dozen rotations of the pulsar -- to allow for the average pulse profile to be stable. 

\subsubsection{PSR~J1858+0239} \label{sec:1858}
\begin{figure}
	\centering
    \includegraphics[scale=0.73]{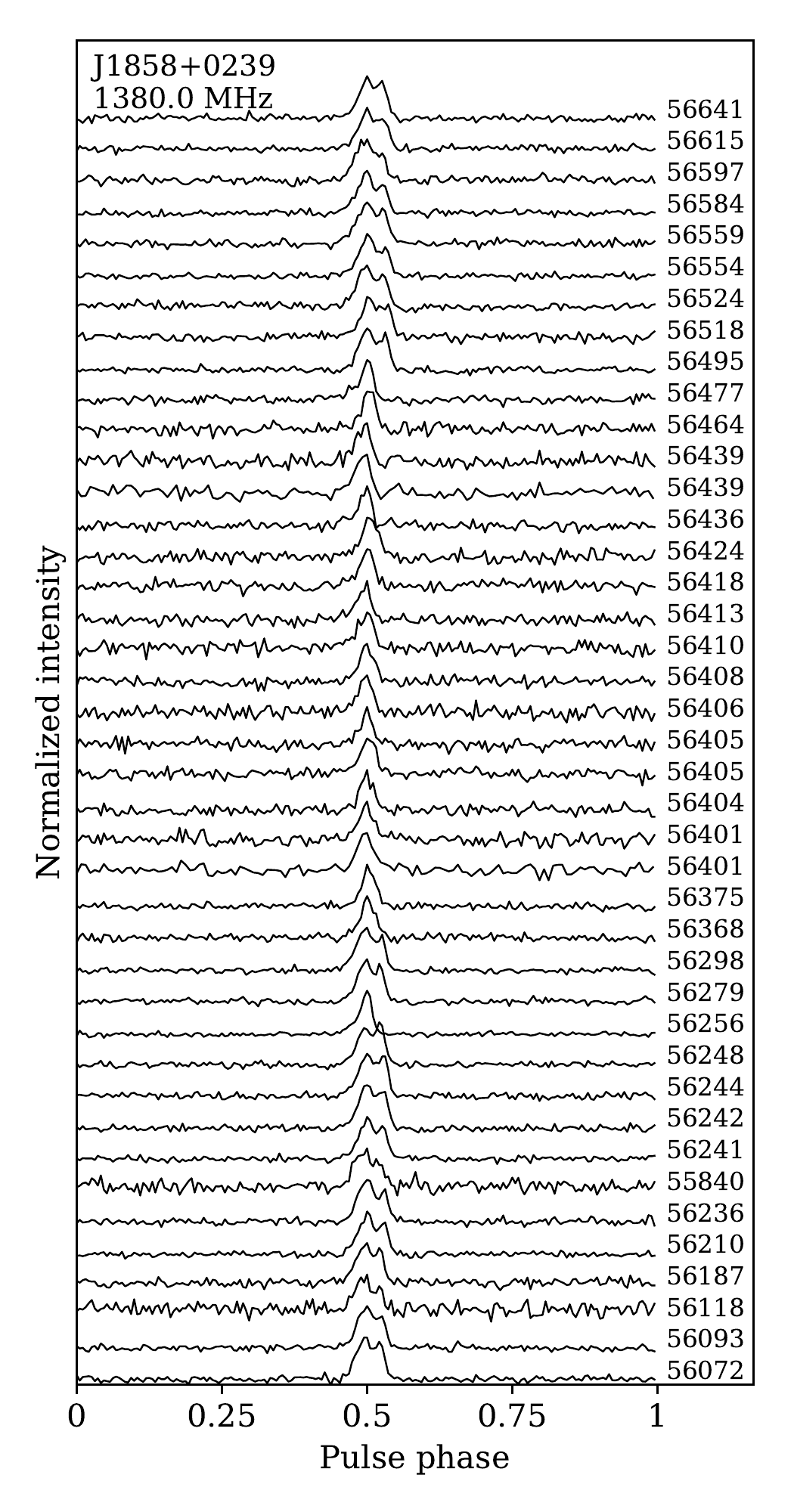}
    \vspace*{-3.5mm}
    \caption{ PSR~J1858$+$0239 average pulse profiles. All data presented here were collected at AO. Because we do not have a phase-connected ephemeris for this source, the profiles are not phase-aligned but simply have their largest-amplitude peak centered at pulse phase 0.5. Profiles are stacked in chronological order, with the MJD listed on the left of each profile.  Partial coherence was achieved for TOAs extracted for MJDs between 56401 and 56439.}
    \label{fig:j1858}
\end{figure}

PSR~J1858$+$0239 ($P$\,=\,197.6\,ms) exhibits significant jitter noise in its integrated profiles. It appears to switch between two equally prevalent emission modes, displaying either a faint and narrow ($S_{1400}\,\approx\,65\,\mu$Jy, $W50\,\approx\,5$\,ms) roughly single-peaked profile or a brighter and wider ($S_{1400}\,\approx\,160\,\mu$Jy, $W50\,\approx\,11$\,ms) profile having two peaks. In the latter mode, we have seen large fluctuations in the relative intensity of the leading and trailing peaks from one epoch to the next (see integrated profiles in Figure~\ref{fig:j1858}). In most (but not all) observations, the leading peak is stronger than the trailing peak. Our short 5-min timing observations do not allow us to determine whether profile variations in the double-peak mode are caused by self-noise in the pulsar emission mechanism \citep{k89,gjs+11,jg12} or if they are associated with distinct magnetospheric states with different period derivatives. The lack of a coherent timing solution further prevents us from comparing the pulsar rotation phase of the emission, which could have provided the means for determining whether the dissimilarities in profile shapes between the emission modes are due to jitter or if they are more likely linked to changes in the configuration of the radio beam. 

In addition, during one observation (MJD 56186) we observed the pulsar emitting in the bright, double-peak mode ($S_{1400}\,\approx\,143\,\mu$Jy) for the first half of the observation before transitioning to a nulling state (see Section~\ref{sec:nullers}), where emission was off for the remaining half of the scan (no detectable emission above $\sim15\,\mu$Jy). \cite{vkr+02} found that PSR~B0809+74 emits in a distinct mode following null episodes. In our case however, the observation on MJD 56186 unfortunately ended before pulses could be detected again, and the following observation was carried out only three weeks later, at which point the pulse profile was double-peaked. Thus, we are not in a position to examine the immediate impact of the nulling episode on the pulsed emission. Yet, the presence of nulling in PSR~J1858$+$0239 demonstrates that magnetospheric state transitions are taking place.

As mentioned earlier in Section \ref{sec:psrtiming}, we have been unable to produce a fully coherent timing solution for PSR~J1858$+$0239. The potentially large $\dot{P}$ -- suggesting a relatively young characteristic age ($\sim 10^5$\,yrs) -- and possibly glitch activity could explain why phase connection could not be achieved. Another possibility is that the changes in the pulsar magnetosphere causing the mode-changing and nulling behavior also affect the torque on the neutron star, which would be detectable as variations in the pulsar period derivative. Correlations between average pulse profiles and spin-down rate have been reported in a number of pulsars (e.g., \citealt{lhk+10,crc+12,llm+12,lsf+17,slk+19}), the most notable being PSR~B1931$+$24, for which weeks-long nulls are associated with a 50\% decrease in $\dot{P}$ \citep{klo+06}. We note that the two aforementioned possible explanations are not mutually exclusive -- \cite{wje+11} showed that intermittent and erratic emission events in PSR~J1119$-$6127 were preceded by a large amplitude glitch, whereas timing observations of PSR~B1828$−$11 revealed glitch activity \citep{elsk11} in addition to variations in spin-down rate correlated with observed pulse shape \citep{sls00,slk+19}. Interestingly, the latter is a relatively young pulsar ($\tau_c\sim110$\,kyr) whose emission switches between a narrow and brighter profile and a wide and fainter profile \citep{sls00,slk+19}. This is similar to what we observed in PSR~J1858$+$0239, but contrarily to PSR~B1828$−$11, the former's narrow profile mode is associated with fainter emission. PSR~J1858$+$0239 is thus an interesting object for future studies of physical processes at play in pulsar magnetospheres, the origin of mechanisms that trigger emission state transitions and their impact on timing behaviors.
\subsubsection{PSR~J1914+0625} \label{sec:1914}
\begin{figure}[b]
	\centering
	\vspace*{3.0mm}
    \includegraphics[scale=0.665]{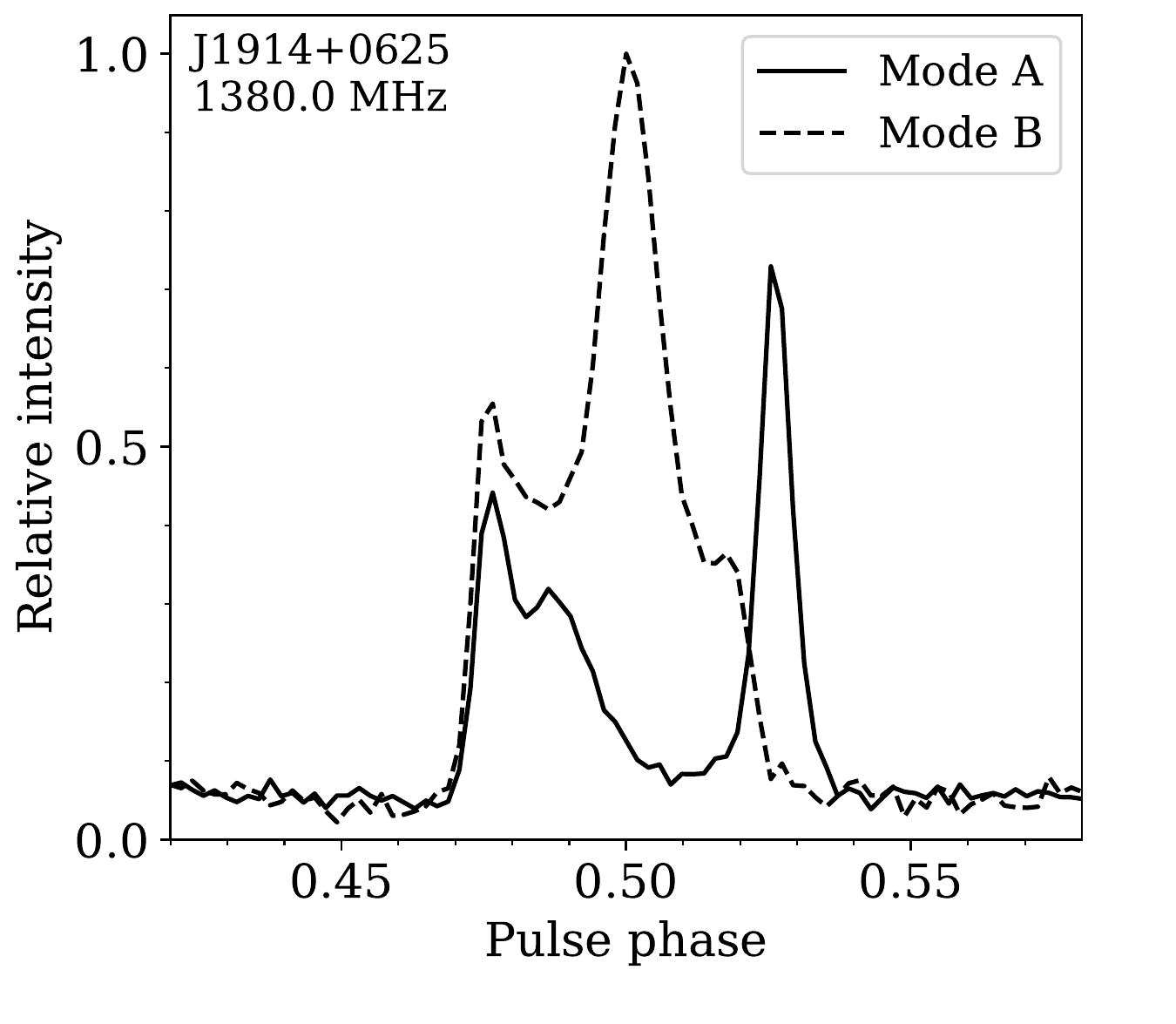}
    \vspace{-8.0mm}
    \caption{ Phase-aligned pulse profiles of the two modes displayed by PSR~J1914$+$0625, calibrated based on their relative fluxes. The data used to generate these integrated profiles were collected with the L-Wide backend at AO.}
    \label{fig:j1914}
\end{figure}

\cite{r93} suggested an empirical classification for radio pulsar pulse profiles that is based on a core/cone emission model. In this model, the radio beam consists of a central core component with nested pairs of conal emission. The intensity and phase separation of components of an observed integrated profile depend on the sightline traverse geometry and the intrinsic shape of the emission beam. Following that classification scheme, PSR~J1914$+$0625 ($P$ = 0.879\,s) has a five-component (class M) average pulse profile, where pairs of inner and outer conal emission surround a core component. We observe two distinct emission modes in the integrated profiles, each exhibiting three of the five components. Figure~\ref{fig:j1914} shows the phase-aligned pulse profile of both modes and their relative intensity. 

In $\sim$ 90\% of our observations, the pulsar emitted radiation in the normal mode (mode A, plotted with a solid line in Figure~\ref{fig:j1914}). Mode A is characterized by emission in both the leading and trailing components of the  outer cone, which are separated by 17.6$\degrees$ in pulse longitude, where the leading peak intensity is roughly 60\% that of the trailing peak. There is also fainter emission in the leading inner cone, which is $\sim$3.5$\degrees$ from the leading outer cone in longitude, and there is no discernible core structure. Only the trailing cone is resolved at 50\% level of peak intensity, and the corresponding width ($W50$) we measure  is 6.2\,ms. The width at 10\% of the peak intensity ($W10$) for the entire profile in mode A is 56.2\,ms, and the average pulsed flux density $S_{1400}$\,=\,52\,$\mu$Jy.

The brightest configuration is the abnormal mode (mode B), where the strongest emission originates from the core component. The early part of the profile in mode B is illuminated, but the core and the two leading conal components are unresolved. Therefore we do not know  whether the inner conal emission is active in this mode. The amount of energy coming from the trailing inner cone in mode B is similar to that of the inner leading cone in mode A,  but there is no detectable emission from the trailing outer cone. 
In mode B, the core is separated by 8.4$\degrees$ from the leading outer cone and by 6.3$\degrees$ from the trailing inner core. The core has a peak width at half maximum $W50$\,=\,14.2\,ms and it is the only resolved structure at that intensity level. The entire emission of the mode-B profile has a width $W10$\,=\,48.3\,ms and $S_{1400}$\,=\,94\,$\mu$Jy. 

Individual pulses are too faint to investigate potential pulse nulling or sub-pulse drifting. No transition has been observed within one of our 900-s timing observations (all carried out with Arecibo), implying that both modal emissions are stable on timescales longer than 900\,s.  Polarization data is unavailable for this source as timing data were only recorded in total intensity mode.

\subsection{Nulling pulsars} \label{sec:nullers}
First reported by \cite{b70a}, the nulling phenomenon is a sudden cessation of detectable pulsed emission that lasts for one or more pulsar cycles. RRATs and intermittent pulsars (discussed in the next sections) are extreme manifestations of pulsar nulling. The triggering mechanism responsible for nulling remains largely unknown, but it is believed to be intimately related with mode changing (e.g., \citealt{vkr+02,rwr05,wmj07}). In general terms, most interpretations of the absence of emission in nulling pulsars invoke processes occurring in the magnetosphere, for example a loss of plasma conditions required for coherent emission \citep{fr82}, or intense time-varying pulse modulations where the radio flux density drops below the detection threshold (e.g., \citealt{elg+05}). 

Nulling behavior has been reported in approximately 10\% of the radio pulsar population, but the proportion of pulsars experiencing nulling could be much larger since most known pulsars are not monitored regularly for the purpose of identifying and studying nulling behavior. For instance, through regular monitoring with CHIME/Pulsar, \cite{nwm+20} recently reported on the first detections of nulls from bright pulsars discovered decades ago, suggesting that nulling is a phenomenon much more common than previously thought. On the other hand, misinterpreting undetectable emission in data from a given (sensitivity-limited) instrument as a null, as opposed to an extrinsic reduction in the radio flux density (e.g., due to a reduction in intrinsic luminosity or due to scintillation), results in an overestimation of the size of this sub-population. 

\begin{figure*}[b]
    \begin{center}
        \centerline{\includegraphics[scale=0.5]{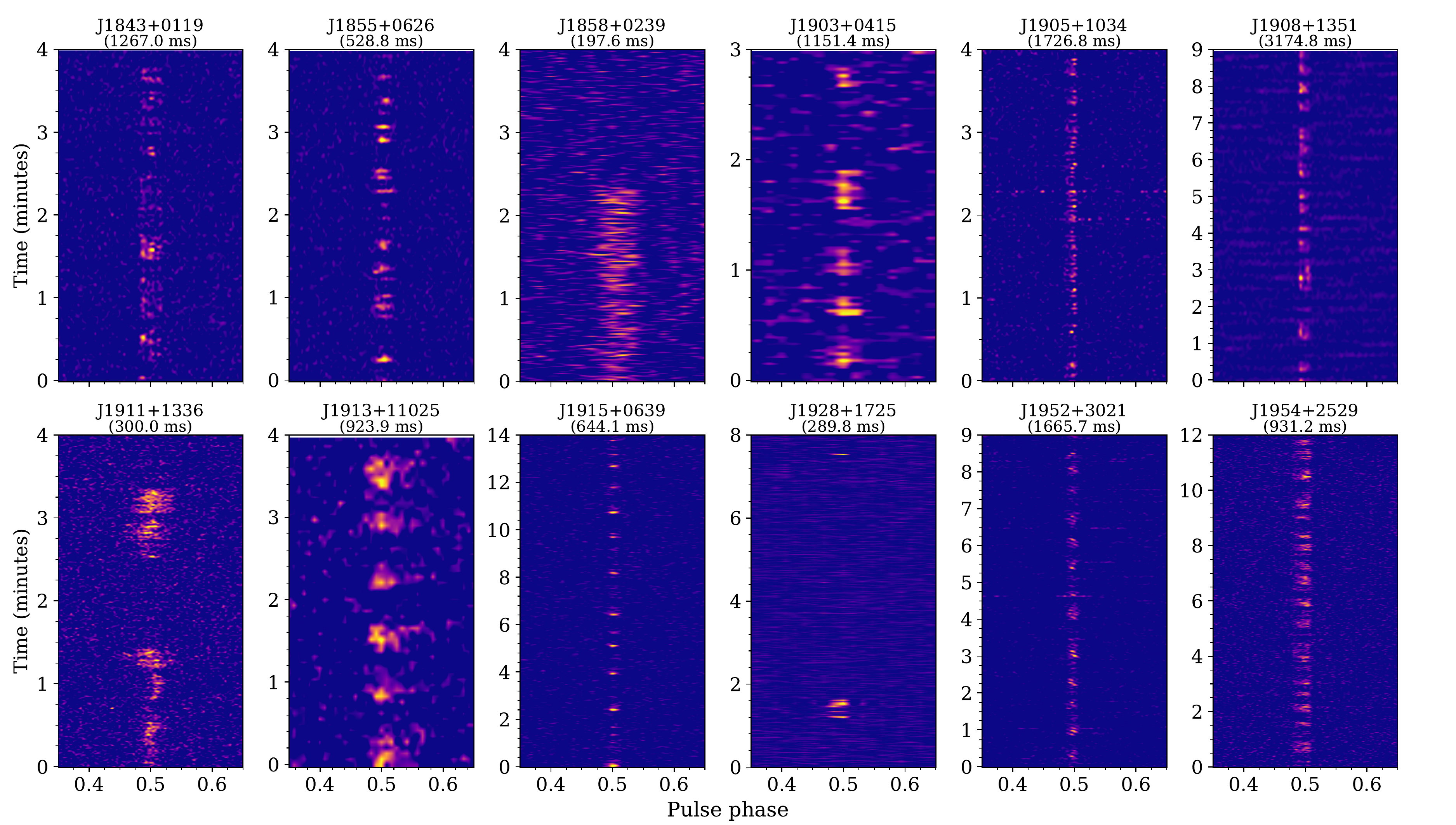}}
    \end{center}
    \vspace*{-9mm}
    \caption{Phase-time plots showing the variation in intensity at 1.4\,GHz in single Arecibo observations for the 12 pulsars that display nulling behavior. }
    \label{fig:nullers}
\end{figure*}
Among the pulsars presented in this work, we identify 12 pulsars that displayed discernible nulls. We show examples of the emission intensity as a function of time in Arecibo observations for each pulsar in Figure~\ref{fig:nullers}. Because of the low flux densities of these pulsars, data had to be folded into subintegrations ranging from 2 to $\sim$20 pulsar cycles in length. Only the RRAT, PSR~J1928+1725 (whose properties are described separately in Section \ref{sec:1928}), has individual pulses with signal-to-noise ratios sufficiently high that averaging over multiple rotation is not required.

Dedicated observations with long integrations are preferable for characterizing nulling properties such as nulling fractions and average duration of nulling episodes. This is especially important for pulsars showing long nulling episodes. Such observations were not carried out for this work; all the pulsars here were observed through the same follow-up program at AO for timing purposes, with scans duration ranging from 5 to 15 minutes. Thus, we do not attempt to determine nulling parameters, except for RRAT J1928$+$1725 which we discuss in Section~\ref{sec:1928}. 

\subsection{Intermittent pulsars} \label{sec:intermit}

While the distribution of nulling fractions in intermittent pulsars can be similar to that in nulling pulsars (e.g., \citealt{gjk12}), the cessation of pulsed emission in intermittent pulsars is seen on timescales lasting from days to years (e.g., \citealt{klo+06,lsf+17}), orders of magnitude longer than nulling pulsars. Below we discuss the properties of two intermittent pulsars identified in our sample: PSRs~J1855$+$0626 and J1925$+$2513. 

\subsubsection{PSR~J1855+0626}  \label{sec:1855}
\begin{figure*}[bt!]
    \begin{center}
        \centerline{\includegraphics[scale=0.63]{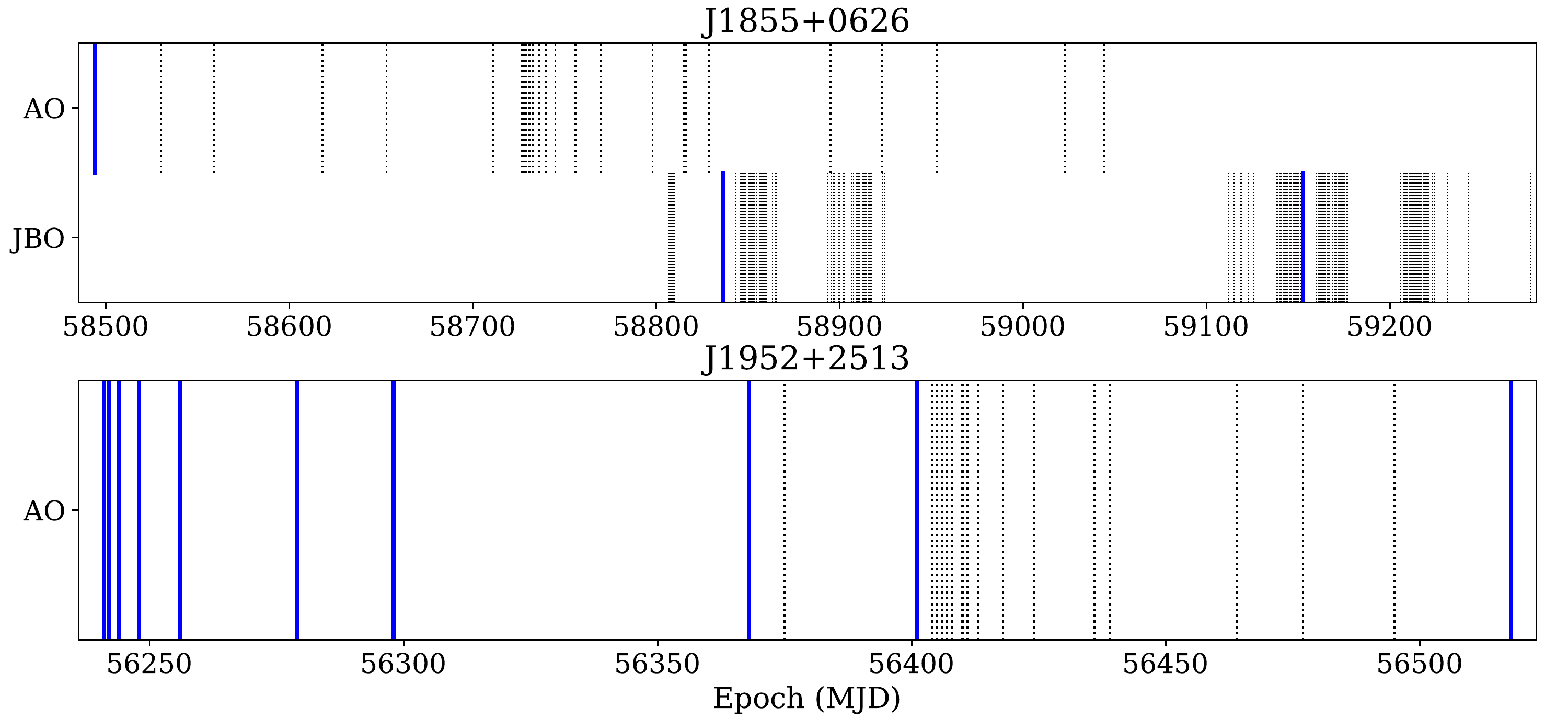}}
    \end{center}
    \vspace*{-8mm}
    \caption{Epoch of observations where pulsations were detected (solid blue) and were not detected (dotted black) in observations of PSRs~J1855$+$0626 (\textit{top}) and J1952$+$2513 (\textit{bottom}). For PSR~J1855$+$0626, AO and JBO data sets are shown on the upper and lower rows, respectively. PSR~J1952$+$2513 was observed at AO exclusively.
    }
    \label{fig:intermit}
\end{figure*}

PSR~J1855$+$0626 ($P$\,=\,528.8\,ms) was discovered in 2018 May survey data as a relatively bright pulsar ($S_{1400}$=130\,$\mu$Jy) exhibiting short (average duration $\sim$\,16 \,rotations) but frequent (nulling fraction of $\sim$ 39\%) nulling episodes. From the time of discovery until 2020 August, we carried out 36 300-s follow-up observations of the pulsar at AO, for a total of 3 hours. The pulsar was only redetected once; it was in a quiescent state the rest of the time (upper limit $S_{1400}\,\lesssim$\,12\,$\mu$Jy). An extensive follow-up campaign was also carried out at JBO. A total of 148 30-min observations at JBO between 2019 November and 2021 March led to only two detections of the pulsar. Hence PSR~J1855$+$0626 was inactive in 98.4\% of our observations. The top panel of Figure~\ref{fig:intermit} shows the timeline of the observations and epochs where the pulsar was active are highlighted in blue. 

In 2020 June, we began observing  PSR~J1855$+$0626 with CHIME/Pulsar on a regular basis, but to date we have not been able to detect the pulsar. We note however that it is possible that radio pulsations were being emitted during an observation but that their flux density in the CHIME band was below the instrument sensitivity. Due to the rarity of  detections, we do not have a phase-coherent timing  solution for PSR~J1855$+$0626. 
\subsubsection{PSR~J1952+2513}  \label{sec:1952}
We have observed PSR~J1952$+$2513 ($P$\,=\,1077.6\,ms) for a total of 2.6 hours with AO in 2012 and 2013 (26 sessions). The pulsar was observed in an active state in 39\% of our observations. With an average pulsed flux density of only 36\,$\mu$Jy, the pulsar is one of the faintest in our sample. In some of the follow-up data, measured $S_{1400}$ values have been as low as $\sim$15\,$\mu$Jy,  near the sensitivity limit for most of our timing observations at AO. These variations are not caused by diffractive scintillation - the DM (246.9\,pc\,cm$^{-3}$) of the pulsar is too large for that - but instead caused by variations in the telescope sensitivity and the RFI environment, which may have hindered our ability to identify weak emission during what we considered inactive states. Thus, it is possible that we are overestimating the fractional time spent in the inactive state. Another consequence of the low brightness of PSR~J1952$+$2513 is that, unlike PSR~J1855$+$0626 we are unable to identify and characterize potential nulling behaviors when the pulsar is active. 

One interesting feature of this pulsar is that, if it is indeed an intermittent pulsar, its $\dot{P}$ is one order of magnitude smaller than expected from the observed relation between
$P$ and $\dot{P}$ for other intermittent pulsars by \cite{lsf+17} (see their Fig. 7). That relation suggested that all intermittent pulsars have the same spin-down energy and the same acceleration potential above their polar caps. Thus, PSR~J1952$+$2513 might be indicating that intermittent pulsars have a wider distribution of spin-down energies than previously recognized.

\subsection{RRAT~J1928+1725} \label{sec:1928}
\begin{figure}[bt!]
    \begin{center}
    \includegraphics[scale=0.58]{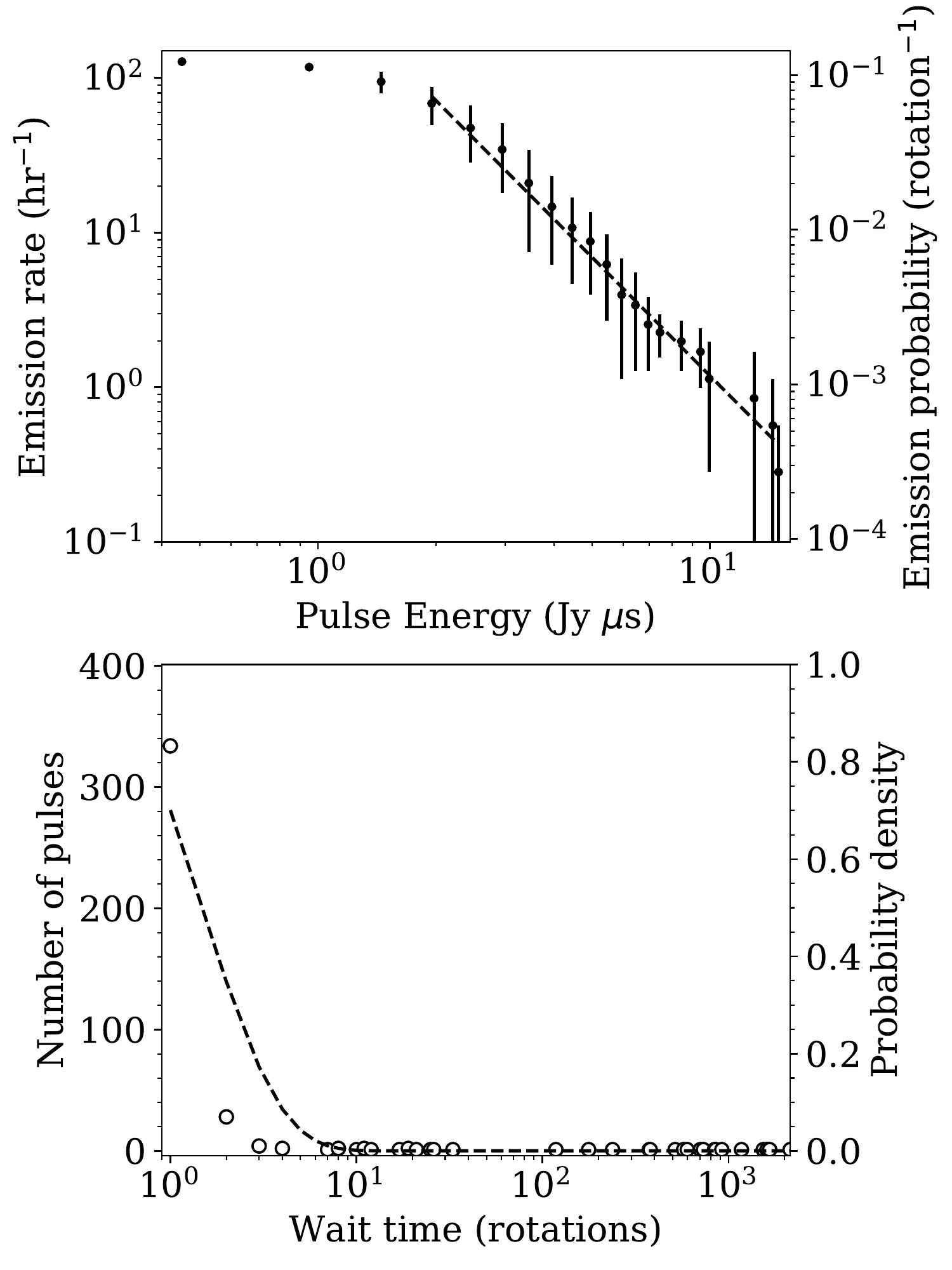}
    \end{center}
    \vspace*{-3.5mm}
    \caption{ Single pulses statistics for RRAT~J1928$+$1725. \textit{Top}: Cumulative emission rate and emission probability of single pulses exceeding a given pulse energy, binned in intervals of 0.5\,Jy\,$\mu$s. The dashed line shows the power law that best fits the pulse energy distribution, which has an index of --2.6(2). \textit{Bottom}: Distribution of wait times between consecutive single pulse detections, which shows that during the short, bright episodes where the RRAT is active, most detectable pulses are emitted consecutively or within very few rotations of the pulsar. The dotted curve is an exponential fit to the wait-times distribution, i.e., the expected distribution if the emission of pulses is a Poisson process. The emission mechanism responsible for RRAT~J1928$+$1725's bright single pulses produces pulses that are more clustered than a Poisson process.     
    }
    \label{fig:J1928}
 \end{figure}

PSR~J1928$+$1725 ($P\,=\,289.8\,$ms) is a RRAT that emits bright, heavily clustered pulses followed by long (hundreds of rotations) nulls. Other than the single pulses themselves, no underlying signal is visible when folding the data. We analyzed the distribution of wait times between consecutive pulses, and find that this distribution is inconsistent with a Poisson process (see Figure~\ref{fig:J1928}). The longest active phase we observe lasts 29 consecutive pulsar cycles, while the longest inactive phase lasts 1662 cycles ($\sim$482 s). Active phases are also rare -- we detect neither single-pulse or periodic emission in 3/17 of our 15-min observations of PSR~J1928$+$1725. The average detection rate is 78 pulses per hour.

Despite having detected numerous bright pulses, the long time gaps between observations during which pulses are detected makes it difficult to maintain timing coherence over a few epochs, and thus we have been unable to solve this RRAT.  PSR~J1928$+$1725 could be similar to the young and energetic ($\tau_c\,=\,865$\,kyr and $\dot{E}\,=\,4.6\times10^{34}$\,erg\,s$^{-1}$) 125-ms PSR~J1554$-$5209 \citep{kle+10}, one of the very few RRATs having a spin period $<\,500\,$ms for which $\dot{P}$ has been measured\footnote{Based on the ATNF catalog, \cite{mht+05}}. \cite{kkl+11} conducted a timing analysis on  PSR~J1554$-$5209 along with a dozen more RRATs with longer periods and lower $\dot{E}$, and noted that PSR~J1554$-$5209 displays a far larger scatter in its timing residuals than the other, lower-$\dot{E}$ RRATs. Significant jitter in pulse phase similar to PSR~J1554$-$5209 could explain why we have been unable to derive a coherent timing ephemeris for PSR~J1928$+$1725. During our phase connection attempt, it also appeared that the pulsar may be suffering from  significant glitch activity, although we cannot confirm this.  Of the 34 known RRATs with reported spin-down rate measurements \citep{mht+05}, PSR~J1819$-$1458 \citep{mll+06} is the only one for which glitch activity has been reported \citep{bls+18}. While the former has a much longer rotation period ($P$\,=\,4.263\,s) than PSR~J1928$+$1725, it has a relatively small characteristic age of 120\,kyr. If PSR~J1928$+$1725 is indeed young and energetic, the presence of glitches could very well explain why we have been unable to solve it.
Follow-up observations of PSR~J1928$+$1725 were carried out with CHIME/Pulsar from 2020 October to 2021 April, but no pulses were detected. Establishing phase coherence would require regular and long-integration observations with an instrument whose sensitivity is similar to that of Arecibo. 

\subsection{Glitching pulsars} \label{sec:glitch}
Rotational instabilities in pulsars are generally explained by either timing noise or glitches. Whereas timing noise arises from random spin fluctuations over long timescales (e.g., one of our pulsars, PSR~J0608$+$1635, displays intense timing noise; see residuals in Figure~\ref{fig:residuals}), a glitch is a discrete transition in the pulsar rotational state marked by an abrupt increase in spin frequency, typically with frequency jump magnitudes between $10^{-6}\,\nu$ and $10^{-10}\,\nu$, and often accompanied by a decrease in frequency derivative $\dot{\nu}$ \citep{elsk11}. Glitches are believed to be caused by erratic transfers of angular momentum from the superfluid inside the star to the more slowly rotating (and cooling) crust \citep{ai75,rzc98}. Hence their study represents a unique opportunity to gain insight into the internal structure of neutron stars. 

\begin{figure*}[t]
    \begin{center}
        \centerline{\includegraphics[scale=0.7]{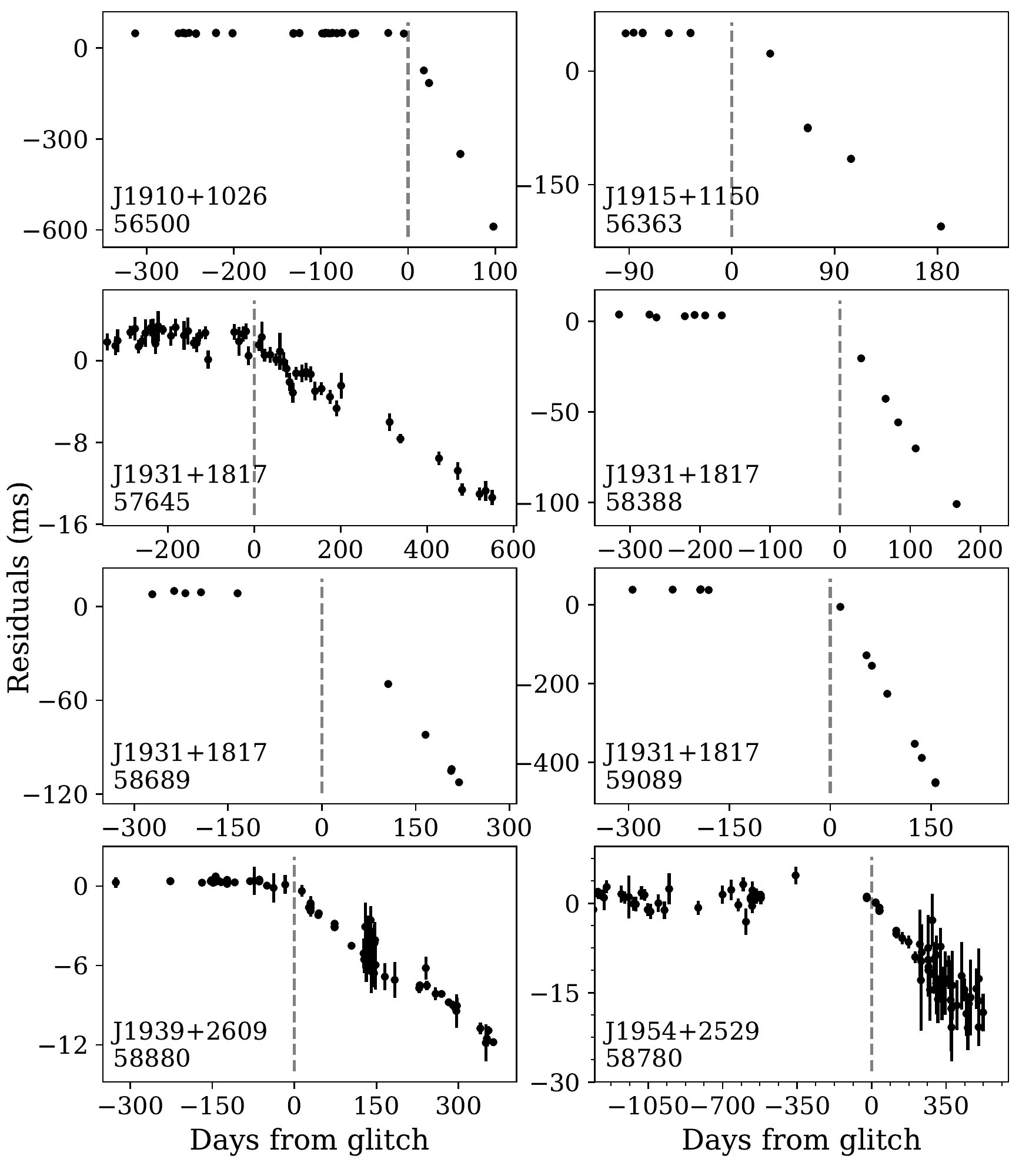}}
    \end{center}
    \vspace*{-10mm}
    \caption{Glitch detections in the five pulsars exhibiting glitch activity, showing the residuals obtained when not fitting for glitch parameters (Table~\ref{tab:glt}) from the timing model and keeping best-fit rotation and position parameters fixed (Table~\ref{tab:timing}). Residuals for the four glitches detected in PSR~J1931$+$1817 are shown separately, each time by using a timing solution that includes all glitch parameters (fixed) and excluding only the ones associated with the glitch being displayed. Best estimates for the MJD of the glitch epochs are listed below the pulsar name in each panel.  \\}
    \label{fig:glt}
\end{figure*}

Eight glitches of moderate and small magnitude were detected in five of the 72 pulsars presented here during the timing program. Figure~\ref{fig:glt} illustrates the glitch signatures in timing residuals when they are not included in the timing model, signatures that are characterized by the sudden onset of a steady decrease towards negative residuals. We are able to maintain phase coherence over the time gap between observations around the glitch epochs when using the best-fit glitch parameters listed in Table~\ref{tab:glt} in our timing models. Residuals corresponding to the ephemerides that contain the best-fit glitch parameters are the ones shown in Figure~\ref{fig:residuals}. As a result of sparse observation cadences and/or high levels of timing noise and/or large TOA uncertainties, we are unable to detect the changes in the spin-down rate ($\Delta \dot{\nu}$) during recovery for three glitch events. 

The three youngest pulsars presented in this work, PSRs~J1910$+$1026 ($\tau_c$ = 33\,kyr), J1931$+$1817 ($\tau_c$ = 35\,kyr), and J1915$+$1150 ($\tau_c$ = 116\,kyr), have all exhibited glitch activity. This is not surprising, as young pulsars are known to display higher glitch activity (e.g., \citealt{ml90,lss00,js06,elsk11}). 

The largest glitch we observed occurred in PSR~J1910$+$1026, the youngest pulsar in our set, with a fractional step in spin frequency of $78.4 \pm 0.8 \times 10^{-9}\,\nu$. Due to its low flux density (58\,$\mu$Jy at 1.4\,GHz), timing data were  solely collected at AO and the pulsar was observed roughly twice a month for $\sim$\,1.5\,years. The sparse observations combined with the relatively large timing residuals ($\sim\,1\,$ms) makes the identification of smaller glitches difficult. For the aforementioned reasons, the glitch activity of PSR~J1910$+$1026 is not well constrained and could be much higher than what our dataset reveals. 

\begin{table}[t]
  \centering
  \caption{Glitch parameters for the five pulsars that displayed glitch activity. For each glitch, we list its epoch and the fractional change in spin frequency $\Delta\nu/\nu$ and in frequency derivative $\Delta\dot{\nu}/\dot{\nu}$. Numbers in parentheses are the 1$\sigma$ uncertainties on the last digit reported by \tempo\, after weighting the TOAs such that $\chi^2$ = 1.} 
  \setlength{\tabcolsep}{1.2mm}
  \begin{tabular}{l c c r r }
\hline
PSR & Glitch No. & Epoch & $\Delta \nu / \nu$ & $\Delta \dot{\nu} / \ \dot{\nu}$ \\ 
(J2000) &        & (MJD) & ($10^{-9}$)        & ($10^{-3}$) \\ 
 \hline \hline
J1910$+$1026  &   1 &    56500(2)  &   78.4(8)   &    1.6(3) \\[-0.05em] 
J1915$+$1150  &   1 &    56363(4)  &   29.19(7)   &    $-$20.51(7) \\[-0.05em] 
J1931$+$1817  &   1 &    57645(9)  &   0.314(9)   &     $-$ \\[-0.05em] 
            &   2 &    58388(2)  &   8.8(3)    &     0.49(7) \\[-0.05em]
            &   3 &    58689(6)  &   6.3(7)    &     $-$ \\[-0.05em]
            &   4 &    59089(2)  &   36.5(7)   &     0.22(9) \\[-0.05em]
J1939$+$2609  &   1 &    58880(2)  &   0.333(3)   &     $-$ \\[-0.05em]
J1954$+$2529  &   1 &    58780(9)  &   0.61(5)   &     7(2) \\
\hline\\[-0.6em]
  \end{tabular}
  \label{tab:glt}

\end{table}

On the other hand, PSRs~J1931$+$1817, J1939$+$2609 and J1954$+$2529 were bright enough to be followed up at JBO and CHIME/Pulsar, and benefited from more regular timing observations over a longer time span. Over the course of our 5-yr follow-up campaign of PSR~J1931$+$1817, we detected four glitches ranging from 3.3$\times10^{-10}$ to $3.7\times10^{-8}\,\nu$ in size. 
A small glitch ($\Delta \nu/\nu=6.0\times10^{-10}$) was observed in PSR~J1954$+$2529, a relatively old ($\tau_c\,=\,$11.7\,Myr)  non-recycled pulsar in a binary system which we discuss in Section \ref{sec:1954}. 

Even more surprising is that PSR~J1939$+$2609, an old pulsar with characteristic age $\tau_c\,\sim\,1.4\,$Gyr, suffered a glitch, albeit of small magnitude 3.3$\times10^{-10}\,\nu$. Only two other glitching pulsars with ages $>\,100\,$Myr are known in the Galactic field: PSRs~B1913$+$16 with $P\,=\,59\,$ms and $\tau_c\,=\,109\,$Myr \citep{ht75} and J0613$-$0200, a 3.1-ms MSP with $\tau_c\,=\,5\,$Gyr \citep{lnl95}. 
Similarly to PSR~J1939$+$2609, only one glitch of small magnitude has been detected in each system, with $\Delta \nu/\nu=3.7\times10^{-11}$ for PSR~B1913$+$16 \citep{wnt10,wh16} and 2.5$\times10^{-11}$ for PSR~J0613$-$0200 \citep{mjs+16}.  

We note however that for PSRs~J1939$+$2609, J1954$+$2529 and J1931$+$1817 (first glitch; MJD 57645), we cannot positively rule out the possibility that the features in the timing residuals are caused by timing noise rather than glitch activity. While the latter explanation minimizes the post-fit RMS residuals in these systems and is thus currently preferred over the former, future modelling of their long-term timing behavior will be needed to be certain that those are real glitches and not part of large timing noise features. 

\subsection{Young Pulsars}  \label{sec:young}
Among the 69 pulsars with coherent timing solutions, only two are young ($\tau_c < 10^5$\,yr) objects: PSRs~J1910$+$1026 ($P\,=\,531.5\,$ms) and J1931$+$1817  ($P\,=\,234.1\,$ms) with characteristic ages of 33 and 35\,kyr, respectively. Pulsars of such young age are often found in supernova remnants (SNR). Despite the lack of proper motion measurements -- important to confirm any potential pulsar/remnant association -- we searched for coincident SNR by cross-matching the pulsar positions against Green's SNR Catalog \citep{g19}. Accounting for potential angular offsets that could arise from large post-supernova tangential velocities, we search over a conservative region of 0.5$\degrees$ radius around each pulsar. No remnant coincident with PSR~J1910$+$1026 was identified, which is not surprising given its very large inferred distance ($\gtrsim$ 15\,kpc). One catalogued object, SNR~G53.41+0.03 \citep{awb+17}, is found 28$'$ away from the timing position of PSR~J1931$+$1817. \cite{ddv+18} carried out a deep radio search for a pulsar associated with G53.41+0.03, covering a 10$'$ region around the center of the SNR (the estimated size of the remnant), but did not detect pulsations. They also investigated and characterized G53.41$+$0.03 using multi-wavelength data, and estimated its distance at roughly 7.5\,kpc and its age at $\sim$ 1000 to 8000\,yrs. Given the 28$'$ angular offset and adopting a distance of 7.5\,kpc and the most conservative age of 8000\,yrs, an unreasonably high spatial velocity ( $>\,12\times10^3$\,km\,s$^{-1}$) would be required for PSR~J1931$+$1817 to have traveled from its birth location near the center of the remnant to its current position. We therefore rule out SNR~G53.41$+$0.03 as being associated  with PSR~J1931$+$1817. 

Due to their large spin-down power $\dot{E}$, young pulsars sometimes have high-energy counterparts. However, we did not find any gamma-ray point sources or pulsations in \textit{Fermi} LAT data for these pulsars (analysis described in Section \ref{sec:fermi}). This is also not surprising given the large distances of the pulsars and the corresponding ``heuristic'' energy fluxes $\sqrt{\dot{E}}/d^2$  of $\sim 3\times 10^{15}$ and $\sim 5\times  10^{15}$\,(erg\,s$^{-1} $)$^{1/2}$\,kpc$^2$ for PSRs~J1910$+$1026 and J1931$+$1817, respectively, below the approximate LAT threshold of $10^{16}$\,(erg\,s$^{-1}$)$^{1/2}$\,kpc$^2$ \citep{aaa+13,sbc+19}.

\subsection{An unusual binary system: PSR~J1954+2529} \label{sec:1954} 
Of all the new pulsars presented here, only one, PSR~J1954+2529, is a member of a binary system. It has an orbital period of 82.7 days and, interestingly, a system eccentricity of 0.11. The system has been monitored with Arecibo and CHIME, but mostly with the Lovell telescope. PSR~J1954+2529 is also among our five pulsars that displayed a glitch. The Keplerian parameters of the pulsar's orbit and some derived quantities are presented in Table~\ref{tab:binary}. Included in our timing model is the  relativistic advance of the angle of periastron, $\dot{\omega}$, which in GR is a function of the total mass of the system, $M_{\rm tot}$. Unfortunately our timing data do not yet provide a significant measurement (see upper limit in Table~\ref{tab:binary}), and the 3-$\sigma$ upper limit on$\dot{\omega}$ dot doesn't yield a meaningful constraint on $M_{\rm tot}$.

The pulsar does not appear to be recycled: with $P \, = \, 0.931\, \rm s$, a characteristic age of about 12 Myr and a magnetic field at the surface of about $10^{12}\, $G, it is near the center of the cloud of ``young'' pulsars in the $P$-$\dot{P}$ diagram (see Figure~\ref{fig:ppdot}). Such middle-aged radio pulsars in binary systems are rare but interesting from the point of view of stellar evolution.

In what follows, we discuss three possibilities for the formation and nature of this system. At the moment, we have no data to decide in favor of any hypothesis, but the last hypothesis is currently preferred based on theoretical arguments from stellar evolution theory. 

\begin{table}
\begin{center}
\caption{ PSR~J1954+2529's Keplerian orbital parameters. Numbers in parentheses are \texttt{TEMPO}-reported uncertainties in the last digit quoted. Minimum and median companion masses are calculated assuming a pulsar mass of $1.4 \, M_{\odot}$. We also provide the 3$\sigma$ upper limit on the relativistic advance of periastron produced by our timing model.   \label{tab:binary}}
\setlength{\tabcolsep}{0.95mm}
\begin{tabular}{l c}
\hline
\multicolumn{2}{l}{Orbital parameters} \\
\hline\hline
Binary model                                                &   DD       \\[-0.25em]
Orbital period, $P_b$ (d)                                   &   82.71733(5)   \\[-0.25em]
Proj. semi-major axis, $x=a$sin$i/c$ (lt-s)             &   71.62485(9) \\[-0.25em]
Time of periastron, $T_{0}$ (MJD)                           &   58094.3699(6)  \\[-0.25em]
Eccentricity, $e$                                           &   0.114023(3)  \\[-0.25em]
Longitude of periastron, $\omega$ (deg)                     &   78.545(2)  \\[0.4em]
\hline
\multicolumn{2}{l}{Derived Parameters} \\
\hline\hline
Mass function ($M_\odot$)                          &   0.0576608(2)       \\[-0.25em]
Minimum companion mass ($M_\odot$)        	       &   0.62         \\[-0.25em]
Median companion mass  ($M_\odot$)         	       &   0.74         \\[-0.25em]
Advance of periastron, $\dot{\omega}$ (deg/yr)     &   $<$0.003     \\
\hline
\end{tabular}
\end{center}
\end{table}

\subsubsection{A progenitor to a low-mass X-ray binary}

Although most massive stars that eventually form pulsars are born in binary systems (e.g., \citealt{a83,dse+01,se11}), when one of the stars in these systems (generally the most  massive) undergoes a supernova (SN), they will in most cases become unbound. This is why, among normal pulsars, isolated pulsars are hundreds of times more prevalent than binaries. 

\begin{table*}[t]
\begin{center}
    \caption{Comparison between the pulsars with massive main sequence (MS) companions and systems like PSR~J1954+2529, ordered by increasing characteristic age ($\tau_c$). Minimum companion masses $M_c$ assume a pulsar mass of 1.4\,$M_\odot$. References listed in the last column are, by order of publication: 1: \cite{lm89}, 2: \cite{jml+94}, 3: \cite{bbs+95}, 4: \cite{kbm+96}, 5: \cite{hlk+04}, 6: \cite{lfl+06}, 7: \cite{bbn+11}, 8: \cite{msk+12}, 9: \cite{sjm14}, 10: \cite{lsk+15}, 11: \cite{hnl+17}, 12: \cite{lsb+17}, 13: \cite{bsb+19}. 
    Values for PSR~J1954+2529 are from this work. All low-eccentricity systems similar to PSR~J1954+2529 are older than pulsars with MS companions, their magnetic fields are only slightly smaller.}
    \label{tab:comparison}
    \begin{tabular}{l c r c c r c r c c c}
    \hline
    Pulsar  & $P$ & $\dot{P}$    &  $\tau_c$ & $B$            & $P_b$ & $e$ & Min. $M_c$ & Mass function  & Comp. & Ref. \\
        & (s) & ($10^{-15}$) &  (Myr)    & ($10^{12}\,$G) & (days)&     & ($M_\odot$)&  ($M_\odot$)   &      &      \\
    \hline\hline
    J2032+4127    & 0.1432   & 11.3   & 0.21   & 1.29 & 16835	 & 0.964	& 11.86	& 9.488	& MS  & (10,11) \\[-0.15em]
    B1259$-$63    & 0.0478   & 2.28   & 0.33   & 0.33 & 1236.7   & 0.870	& 3.17	& 1.529	& MS  & (2,9) \\[-0.15em]
    J1740$-$3052  & 0.5703   & 25.5   & 0.35   & 3.86 & 231.0	 & 0.579	& 11.07	& 8.723	& MS  & (7,8) \\[-0.15em]
    J1638$-$4725  & 0.7639   & 4.8    & 2.53   & 1.93 & 1940.9   & 0.955	& 5.90	& 3.852	& MS  & (6) \\[-0.15em]
    B1820$-$11    & 0.2798   & 1.38   & 3.22   & 0.63 & 357.8	 & 0.795	& 0.66	& 0.068	&          & (1,5) \\[-0.15em]
    J0045$-$7319  & 0.9263   & 4.46   & 3.29   & 2.06 & 51.2	 & 0.808	& 3.97	& 2.170	& MS  & (3,4) \\[-0.15em]
    \bf{J1954+2529} & 0.9312 & 1.26   & 11.7   & 1.10 & 82.7     & 0.114    & 0.62  & 0.058 &           & This work \\[-0.15em]
    J1932+1500    & 1.8643   & 0.459  & 64.4   & 0.94 & 198.9	 & 0.029	& 0.33	& 0.012	&           & (12) \\[-0.15em]
    J1822$-$0848  & 0.8348   & 0.135  & 97.7   & 0.34 & 286.8	 & 0.059	& 0.33	& 0.012	&           & (6) \\[-0.15em]
    J1837$-$0822  & 1.0992   & 0.121  & 144    & 0.37 & 98.4	 & 0.024	& 0.28	& 0.008	&           & (13) \\
    \hline
    \end{tabular}
\end{center}
\end{table*}
Systems that survive the first SN will consist of a young pulsar in an eccentric orbit with a bright, massive main-sequence (MS) star. There are six such systems in the literature, all with massive companions, they are listed in Table~\ref{tab:comparison}. The orbits of most of these systems are quite unlike that of PSR~J1954+2529; not only are their eccentricities much higher ($e \, > \, 0.5$) -- so high in some cases that they suggest near disruption of the system -- but their companions are much more massive ($M_c \, > \, 5 \, M_{\odot}$). 

If the companion to PSR~J1954+2529 is a MS star, it has to be much less massive (consistent with the system's mass function), and therefore much cooler, consistent with the non-detection of an optical counterpart. An examination of existing point-source catalogs revealed no point source within 5$''$ of the position of the pulsar, which has only a 0.15$''$ uncertainty.
The DM-estimated distance of PSR~J1954$+$2529 is 6.8 and 7.8\,kpc for the NE2001 and YMW16  models, respectively. The magnitude limits for a 10$-\sigma$ detection in the 2MASS point-source catalog are 15.8, 15.1 and 14.4 for J, H and K s bands respectively \citep{scs+06}. We use this survey because these near-infrared bands are least affected by reddening, up to 2.33, 1.484 and 1.000 magnitudes of absorption in these bands \citep{2011ApJ...737..103S}\footnote{ \url{https://irsa.ipac.caltech.edu/applications/DUST/}}. Given the pulsar's distance estimates, the 2MASS magnitude limit, and the lack of a detected point source at the pulsar position, a MS companion to PSR~J1954+2529 must have a mass smaller than $2 \, M_{\odot}$ and therefore a spectral type later than F0.

The problem with such an identification is that no such systems have been otherwise identified: all MS companions to unrecycled pulsars are quite massive. However, such systems should exist, because they are the theoretical
progenitors to low-mass X-ray binaries (LMXBs). The hypothesis of a MS companion can be investigated further via deeper optical observations of the position of the system. 

\subsubsection{Is the pulsar the younger degenerate object in the system?}

Most known binary pulsars evolve as millisecond pulsars - following one (or more) episode of mass transfer, the neutron star is spun up to short spin periods and the orbit is circularized. If the donor has a low mass, it evolves into a white dwarf (WD). This clearly does not describe the evolution of PSR~J1954+2529. However, if the donor is massive enough, it may undergo a SN and become a NS, possibly causing the orbit to be eccentric again. The companion would then be observed as a young pulsar. 

\begin{figure}[t]
    \vspace*{-4.5mm}
    \begin{center}
        \centerline{\includegraphics[scale=0.64]{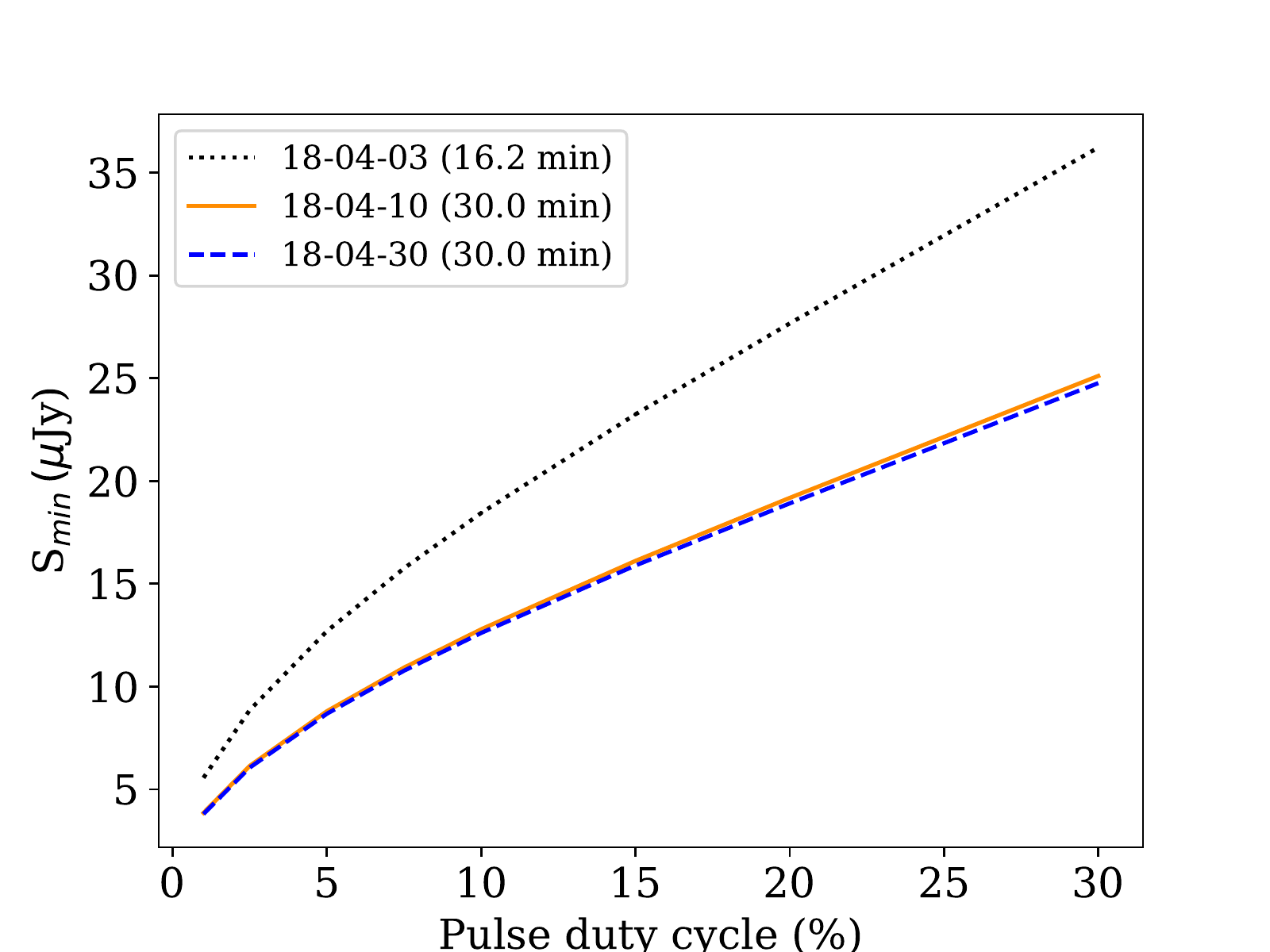}}
    \end{center}
    \vspace*{-6mm}
    \caption{Upper limits for the flux density at 1.4\,GHz of a possible radio pulsar companion to PSR~J1954+2529. \label{fig:flux_density_companion}}
\end{figure}

The only confirmed case of a young pulsar observed in a double NS system is that of PSR~J0737$-$3039B, the second pulsar of the double pulsar system \citep{lbk+04}. There are other candidates, like the young pulsars J1906+0746, the first binary system found in the PALFA survey \citep{lsf+06,lks+15} and J1755$-$2550 \citep{nkt+18}, however, no pulsations from a recycled pulsar companion were detected in either system. For PSR~J1954+2529, we obtained several long observations in search mode with the Arecibo telescope. Because of heavy RFI, only three proved usable, with integration lengths of 16, 30 and 30 minutes. Dedispersing the data from these observations at the DM of PSR~J1954+2529, and then doing a simple Fourier transform (no acceleration is expected given the long orbital period) did not reveal any other radio pulsars in the same observations. Assuming, conservatively, a minimum S/N of 10, we can derive, based on the radiometer equation, upper limits for the flux density of the companion as a function of the assumed pulse duty cycle. These are presented in Figure~\ref{fig:flux_density_companion}. At the assumed distance of 7.6\,kpc, these translate (assuming a pulse duty cycle of 5\%) to a pseudo-luminosity of $\sim$0.45\,mJy\,kpc$^2$, placing it within the bottom 7.5\% of all radio pulsars with reported pseudo-luminosities at 1400 MHz in the ATNF pulsar catalog. This does not by itself exclude the possibility of a double neutron star system -- the first pulsar could be pointed away from our line of sight, or it could just be very faint -- but does not support it either.


The older degenerate objects in the PSR~J1906+0746 and PSR~J1755$-$2550 systems can alternatively be massive WDs, as in the case of PSR~B2303+46 \citep{tamt93,kk99}. The formation and evolution of these systems from nearly equal-mass MS binaries is described in detail by \cite{ts00}, who predicted that PSR~J1141$-$6545 \citep{klm+00} also has a massive WD companion. This was confirmed by the measurement of the companion mass, $1.02 \, \pm \, 0.01 \, M_{\odot}$ \citep{bbm+08} and by optical observations \citep{abw+11}. The detailed evolution of the system was finally proven when the companion WD showed clear signs of having been spun up by accretion of mass from the progenitor of the pulsar \citep{vbs+20}. One apparent advantage of this hypothesis is that the companion mass estimates of PSR~J1954+2529 are, indeed, in better agreement with the possibility of a massive WD.

Let us look in more detail into this possibility. During the giant phase of the pulsar progenitor, there was likely mass transfer to the companion of PSR~J1954+2529, whether a WD or NS. This would have circularized the orbit.
After the SN, the mass loss and kick would greatly increase the the orbital eccentricity, to values that are, as we will see below, likely much higher than 0.11. However, it is possible the PSR~J1954+2529 formed in a electron capture supernova (ECSNe) - which could in principle occur for companions with initial masses near $8 M_{\odot}$. If these are perfectly symmetric, with no associated kick, then the post-SN eccentricity will be given by eq. (20) of \cite{tkf+17}:
\begin{equation}
e = \frac{\Delta M}{M_{T}},
\end{equation}
where $\Delta M$ would be the mass loss during the SN (which includes the binding energy of the NS) and $M_T$ is the total mass of the system after the SN. For the double pulsar system, we know the second SN had a kick between 0 and 50 km s$^{-1}$, which is small compared to the orbital velocities in that system, thus without significant effect on the eccentricity or even the orbital inclination of the system (\citealt{fsk+13}, see discussion in \citealt{tkf+17} and references therein). This means that we can estimate the mass loss associated with the second SN: it was such that the post-SN eccentricity was $\sim 0.1$. This is similar to what we observe for PSR~J1954+2529.

However, the problem for PSR~J1954+2529 is that it has a much wider orbit.
Extensive simulations carried out by \cite{nkt+18} and \cite{ktl+18} show very few binaries with $e \, < \, 0.5$ for orbital periods of 80 d or larger (see their Fig. 7).
The reasons for this are twofold: first, for more widely separated binaries, mass stripping - which determines, in these simulations, the kick magnitude for iron core collapse SNe - is less effective, and the larger envelopes result in larger kicks. Second, the simulation also includes ECSNe, with kick velocities between 0 and $50 \rm \, km \, s^{-1}$ with a flat probability distribution \citep{ktl+18}; these are consistent with the estimates for the second NS in the double pulsar \citep{tkf+17}.
Even such small kicks will, when added to the small orbital velocities of wider systems, significantly raise their orbital eccentricities. In a few cases where the kicks point nearly 180 degrees away from the orbital motion of the system the resulting orbital eccentricities are low, however such cases are relatively rare.
Thus, if one assumes these simulations - and in particular, their assumptions regarding SN kick distribution - faithfully capture reality, then it is highly unlikely (but not impossible) that PSR~J1954+2529 is the second degenerate object in the system.

\subsubsection{A non-recycled pulsar - massive WD system?}

A third possibility is discussed in detail by \cite{tlk12}: as described above, after the SN that formed the first NS, the latter was in an eccentric and wide orbit with its massive MS star companion. The companion then evolves into the asymptotic giant branch, with a weakly bound envelope and a massive stellar wind. These circularize the orbit either partially or, if the orbit is very wide, not at all. These stellar winds will also cause very mild accretion into the pulsar, barely spinning it up. 
If its initial mass is below $\sim 8 \, M_{\odot}$, the companion will eventually evolve into a massive white dwarf. Since the system's mass loss is slow, there would be no further changes to the orbital eccentricity of the system.

There are four other binary pulsars thought to have formed in this way: three with partially circularized orbits (PSRs~J1822$-$0848, J1837$-$0822 and J1932+1500, the latter found in PALFA), and one with a wider, eccentric orbit, PSR~B1820$-$11 (see Table~\ref{tab:comparison}). The orbital characteristics and B-field of PSR~J1954+2529 are similar to those of the three partially circularized systems. Finally, the mass functions for PSR~J1954+2529 and B1820$-$11, and the minimum companion mass estimates derived from them are in good agreement with the possibility of a massive WD companion, and significantly larger than the mass for a He WD predicted for their orbital periods by \cite{ts99}. This is not the case for the other three partially circularized systems, which have smaller mass functions and minimum masses of about $0.3 \, M_{\odot}$; this does not exclude the possibility of massive WD companions, as it could be caused by lower orbital inclinations. From this point of view, PSR~J1954+2529 and to some extent PSR~B1820$-$11 provide the best evidence that at least some of these systems are associated with massive WDs.

There is a simple consistency test for this hypothesis. If it is true that the systems similar to PSR~J1954+2529 started as a PSR -- O/B systems, and later evolved to PSR -- massive WD systems, then the main difference between them should be, apart from the companion mass and orbital eccentricity, the characteristic age: the systems with O/B companions must be young enough to still have unevolved O/B companions, whereas the systems like J1954+2529 should be older. The magnetic fields of the latter pulsars should be similar, but slightly smaller, not only because of possible minor accretion, but also because of observational biases: larger B-fields cause a larger braking torque, which means that a pulsar will reach the death line sooner. Ordering Table~\ref{tab:comparison} by characteristic age, we see that the results are in agreement with this expectation. Interestingly, PSR~B1820$-$11 is slightly younger than PSR~J0045$-$7319. Assuming that the characteristic age is a close indicator of actual age, this would suggest that the companion of PSR~B1820$-$11 was slightly more massive than the companion of PSR~J0045$-$7319, which made it evolve faster into a massive WD.

B1820$-$11 and J1954+2529 are the youngest systems within their class. This raises the possibility that, if their companions really are massive WDs, they might still be hot and bright enough for detection in deeper optical observations. 

\section{Search for gamma-ray associations}  \label{sec:fermi}
Where most non-recycled pulsars have spin-down powers  $\dot{E}\gtrsim10^{34}$\,erg\,s$^{-1}$ \citep{aaa+13}, gamma-ray pulsations have been found in low-$\dot{E}$ radio pulsars by folding \textit{Fermi} LAT data using their ephemerides. For instance,  \cite{sbc+19} found pulsations in PSR~J2208+4056 \citep{slm+14}, a pulsar with $\dot{E}=8\times10^{32}$\,erg\,s$^{-1}$, a spin-down power well below the ``deathline'' for gamma-ray emission \citep{a96,aaa+13,khk+17}. Hence, we search for gamma-ray counterparts to all pulsars presented here. 

First, we attempt to identify gamma-ray point sources coincident with our pulsars by cross-matching the \textit{Fermi} Large Area Telescope fourth catalog\footnote{\url{https://heasarc.gsfc.nasa.gov/W3Browse/fermi/fermilpsc.html}} of gamma-ray sources \citep{fermi_10yr} against the timing positions (or discovery positions, if unsolved) of our 72 pulsars. 
We find three unidentified \textit{Fermi} point sources for which the reported semi-major axis of the error ellipse at the 95\% confidence level is coincident with our sources and no other known radio pulsars. 

The first is 4FGL J1912.0$+$0927, coinciding with PSR~J1911$+$0925, a pulsar with $\tau_c\,=\,1.4\,$Myr and $\dot{E}\,=\,4\times10^{33}$\,erg\,s$^{-1}$. The \textit{Fermi} source was detected at a 7.0$-\sigma$ significance level and has a log-normal spectrum, common in gamma-ray pulsars. NE2001 and YMW16 suggest distances of 6.9\,kpc and 8.1\,kpc for this pulsar, corresponding to energy fluxes $\sqrt{\dot{E}}/d^2$ of $\sim 1.3\times 10^{15}$ and $\sim 1.0\times  10^{15}$\,(erg\,s$^{-1} $)$^{1/2}$\,kpc$^2$, roughly one order of magnitude below the LAT $10^{16}$\,(erg\,s$^{-1} $)$^{1/2}$\,kpc$^2$ threshold. However, it is possible that the DM distances are overestimated and thus we cannot rule out the association based on $\sqrt{\dot{E}}/d^2$ alone.    

The second possible association is 4FGL J1911.3$+$1055, a low-significance (4.32\,$\sigma$) point source that has a power-law spectrum, also consistent with gamma-ray pulsar spectra, and a position matching that of PSR~J1911$+$1051, one of our glitching pulsars. That pulsar is relatively young ($\tau_c\,=\,249\,$kyr), has a spin-down power of $7\times10^{34}$\,erg\,s$^{-1}$ and its DM-derived distances are 9.0 and 10.1\,kpc for the NE2001 and YMW16 models, respectively, corresponding to low $\sqrt{\dot{E}}/d^2$ values of $\sim 3.3\times 10^{15}$ and $\sim 2.6\times  10^{15}$\,(erg\,s$^{-1} $)$^{1/2}$\,kpc$^2$. 

Finally, 4FGL J1929.0$+$1729 is coincident with the discovery position of our RRAT, PSR~J1928+1725 ($P\,=\,289.8\,$ms). The detection significance of this \textit{Fermi} source is $\sigma\,=\,14.8$ and it has a log-normal spectrum. Because we were unable to achieve phase connection for PSR~J1928+1725, its age and spin-down power (and thus $\sqrt{\dot{E}}/d^2$) are unknown. As discussed in Section~\ref{sec:1928}, we suspect that our inability solve PSR~J1928$+$1725 is a result of the pulsar being young, energetic, and potentially suffering from glitch activity. It is common for young gamma-ray pulsars to exhibit glitches (e.g., \citealt{rkp+11,rap+12,pga+12,abb+13,ggy+20}). Thus 4FGL J1929.0$+$1729 may indeed be associated with PSR~J1928$+$1725, but this cannot be confirmed without a coherent timing solution.

In an attempt to confirm the potential \textit{Fermi} point sources associations with PSRs~J1911$+$0925 and J1911$+$1051 as well as to detect pulsations from any other sources that would not have met the various selection criteria used in constructing the \textit{Fermi} 4FLG catalog \citep{fermi_10yr}, we use the timing ephemerides of solved pulsars to fold $\sim$11\,years of \textit{Fermi} LAT data. We extract photons in the energy range 100\,Mev$\,<\,E_\gamma\,<\,500$\,GeV and follow a similar procedure as \cite{sbc+19}, but because our pulsars are located in the Plane where the background level is high, we use a high energy scale value $\mu_E = \log_{10}(E_{\rm ref}/1\,\rm{MeV})$ of 4.1 rather than a value closer to $\sim$3.6, which generally fits well most pulsar spectra \citep{sbc+19}. No pulsations were identified from any of our sources above a detection threshold of 5\,$\sigma$, including PSRs~J1911$+$0925 and J1911$+$1051. We note that non-detections are somewhat anticipated here given that all but three pulsars have inferred distances $>\,3\,$kpc and are located in highly confused regions.

\section{PALFA and the Galactic plane population} \label{sec:discussion}
Most neutron stars in the Galactic field lie in the Plane; however, the higher background sky temperature and plasma density have hindered pulsar searches in low-latitude regions. One of the primary goals of the PALFA survey had been to probe the true population deep in the Galactic plane in order to disentangle the spatial distribution and ISM from pulsar properties -- essential for an accurate modeling of the underlying population. A thorough population synthesis is beyond the scope of this paper. Here, we compare some of the observed properties of the 206 pulsars discovered by PALFA to those of the population that was known up until PALFA observations ceased in 2020 August. Consequently, we do not take into account recent discoveries by other surveys, for example those reported by the FAST Galactic Plane Pulsar Snapshot survey \citep{hww+21}.  To limit biases due to selection effects in our comparison, we exclude sources located in globular clusters and the Magellanic Clouds and only consider field pulsars within 10$\degrees$ of the Galactic plane that were discovered in non-targeted searches, resulting in a set of 1624 (non-PALFA) radio pulsars. Their properties were taken from the ATNF catalog (version 1.64). The same set of 1624 known sources is used throughout the analysis described below, except when the parameter being examined is not available for a given object, in which case the pulsar is excluded from the set.  


\subsection{Flux density}
\begin{figure*}[t]
	\centering
    \includegraphics[scale=0.70]{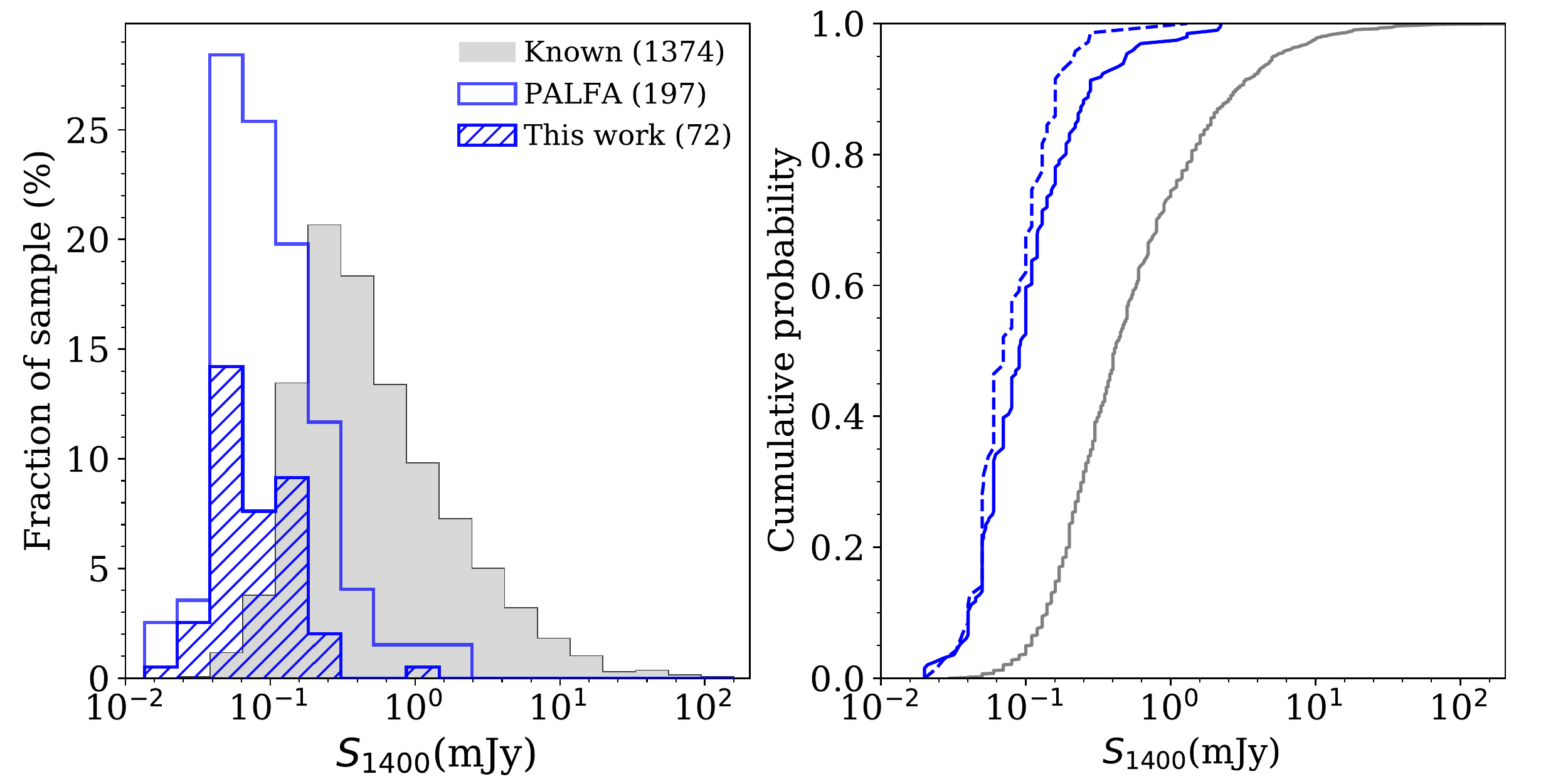}
    \caption{ Probability density (\textit{left}) and cumulative (\textit{right}) distributions of average pulsed flux densities at 1400\,MHz ($S_{1400}$) of field pulsars within $|b|<10\degrees$ discovered in radio surveys other than PALFA (grey). PALFA sources are shown in blue, and the hatched pattern (\textit{left})/ dashed line (\textit{right}) corresponds to the pulsars featured in this work. Values of $S_{1400}$ for non-PALFA objects were taken from the ATNF catalog (v. 1.64). 
    }
    \label{fig:flux}
\end{figure*}

As a result of Arecibo's large collecting area, the PALFA survey was able to detect pulsars to much lower flux densities than the average known population. In Figure~\ref{fig:flux}, we compare the distribution of average pulsed flux densities $S_{1400}$ of the PALFA sample to the known set. For pulsars presented in this work, we used the $S_{1400}$ values presented in Tables~\ref{tab:new} and \ref{tab:derived}. Sources with no reported $S_{\rm 1400}$ in the ATNF catalog were excluded from the sample. The median $S_{\rm 1400}$ value for PALFA discoveries is 100\,$\mu$Jy (70\,$\mu$Jy for sources presented in this work), compared to 480\,$\mu$Jy for the other known pulsars in the Plane. 

We are interested in the null hypothesis that the PALFA sample is statistically consistent with being the lower end of the brightness distribution of the known population in the Plane. The null hypothesis is tested using a two-sided Anderson-Darling (A-D) test. We chose the A-D statistics over the commonly used Kolmogorov–Smirnov test as the former is more sensitive to deviations in the tails of distributions that depart from Gaussianity \citep{s74}, which is where our interest lies. To extract accurate and stable statistics, empirical distribution functions for both the PALFA and known pulsar samples were generated through 10$^3$ iterations of bootstrap resampling. We find that the probability $p$ of $S_{\rm 1400}$ values for PALFA discoveries being drawn from the known population is $< 0.001$. Combined with the distribution of average pulsed flux densities shown in Figure~\ref{fig:flux}, we thus conclude that PALFA was successful in probing a fainter population.

\subsection{Distance and luminosity}
We are now interested in determining whether the low flux density of PALFA discoveries is a result of the pulsars being located deep within the Plane or if they are low-luminosity objects. This essentially consists of examining the sample's distribution in terms of distances, $D$, and inferred luminosity at 1400\,MHz, $L_{\rm 1400}$, given the inverse square law. We derive distances using both the NE2001 ($D_{\rm NE2001}$) and YMW16 ($D_{\rm YMW16}$) Galactic electron density models, except for 97 pulsars for which precision distance measurements have been reported. For the remaining sources, luminosities ($L_{\rm 1400} = 4\pi S_{\rm 1400} D^2$) are calculated on a per-model basis. Pulsars with DM values exceeding the maximum Galactic contribution to the DM predicted along their respective lines of sight have poorly constrained distances, and are thus removed from their respective sample. That is the case for two PALFA-discovered pulsars whose DMs exceed the maximum Galactic DM predicted by the NE2001 model (PSR~J1901$+$0459, DM\,=\,1108.0\,pc\,cm$^{-3}$, \citealt{lbh+15};  PSR~J2005$+$3547, DM\,=\,401.6\,pc\,cm$^{-3}$,  \citealt{nab+13}). We note however that, in both cases, the amount of excess DM is small ($\sim$ 30 \,pc\,cm$^{-3}$) and within the typically assumed 25\% uncertainty level for NE2001 predictions. 
As for the known pulsars sample, none have DMs larger than the maximum DM predicted by NE2001 but six do exceed the YMW16 predictions (PSR~B1714$-$34, DM=587.7\,pc\,cm$^{-3}$, \citealt{jlm+92}; PSR~J0837$-$24, DM=142.8\,pc\,cm$^{-3}$, \citealt{bbj+11}; PSR~J1305$-$6256, DM=967\,pc\,cm$^{-3}$, \citealt{mlc+01}; PSR~J1321$-$5922, DM=383\,pc\,cm$^{-3}$, \citealt{kel+09}; PSR~J1324$-$6146, DM=828\,pc\,cm$^{-3}$ and PSR~J1637$-$4335, DM=608\,pc\,cm$^{-3}$, \citealt{kbm+03}). 
\begin{figure*}[t]
	\centering
	\vspace*{-3.0mm}
    \includegraphics[scale=0.52]{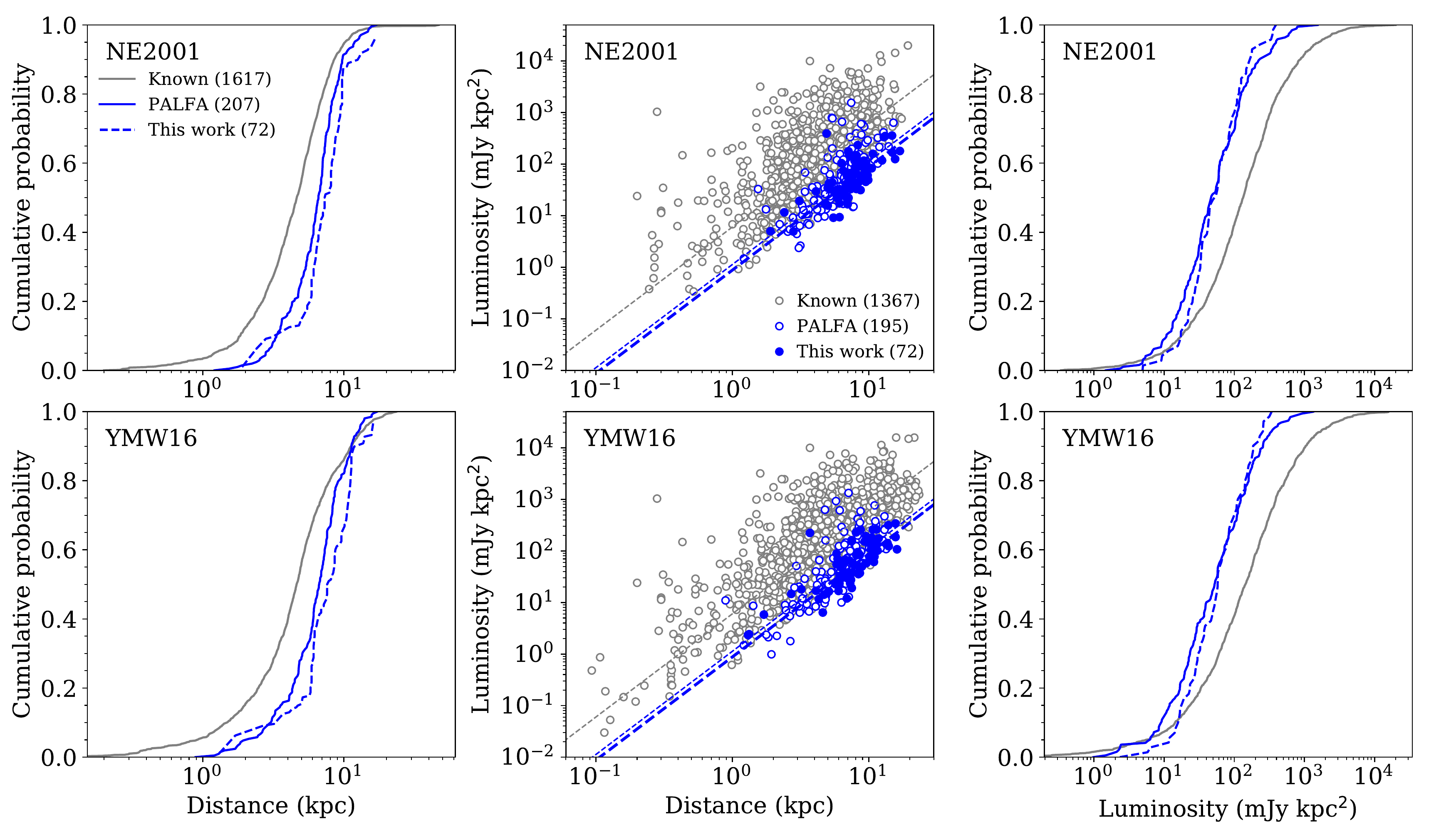}
    \vspace*{-5.0mm}
    \caption{ Distribution of DM-predicted distances and inferred luminosities at 1400\,MHz of known pulsars within $10\degrees$ of the Galactic plane discovered in non-targeted radio searches, computed for both the NE2001 (\textit{top}) and YMW16 (\textit{bottom}) models. Luminosities were estimated using DM distances and $S_{1400}$ values, and should therefore be interpreted with caution. In blue are PALFA discoveries, and sources discovered by other surveys are shown in grey. Lines in the middle panels correspond to lines of constant flux densities, valued at the median $S_{1400}$ for each sample (see text). Values of distances and luminosities for non-PALFA objects were computed from the DM, position and $S_{1400}$ measurements reported in the ATNF catalog (see text).}
    \label{fig:lum_dist}
\end{figure*}

In Figure~\ref{fig:lum_dist}, we compare the distribution in distances and luminosities of PALFA discoveries against the known pulsars. PALFA-discovered objects have a median distance of 6.8\,kpc (for both electron density models), about 1.5 more distant than the median value of 4.6\,kpc for the other radio pulsars in the Plane. Pulsars featured in this work are even more distant, with median distances of $D_{\rm NE2001}\,=\,7.3\,$kpc and $D_{\rm YMW16}\,=\,7.6\,$kpc. 

We perform A-D tests again with the bootstrap resampling method for the NE2001 and YMW16 models separately and find that in both cases, the probability that the DM-derived distances of PALFA sources are drawn from the same distribution as the known sample is $p<0.02$. Unsurprisingly, the A-D test also rejects the null hypothesis 
for luminosities ($p<0.01$). We therefore conclude that PALFA has indeed uncovered a distant population deep in the Galactic plane, but cannot independently evaluate the contribution of the intrinsic luminosity distributions to the observed flux densities. However it is important to stress here that both parameters used in estimating luminosities have large uncertainties, and distances inferred from DMs can be incorrect by factors of 5 or more (e.g., \citealt{dgb+19}). The above results should therefore be interpreted with caution.  

\subsection{Propagation effects}
\begin{figure*}[ht!]
	\centering
    \includegraphics[scale=0.65]{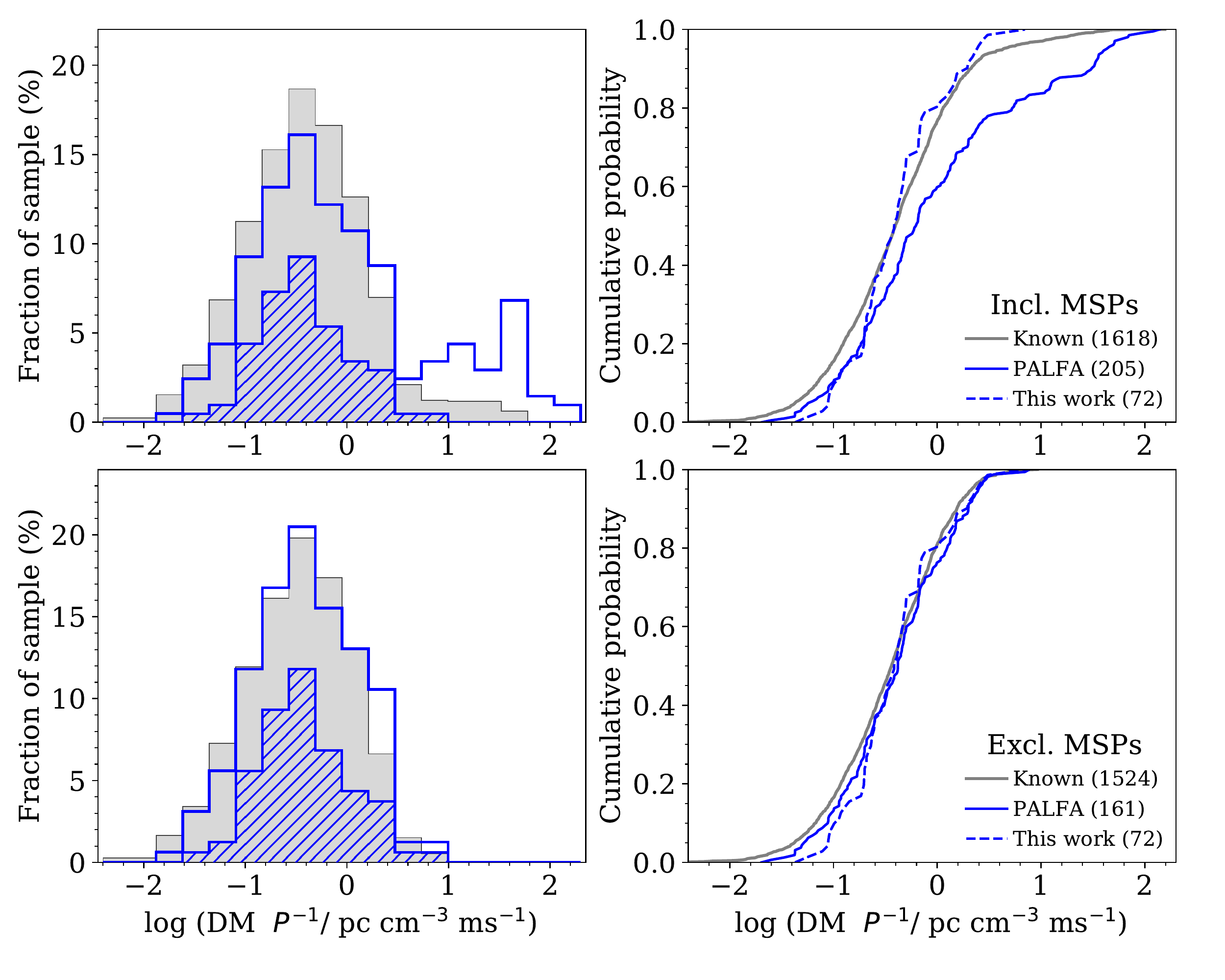}
    \vspace*{-5mm}
    \caption{Distribution of DM/$P$ ratios of pulsars, including (\textit{top}) and excluding (\textit{bottom}) MSPs, within $|b|< 10\degrees$ of the Galactic plane discovered in non-targeted radio searches. PALFA discoveries are shown in blue and in grey are sources found by surveys other than PALFA. The hatched regions/dashed lines in blue are pulsars presented in this work. RRATs with no available $P$ measurements were excluded from both samples.\\
    }
    \label{fig:dm_over_p}
\end{figure*}
We now explore how propagation effects have impacted the observed population in the Plane. To mitigate the effect of dispersion smearing, PALFA has observed the Galactic plane at 1400\,MHz with high spectral (0.33\,MHz) and temporal (64\,$\mu$s) resolutions.  Because the PALFA sample has very few sources showing significant scattering, we only investigate the effect of pulse dispersion. In terms of DM distribution, an A-D test indicates that the DMs of PALFA sources are consistent with the known population ($p\sim$ 0.1). However this is not very informative since selection biases arising from propagation in the ISM only become significant when pulse broadening becomes comparable to the intrinsic pulse width. The irreversible intra-channel smearing due to dispersion is detrimental to survey sensitivities, even more so for fast-spinning (i.e., highly recycled) pulsars because of their small pulse widths. Hence, examining the measured DMs versus $P$s of known pulsars provides insight into observational biases caused by the ISM affecting a survey. 

In Figure~\ref{fig:dm_over_p}, we show the distribution of the DM/$P$ ratios (where DM is in pc\,cm$^{-3}$ and $P$ is in ms) of PALFA and non-PALFA pulsars discovered in the Galactic plane. 
One clear distinction between the two sets (in the top panels) is the presence of a significant second mode at high DM/$P$ values in the PALFA sample, which are primarily MSPs and make up for $\sim$\,20\% of the entire survey yield whereas MSPs found through blind radio surveys (excluding PALFA sources) represent 6\% of the known pulsars in the Plane\footnote{Radio searches targeting the position of unidentified \textit{Fermi} point sources have led to an increase in MSP discoveries this past decade (e.g., \citealt{rrc+11,kcj+12,ckr+15,fcp+21}).}. Moreover, of the three known pulsars with $\log{{\rm DM}/P}\,>2$, two were discovered by PALFA (J1850$+$0244; \citealt{skl+15} and J1903$+$0327; \citealt{crl+08,fbw+11}).  

When MSPs are excluded (bottom panels), the A-D test suggests that there is no difference in the DM/$P$ distributions of the PALFA and known samples ($p\,\gtrsim$\,0.25), whereas when MSPs are included, the A-D test probability drops to $p\,<\,0.01$. Hence, the superior ability of PALFA to mitigate pulse dispersion and hence probe deep into high electron density regions has only had a statistically significant impact in uncovering the highly dispersed MSP population within the Plane. \\
\section{Survey status} \label{sec:srv_status}
On 2020 August 10, operations at Arecibo stopped when an auxiliary cable slipped out of its socket atop one of three towers supporting the platform, causing damage to the dish below. This was the first cable failure of a series that led to the collapse of the platform on 2020 December 1. PALFA therefore terminated\footnote{PALFA observations were last conducted on 2020 August 9, a few hours prior to the first cable break.} before completion. Here, we provide an updated accounting of the survey sky coverage and completeness for potential use in future population synthesis work.

To optimize the use of telescope resources, PALFA observed commensally with the Zone of Avoidance \citep{hsm+10} and the radio recombination lines \citep{lmt+13} projects. Observations in the outer Galaxy were led by our commensal partners, while we led observations in the inner Galaxy. Our pointing strategy for the inner Galaxy region prioritized $| b |\,<\,3\degrees$ regions before moving on to higher Galactic latitudes. To maximize efficiency and sensitivity, our pointing grid sampled the sky by interleaving three ALFA pointings such that the gain at any sky position was equal to or greater than the receiver half maximum gain (see \citealt{cfl+06}). Rather than maximizing the sky coverage, our commensal partners adopted a strategy where certain longitude/latitude ranges are densely sampled -- most pointings in the outer galaxy region were re-observed several times. While that approach is not optimal for deep pulsar surveying in terms of survey speed, searching those data sets has benefits for the detection of erratic sources. For instance, one of our FRBs and four RRATs were detected only once in outer galaxy pointings that were observed three times or more. 
\begin{table*}
    \centering
    \caption{Census of PALFA survey data. Only Mock beams that were successfully processed and analyzed are listed. Likewise, the WAPP data sets are excluded from the sky coverage and completeness values. We note that each ALFA pointing consists of seven beams that are processed independently. As such, the values provided here represent the number of searched beams rather than pointings. That provides an accurate accounting by excluding failed beam(s) from a pointing. The difference in the number of beams versus unique sky positions are due to re-observations of a pointing, which is especially significant for outer galaxy data sets.  }
    \setlength{\tabcolsep}{2.5mm}
    \begin{tabular}{l r r c c c c}
    \hline
      & No. beams &No. unique&Sky coverage& Completeness,& Completeness,& Completeness,\\
      &  &sky positions&(sq. deg.)& $|b|< 2^\circ$(\%) & $|b|< 3^\circ$(\%) & $|b|< 5^\circ$(\%) \\ 
 \hline\hline
  Inner Galaxy  & 86705 & 84846 & 218 & 95 & 89 & 71 \\[-0.1em]
  Outer Galaxy  & 123234 & 44465 & 114 & 37 & 30 & 23 \\
 \hline
 
\end{tabular}

\label{tab:srv}

\end{table*}


From 2004 to 2009, PALFA observations were recorded with the Wide-Band Arecibo Pulsar Processor (WAPP) spectrometers \citep{dsh00}, which processed 100-MHz passbands centered at 1.42\,GHz for each of the seven ALFA beams. We found a total of 56 pulsars and covered 200 sq.\,deg. of the sky with WAPP: 110 sq.\,deg. in the inner galaxy region (38 normal pulsars, ten MSPs and four RRATs), and 90 sq.\,deg. in the outer region (three normal pulsars and one RRAT). More details on the WAPP data collection volumes can be found in \cite{akc+13}.

\begin{figure*}
	\centering
    \includegraphics[scale=0.55]{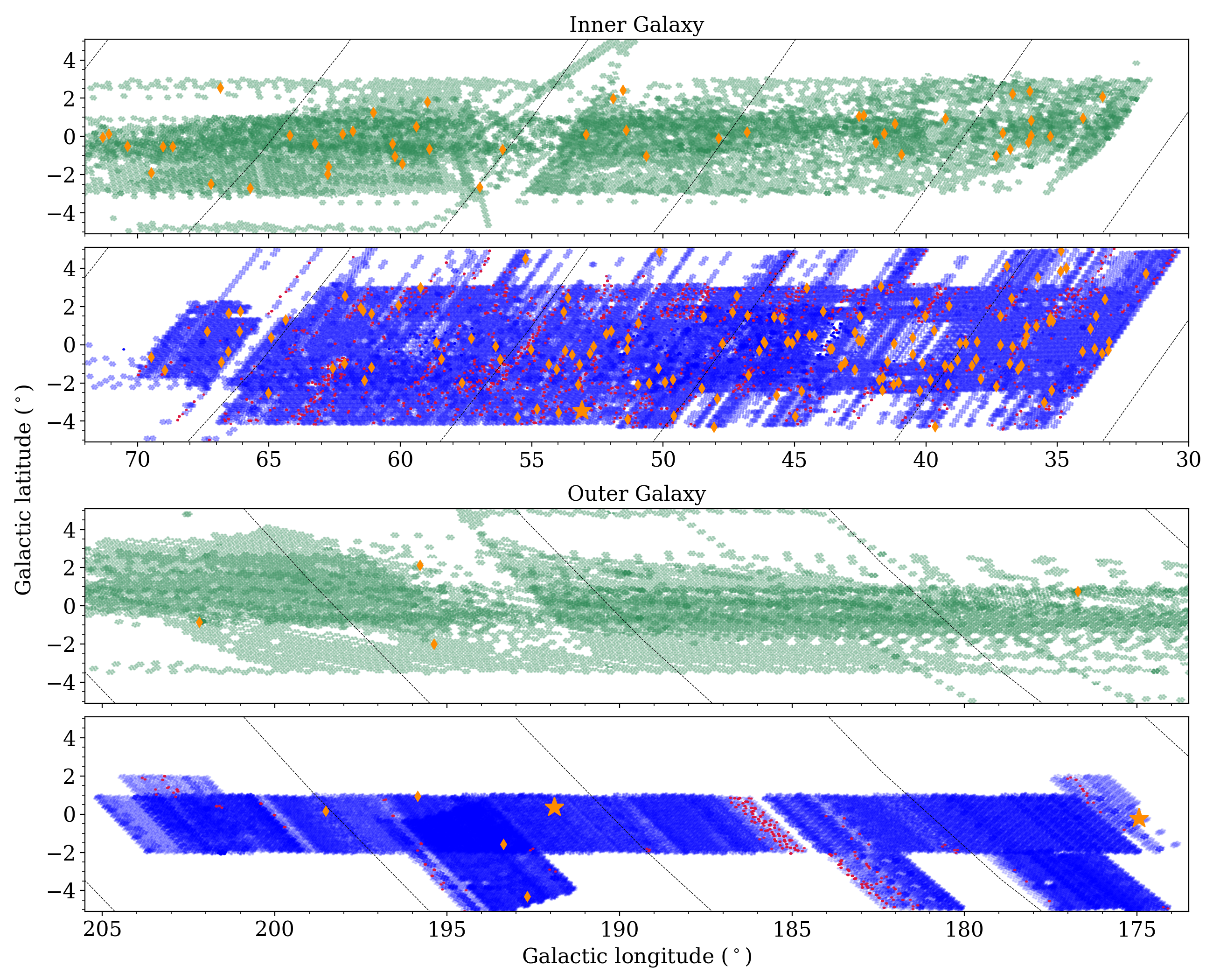}
    \vspace*{-5.0mm}
    \caption{Sky map showing the sky position of analyzed PALFA beams. Beams recorded with the WAPP (first and third panels) and Mock (second and fourth panels) spectrometers are shown in green and blue, respectively, and red points correspond to beams for which processing failed due to high levels of RFI.  Observing density at each sky position scales with the marker opacity. All pulsars discovered by the survey are shown as orange diamonds, while the three orange stars are FRB discoveries. Dashed black lines are lines of constant declination. }
    \label{fig:srv}
\end{figure*}
In mid-2009, we switched to using the Mock spectrometers, which provided an observing bandwidth three times wider than that of the WAPP system (see Section~\ref{sec:obs-disc} for more details on the Mock instrument parameters). The increased bandwidth, multi-bit depth and poly-phase filterbank design of the Mock spectrometers improved the survey sensitivity and increased its robustness to RFI. In fact, our discovery rate was nearly doubled even though there was some overlap in the WAPP and Mock sky coverage. With Mock, we discovered a total of 151 pulsars and three FRBs: 147 pulsars (including 35 MSPs and eleven RRATs) and one FRB in the inner region and four pulsars (including one MSP and two RRATs) and two FRBs in the outer region, after covering 218 and 114 sq.\,deg. in the inner and outer regions, respectively. 

Approximately 16\% of the WAPP pointings were re-observed with Mock during the first year of transition and it was our intention to eventually re-observe all remaining WAPP pointings so that the  sensitivity of the survey would be uniform across the PALFA sky. A breakdown of the Mock survey data, sky coverage and completeness for both the inner and outer regions is provided in Table~\ref{tab:srv}. Values presented in the table only include data sets that have been processed and inspected, which represents approximately 96\% of the total number of beams collected with Mock. The remaining 4\% could not be searched, largely due to severe RFI contamination which  rendered the data unusable. 

Figure~\ref{fig:srv} shows the sky locations of the beams that were searched, collected with either Mock (blue) or WAPP (green), as well as the positions of failed beams (red). It also shows the positions of our pulsar (orange) and FRB (yellow) discoveries. We provide the lists of all processed beams and relevant information as Supporting Information with the online version of the paper.  Further details regarding processing of WAPP and Mock data can be found in \cite{cfl+06} and \cite{lbh+15}, respectively. Information relating to detections of known sources, which we do not discuss here, will be reported in a future publication.

\section{Conclusions} \label{sec:conclusion}

In this work, we have presented the results of long-term monitoring on 72 pulsars found by the PALFA survey, plus estimates of their emission properties: DMs, scattering times, and flux densities. All of these quantities are important for characterizing the population.
Pulsar timing is especially important since it allows estimates of fundamental properties of the neutron star like the dipolar magnetic field strength, characteristic age, spin-down power, glitching behavior and possible membership of binary systems.

Overall, the pulsars characterized here have much lower flux densities than the previously known population in the survey area, a demonstration of the sensitivity of the PALFA survey and also, in part, a consequence of the fact that these pulsars were not discovered by previous surveys that covered the search region. Furthermore, the pulsar sample in this work is also faint compared to the previously published PALFA slow pulsars: those were either found and timed earlier with the less sensitive WAPP spectrometers \citep{nab+13}, or were later timed at Jodrell Bank, which can only follow up the brighter systems \citep{lsf+17,lsb+17}. Thus, the pulsars presented in this work tend to be among the faintest PALFA discoveries, which are already quite faint in comparison to the known population.

These faint pulsars could, {\em a priori}, either be low-luminosity pulsars relatively nearby, or very luminous pulsars at very large distances. 
We find that the PALFA discoveries - and even more the group studied in this work - represent a low-luminosity population in comparison to the population previously discovered in the survey region. From their positions in the $P$ - $\dot{P}$ diagram, we additionally conclude that they represent an older, less energetic population than that previously known in the survey region.
This is not a selection effect against the discovery of young, fast-spinning pulsars: the PALFA survey has found many more MSPs than previous surveys in the survey area. In fact, because of RFI, the PALFA survey is biased against the detection of very slow pulsars \citep{lbh+15}. Without RFI, we probably would have found an even slower, older and less energetic population. Thus, the most luminous pulsars in the survey region were already discovered in previous, less sensitive surveys. A possible interpretation is that, for the population of luminous pulsars, those surveys were already covering the full useful volume of our Galaxy, i.e., the extra survey volume provided by the PALFA survey for this type of pulsar is likely beyond our Galaxy. Another consequence of this is that, for the population of highly luminous pulsars, the average distance has remained the same, since few new ones have been found. It is for the less luminous pulsar population that the extra survey volume provided by the PALFA survey is still within the Galaxy. Since those are being discovered at larger distances than was possible before, this results in significant increases in the average distances: this is  7-8 kpc for all new PALFA slow pulsars, significantly larger than the average 4.6 kpc for the previously known pulsars in the survey region.

Among the sources presented in this work are a wide variety of glitching, mode-changing, intermittent and nulling pulsars; we characterize the occurrence of these phenomena within our pulsar sample in some detail. Intermittent, mode-changing and nulling pulsars are especially common among the slow, older and fainter pulsar population being found by our survey \citep{nab+13,lsf+17,lsb+17}. In particular, the four new intermittent pulsars found by PALFA - J1855+0626 and J1952+2513 presented here, plus J1910+0517 and J1929+1357, described by \cite{lsf+17} - more than double the population of intermittent pulsars known before this survey. More detailed studies of these objects may help to improve our understanding of the physics of pulsar magnetospheres. The statistics of RRATs, nulling and intermittent pulsars are also fundamental for better estimates of the true size of the neutron star population in the Galaxy. Our results indicate that there are many more such objects in the Galaxy than have been detected thus far.

Also among our discoveries is a new binary pulsar, PSR~J1954+2529, with a wide ($P_{\rm b} = 82.7$ d) and eccentric ($e = 0.114$) orbit. This is a member of a relatively rare population of binary pulsars which appear to have not been recycled, and is similar to PSR~J1932+1500, which was also discovered by the PALFA survey \citep{lsb+17}. The origin of these systems, which we discuss in detail in this work, is still uncertain: some of them could be progenitors to LMXBs, while others could result from weakly interacting PSR - main sequence star systems. Understanding this is important: given the small characteristic ages of the pulsars in these systems (of the order of Myr, compared to Gyr for MSPs), it is clear that they form at comparable or even larger rates than the better known MSP binaries.

In this work we also describe the PALFA survey status, and summarize the final sky coverage. The pulsars described in this work represent approximately $\sim$1/3 of PALFA's discoveries to date: 142 normal pulsars, 46 MSPs, 19 RRATs and three FRBs. All data collected after 2015 have been searched with an upgraded full-resolution pipeline \citep{pkr+18,pab+18} -- reprocessing of Mock data collected prior to 2015 is ongoing, as is the Einstein@Home search of the later part of the survey data. Timing observations are being conducted with CHIME/Pulsar and Lovell on $\sim$15 MSPs (Parent et al. in prep, Haniewicz et al. in prep), eight normal pulsars and 15 RRATs (Doskoch et al. in prep) that were last discovered by PALFA (see Table~\ref{tab:new}) and for which timing solutions have not yet been determined. Thus, although the PALFA survey was cut short by the demise of the Arecibo 305-m telescope, we expect to report many further discoveries.
\acknowledgments The Arecibo Observatory is operated by the University of Central Florida, Ana G. Mendez-Universidad Metropolitana, and Yang Enterprises under a  cooperative agreement with the National Science Foundation (NSF; AST-1744119). Pulsar research at Jodrell Bank and access to the Lovell Telescope is supported by a Consolidated Grant from the UK’s Science and Technology Facilities Council. The National Radio Astronomy Observatory is a facility of the National Science Foundation operated under cooperative agreement by Associated Universities, Inc. We are grateful to the staff of the Dominion Radio Astrophysical Observatory, which is operated by the National Research Council of Canada. CHIME is funded by a grant from the Canada Foundation for Innovation (CFI) 2012 Leading Edge Fund (Project 31170) and by contributions from the provinces of British Columbia, Qu\'{e}bec and Ontario. The CHIME/FRB Project, which enabled development in common with the CHIME/Pulsar instrument, is funded by a grant from the CFI 2015 Innovation Fund (Project 33213) and by contributions from the provinces of British Columbia and Qu\'{e}bec, and by the Dunlap Institute for Astronomy and Astrophysics at the University of Toronto. 

Additional support was provided by the Canadian Institute for Advanced Research (CIFAR), McGill University and the McGill Space Institute thanks to the Trottier Family Foundation, and the University of British Columbia. The CHIME/Pulsar instrument hardware was funded by NSERC RTI-1 grant EQPEQ 458893-2014. This research was enabled in part by support provided by WestGrid, Compute Canada and Calcul Qu\'ebec. Work at the Naval Research Laboratory was supported by the NASA Fermi program. This work was supported by the Max Planck Gesellschaft and by NSF grants
0555655, 1104617, 1104902, 1105572 and 1816904.

We thank all Einstein@Home volunteers, especially those whose computers found pulsars with the highest statistical significance. PSR~J1851$+$0241: Mel S.\ Stark, Somonauk, Illinois, USA and ``TRON''. PSR~J1853$+$0029: ``[TiDC] Chulma - S'inergy'' and Rodolfo Manalac, Queens College, CUNY, Flushing, New York, USA. PSR~J1855$+$0306: Jeroen Moetwil, Flagstaff, Arizona, USA and Robert E.\ Inman Jr., Virginia Beach, Virginia, USA. PSR~J1859$+$0345: James S.\ Drews, UW-Madison, Wisconsin, USA and ``juergenstoetzel''. PSR~J1909$+$1205: Alexandr Jungwirth, Prague, Czech Republic and ``Administrator''. PSR~J1910$+$0710: John Murphy, Austin, Texas, USA and ``Grey''. PSR~J1910$+$1017: ``mglogan'' and Dave and Emma Johnston, Jacksonville, Florida, USA. PSR~J1911$+$0925: Matthias Pfister, Gland, Switzerland and Ryan D.\ Morton, Ann Arbor, Michigan, USA. PSR~J1914$+$1428: John-Luke Peck, Bellevue, Washington, USA and Mark Henderson, Russellville, Tennessee, USA. PSR~J1948$+$2819: ``Chuck Claybaugh'' and ``Orange$\_$Crunch''. PSR~J1952$+$2513: ``Joe'' and Jarmo Kahila, Vantaa, Finland. PSR~J1953$+$2819: ``Rojer'' and ``testkermit''. PSR~J1955$+$2930: ``Przemek Wisialski'' and Gary S.~ Grant II, Nelsonville, Ohio, USA.

We thank the anonymous referee for their careful reading of the manuscript and helpful suggestions. We also thank Thomas Tauris for a useful comments on the evolution of PSR~J1954+2529, and Rosie Chen for the discussion on the infrared non-detection of its companion. We also thank the FAST GPPS survey collaboration for noticing that we were reporting harmonic frequencies for two pulsars in the initial version of this manuscript. 

EP is a Vanier Canada Graduate Scholar. MAM, FC, AB, SC, JMC, EF, BWM, SMR, and IHS are members of the NANOGrav Physics Frontiers Center, which is supported by the NSF award PHY~1430284. SC and JMC acknowledge support from the National Science Foundation (AAG~1815242). FAD is supported by the University of British Columbia Four Year Doctoral Fellowship. VMK holds the Lorne Trottier Chair in Astrophysics \& Cosmology and a Distinguished James McGill Professorship and receives support from an NSERC Discovery Grant and Herzberg Award, from an R. Howard Webster Foundation Fellowship from the Canadian Institute for Advanced Research (CIFAR), and from the FRQNT Centre de Recherche en Astrophysique du Qu\'ebec. JWM is a CITA Postdoctoral Fellow and is supported by the Natural Sciences and Engineering Research Council of Canada (CITA~490888-16). SMR is a CIFAR Fellow. 

\software{\texttt{PRESTO} \citep{presto} available at \url{https://github.com/scottransom/presto},
\texttt{PSRCHIVE} \citep{hvsm04} available at \url{http://psrchive.sourceforge.net/index.shtml}, 
\texttt{TEMPO} \citep{nds+15}, ASCL: \url{https://ascl.net/1509.002})}

\bibliography{palfa.bib}{}
\bibliographystyle{aasjournal}

\end{document}